\documentclass[acmsmall]{acmart}
\usepackage{svg}
\usepackage{amsmath}
\usepackage{cleveref}
\usepackage[autostyle]{csquotes}
\usepackage[british]{babel}
\usepackage[most]{tcolorbox}
\newtcbtheorem{Definition}{Definition}{
  enhanced,
  sharp corners,
  attach boxed title to top left={
    yshifttext=-1mm
  },
  colback=white,
  colframe=gray!75!black,
  fonttitle=\bfseries,
  boxed title style={
    sharp corners,
    size=small,
    colback=gray!75!black,
    colframe=gray!75!black,
  } 
}{def}
\newtcbtheorem[no counter]{Proof}{Proof}{
  enhanced,
  sharp corners,
  attach boxed title to top left={
    yshifttext=-1mm
  },
  colback=white,
  colframe=gray!25,
  fonttitle=\bfseries,
  coltitle=black,
  boxed title style={
    sharp corners,
    size=small,
    colback=gray!25,
    colframe=gray!25,
  } 
}{prf}

\AtBeginDocument{%
  \providecommand\BibTeX{{%
    \normalfont B\kern-0.5em{\scshape i\kern-0.25em b}\kern-0.8em\TeX}}}

\setcopyright{acmcopyright}
\copyrightyear{2021}
\acmYear{2021}
\acmDOI{to be replaced when assigned DOI}

\acmJournal{CSUR}
\acmVolume{00}
\acmNumber{0}
\acmArticle{000}
\acmMonth{12}

\begin{document}

\title{Down the Rabbit Hole: Detecting Online Extremism, Radicalisation, and Politicised Hate Speech}


\author{Jarod Govers}
\orcid{0000-0002-7648-318X}
\email{jg199@students.waikato.ac.nz}
\affiliation{
  \institution{ORKA Lab, Department of Software Engineering, University of Waikato}
  \streetaddress{Gate 1, Knighton Road}
  \city{Hamilton}
  \state{Waikato}
  \country{NZ}
  \postcode{3216}
}

\author{Philip Feldman}
\orcid{0000-0001-6164-6620}
\email{philip.feldman@asrcfederal.com}
\affiliation{
  \institution{ASRC Federal}
  \city{Beltsville}
  \state{Maryland}
  \country{US}
}

\author{Aaron Dant}
\orcid{0000-0001-5852-5262}
\email{aaron.dant@asrcfederal.com}
\affiliation{
  \institution{ASRC Federal}
  \city{Beltsville}
  \state{Maryland}
  \country{US}
}

\author{Panos Patros}
\orcid{0000-0002-1366-9411}
\email{panos.patros@waikato.ac.nz}
\affiliation{%
  \institution{ORKA Lab, Department of Software Engineering, University of Waikato}
  \streetaddress{Gate 1, Knighton Road}
  \city{Hamilton}
  \state{Waikato}
  \country{NZ}
  \postcode{3216}
}

\renewcommand{\shortauthors}{Govers et al.}

\begin{abstract}
Social media is a modern person’s digital voice to project and engage with new ideas and mobilise communities---a power shared with extremists. Given the societal risks of unvetted content-moderating algorithms for \textit{Extremism}, \textit{Radicalisation}, and \textit{Hate speech} (ERH) detection, responsible software engineering must understand the who, what, when, where, and why such models are necessary to protect user safety \textit{and} free expression. Hence, we propose and examine the unique research field of \textit{ERH context mining} to unify disjoint studies. Specifically, we evaluate the start-to-finish design process from socio-technical definition-building and dataset collection strategies to technical algorithm design and performance. Our 2015-2021 51-study Systematic Literature Review (SLR) provides the first cross-examination of textual, network, and visual approaches to detecting \textit{extremist} affiliation, \textit{hateful} content, and \textit{radicalisation} towards groups and movements. We identify consensus-driven ERH definitions and propose solutions to existing ideological and geographic biases, particularly due to the lack of research in Oceania/Australasia. Our hybridised investigation on Natural Language Processing, Community Detection, and visual-text models demonstrates the dominating performance of textual transformer-based algorithms. We conclude with vital recommendations for ERH context mining researchers and propose an uptake roadmap with guidelines for researchers, industries, and governments to enable a safer cyberspace.
\end{abstract}

\ccsdesc[500]{Computing methodologies~Discourse, dialogue and pragmatics}
\ccsdesc[500]{Computing methodologies~Lexical semantics}
\ccsdesc[500]{Computing methodologies~Information extraction}
\ccsdesc[500]{Computing methodologies~Machine learning}
\ccsdesc[300]{Applied computing~Sociology}
\ccsdesc[500]{Social and professional topics~Hate speech}
\ccsdesc[500]{Social and professional topics~Technology and censorship}
\ccsdesc[300]{Social and professional topics~Political speech}
\ccsdesc[100]{Social and professional topics~Governmental regulations}
\ccsdesc[500]{Computing methodologies~Natural language processing}
\ccsdesc[500]{Computing methodologies~Knowledge representation and reasoning}
\ccsdesc[100]{Computing methodologies~Image and video acquisition}
\keywords{extremism, radicalisation, machine learning, community detection, natural language processing, neural networks, hate speech, sociolinguistics}

\maketitle
\pagebreak
\tableofcontents

\section{Introduction}

Online social media empowers users to communicate with friends and the wider world, organise events and movements, and engage with communities all at the palm of our hands. Social media platforms are a frequent aid for modern political exchanges and organisation~\cite{tucker17_social_media_democracy}, with extremes amplified by algorithmic recommendation systems, such as on Twitter~\cite{huszar2021algorithmic}, TikTok~\cite{isd21_hatescape_extreme_right_tiktok}, and YouTube~\cite{Lewis18_AlternativeInfluence}. Furthermore, the semantic expression of ideas differs between social media platforms, with Twitter’s short 280 character limit resulting in more narcissistic and aggressive content compared to Facebook~\cite{twitter18_hate_vs_facebook}, and anonymous platforms such as 4Chan instilling a vitriolic \textquote{group think} in political threads/\textquote{boards}~\cite{ludemann18_polemics_4chan}. Hateful, emotive, and \textquote{click-worthy} content permeates virtual discourse, which can radicalise users down an ideological rabbit hole towards real-world violent action. The individuals behind the 2014 Isla Vista and 2019 Christchurch shootings appeared as individual \textquote{lone-wolf attacks} without an allegiance. However, investigations found a deep network of perverse and violent communities across social media~\cite{isla_vista_incel_communities, nzsis_chch_report}. Likewise, exploiting social media to plan politically motivated attacks towards the civilian population to coerce political change (i.e., \textit{terrorism}) delegitimises democracies, social cohesion, and physical/mental health~\cite{Kinnvall21_psychology_extremist_identification, meindl17_mass_shootings_imitation, collins19}.

As a response, social media platforms employ text and visual content-moderation systems to detect and remove hate speech, extremism, and radicalising content. This paper offers a state-of-the-art Systematic Literature Review (SLR) on the definitions, data collection, annotation, processing, and model considerations for \textit{Extremism}, \textit{Radicalisation}, and \textit{Hate speech} (ERH) detection.

\subsection{Motivation and Contributions}

Existing ERH literature reviews exist as independent microcosms, often focusing on specific types of models, typically text-only Natural Language Processing (NLP) models or non-textual network analysis via community detection models. Studies seldom \textit{cross-examine} models and evaluate the performance between non-textual network analysis (a \textquote{who-knows-who} approach), textual, and/or multimedia approaches for ERH detection. While we identified ten prior literature reviews for hateful content detection, none consider the similarities and definitional nuance between ERH concepts and what Extremism, Hate Speech, or Radicalisation means in practice by researchers~\cite{Aldera21_onlineextremism_textual_review, Gaikwad21_online_extremism_detection_review_tools, gaikwad2020bibliometric, Adek21_systematics_radical_extremist_review, Rini20SysLitReviewOfHateTextMining, Istaiteh20_randshatespeechlitreview, agarwal2015_onlineradical_civilunrest_review, Mullah21_ml_hate_areview, stephens21socialsciencelitreview, borum2011radicalizationsocialsciencelitreview}. Evaluating the consensus for ERH definitions, dataset collection and extraction techniques, model choice and performance are all essential to create ethical models without injurious censorship or blowback.

Through understanding the groups, beliefs, data, and algorithms behind existing content-moderation models---we can reliably critique often overlooked social concepts, such as algorithmic bias, and ensure compliance between social definitions and computational practice. Hence, this new field of \textit{ERH context mining} extracts the \textit{context} to classifications---enabling researchers, industries, and governments to assess the state of social discourse. 

Given the rise of state-sponsored disinformation campaigns to undermine democratic institutions and social media campaigns, the time is now for ERH research within politicised discussions.

\textbf{The three core contributions for this paper are:}

\begin{enumerate}
\item The establishment of consensus-driven working definitions for Extremism, Radicalisation, and Hate Speech within the novel field of \textit{ERH context mining}---and a proposed framework/roadmap for future researchers, social media platforms, and government advisors.
\item The critical examination of existing textual (NLP), network (community detection), and hybrid text-image datasets.
\item The identification and cross-examination of the state-of-the-art models' performance on benchmark datasets and relevant challenges with the current ERH detection metrics.
\end{enumerate}

\subsection{Structure}

For a high-level summary of this SLR's findings, refer to Section~\ref{section:rq-summaries} on \textit{Key Research Question Findings}, and Section~\ref{section:futureworkERH} for \textit{Recommendations for Future Work}. These key summaries condense and contextualise the 51 studies observed between 2015-2021, which we use to build our proposed computational ERH definitions and technological roadmap for researchers, industry, and government in Section~\ref{section:futureworkERH}.

For a holistic understanding, we present a social context to our motivations in Section~\ref{section:background}. Related work and areas our SLR improves on are outlined in Section~\ref{section:related-work}. Section~\ref{section:slr-design} outlines the systematic protocol used to collect the 51 studies between 2015-2021. Further to the summaries presented in Section~\ref{section:rq-summaries}, we present an in-depth analysis and cross-examination of studies definitions of ERH concepts in Section~\ref{section:rq1}, approaches for collecting and processing data in Section~\ref{section:rq2}, algorithmic approaches for classification in Section~\ref{section:rq3}, and their performance in Section~\ref{section:rq4}. We conclude with recommendations for future SLRs, and studies in Section~\ref{section:futureworkERH}, and conclusions in Section~\ref{section:conclusion}.

\section{Social Context to Social Network Analysis}
\label{section:background}
Analysing social media requires the socio-technical considerations of what constitutes \textit{hate speech}, \textit{extremism}, and \textit{radicalisation} (ERH). To detect such concepts, computational models can investigate multimodal sources---including textual meaning and intent through \textit{Natural Language Processing (NLP)}, \textit{computer vision} for images, and evaluating user relationships through \textit{community detection}. Hence, this section decouples and analyses ERH's social background and definitions.

\subsection{Extremism and Radicalisation Decoupled}
Extremism's definition appears in two main flavours: politically fringe belief systems outside the social contract or violence supporting organisational affiliation.
The Anti-Defamation League (ADL) frames extremism as a concept "used to describe religious, social or political belief systems that exist substantially outside of belief systems more broadly accepted in society"~\cite{adl_extremism_def}. For instance, under the ADL's definition, extremism can be a peaceful \textit{positive} force for mainstreaming subjugated beliefs, such as for civil rights movements. This construct of a socially mainstream belief constitutes the \textit{Overton window}~\cite{overton_window}---and is not the target for content moderation.

Conceptually, extremism typically involves hostility towards an apparent \textquote{foreign} group based on an opposing characteristic or ideology. Core tenants of extremism can stem from political trauma, power vacuums and opportunity, alongside societal detachment and exclusion~\cite{Kinnvall21_psychology_extremist_identification, mcauley14_terrorist_psychology}. Hence, extremism often relies on defending and congregating people(s) around a central ideology, whose followers and devotees are considered \textquote{in-group}~\cite{turner86_in_and_out_group_psych}. Extremists unify through hostility and a perceived injustice from an \textquote{out-group} of people(s) that do not conform to the extremist narrative---typically in a \textquote{us vs. them} manner~\cite{mcauley14_terrorist_psychology, turner86_in_and_out_group_psych, collins19}. Hence, extremism detection algorithms can use non-textual relationships as an identifying factor via clustering users into communities (i.e., \textit{community detection})~\cite{stephens21socialsciencelitreview, borum2011radicalizationsocialsciencelitreview}. Thus, extremism can simply reduce to \textit{any} form of a fringe group whose identity represents the vocal antithesis of another group.

There is no one conceptual factor to make an extremist. Extremism can also emanate from political \textit{securitisation}–-whereby state actors transform a specific referent object (such as Buzan's five dimensions of society: societal, military, political, economic, and environmental security~\cite{buzan98_securitisation_coppenhagen_book}; or individuals and groups~\cite{collins19}) towards matters of national security, requiring extraordinary political measures~\cite{buzan98_securitisation_coppenhagen_book}. As the state normalises policies into matters of existential national security, society can adapt and ideate decisions to ones of existential ‘life or death’ nature~\cite{buzan98_securitisation_coppenhagen_book}.

For example, the ‘Great Replacement’ conspiracy theory claims that non-European immigrants and children are “colonizers" or "occupiers”, and an “internal enemy”–--with the intent to securitise migration, race, religion, and culture into wars with wording to invoke fears of a fifth column or racial invasion/replacement~\cite{bracke20_great_replacement, nzsis_chch_report}. The Christchurch Shooter took direct interest in securitising migrants as an \textit{extreme} military threat, as far to name his manifesto after the conspiracy~\cite{nzsis_chch_report}.

Extremism is not a strictly demographic ‘majority vs minority’ concern, as it encapsulates movements demanding radical change and earmarked by a sense of rewarding personal and social relationships, self-esteem, and belief of a wider purpose against a perceived adversarial force~\cite{Kinnvall21_psychology_extremist_identification}. Exploiting desires for vengeance and hostility are also key recruitment strategies~\cite{Kinnvall21_psychology_extremist_identification, ludemann18_polemics_4chan, borum2011radicalizationsocialsciencelitreview}.

Outside of political, cultural, and socio-economic factors, mental health and media are intrinsically inalienable contributing factors~\cite{meindl17_mass_shootings_imitation, Kinnvall21_psychology_extremist_identification, collins19}. Likewise, repeated media reports of footage and body counts can gamify and normalise extremism as a macabre sport for notoriety~\cite{meindl17_mass_shootings_imitation, borum2011radicalizationsocialsciencelitreview}.  

Within \textit{industry}, Facebook, Twitter, YouTube, and the European Union frame extremism as a form of indirect or direct support for civilian-oriented and politically motivated \textit{violence} for coercive change~\cite{eu_code_of_conduct_16}. Facebook expands this industry-wide consensus to include Militarised Social Movements and "violence-inducing conspiracy networks such as QAnon"~\cite{facebookhate_stnd}.

Radicalisation focuses on the process of ideological movement towards a different belief, which the EU frames as a "\textit{phased and complex process} in which an individual or a group embraces a radical ideology or belief that accepts, uses or condones violence"~\cite{eu_radicalisation_def}.
Terrorism consists of politically motivated violence towards the civilian population to coerce, intimidate, or force specific political objectives, as an end-point for violent \textit{radicalisation} to project \textit{extremism}~\cite{collins19, buzan98_securitisation_coppenhagen_book}. Borum delineates the \textit{passive ideological movement} of radicalisation from \textit{active} decisions to \textit{engage} in \textquote{action pathways} consisting of physical \textit{terrorism}, or hate crimes~\cite{borum2011radicalizationsocialsciencelitreview}. 

Political radicalisation towards increasingly aggrandising groups can also manifest in Roe’s two sides of nationalism: positive socio-cultural and negative ethnic/racial nationalism~\cite{roe05_ethnic_violence_societal_security_dilemma}. These balancing forces create a form of societal security dilemma whereby the actions of one society to strengthen its identity can cause a reaction in another societal group, weakening security between all groups-–-a radicalising spiral which can manifest into a polarised ‘culture war’~\cite{roe05_ethnic_violence_societal_security_dilemma, ludemann18_polemics_4chan}. However, integration over assimilation can inversely undermining culture, self-expression and group cohesion, leading to alienation and oppression by the dominant political or normative force~\cite{collins19, roe05_ethnic_violence_societal_security_dilemma}.

Nonetheless, ERH detection does not offer a panacea to combating global terrorism, nor does surveillance offer a \textquote{catch-all} solution. In the case of the livestreamed Christchurch shooter, the New Zealand Security Intelligence Service concluded that “the most likely (if not only) way NZSIS could have discovered [the individual]’s plans to conduct what is now known of [the individual]’s plans to conduct his terrorist attacks would have been via his manifesto.”~\cite[p. 105]{nzsis_chch_report}. However, the individual did not disseminate this until immediately before the attack, and his 8Chan posts did not pass the criteria to garner a search warrant~\cite[p. 105]{nzsis_chch_report}. Hence, extremism detection is an evolutionary arms race between effective and ethical defences vs. new tactics to evade detection.

\subsection{Hate Speech Decoupled}
Obtaining viewpoint neutrality to categorise hate speech is challenging due to human biases and the risk of hate speech undermining liberties through mainstreaming intolerance---the paradox of tolerance where a society tolerant without limit may have their rights seized by those projected intolerance~\cite{popper12_intolerance_paradox}. Popper encapsulates this challenge by formulating that "if we are not prepared to defend a tolerant society against the onslaught of the intolerant, then the tolerant will be destroyed, and tolerance with them"~\cite{popper12_intolerance_paradox}. Defining clear hate speech restrictions are needed to protect expression rights \textit{and} victim groups rights and safety~\cite{un_hate_speech_definition, baron04_hate_speech_socialsci_dynamics}.

The European Union defines hate speech as "all conduct publicly inciting to violence or hatred directed against a group of persons or a member of such a group defined by reference to race, colour, religion, descent or national or ethnic origin."~\cite{eu_code_of_conduct_16}. Whereas, the U.S. Department of Justice frames that: "A hate crime is a traditional offence like murder, arson, or vandalism with an added element of bias... [consisting of a] criminal offence against a person or property motivated in whole or in part by an offender’s bias against a race, religion, disability, sexual orientation, ethnicity, gender, or gender identity."~\cite{fbi_hate_crime_def} Notably, governmental laws may differ from industry content moderation policies via the omission of sexual, gender, religious or disability protections, and may include threats of violence \textit{and} non-violent but insulting speech.

The United Nations outlines the international consensus on hate speech as "any kind of communication in speech, writing or behaviour, that attacks or uses pejorative or discriminatory language with reference to a person or a group on the basis of who they are, in other words, based on their religion, ethnicity, nationality, race, colour, descent, gender or other identity factor."~\cite[p. 2]{un_hate_speech_definition}

What all these definitions have in common is that they all involve speech directed at a portion of the population based on a protected class.

\section{Systematic Literature Review Design and Protocol}
\label{section:slr-design}
This SLR investigates the state-of-the-art approaches, datasets, ethical, socio-legal, and technical implementations used for extremism, radicalisation, and politicised hate speech detection. We conduct a preliminary review of prior ERH-related SLRs to establish the trends and research gaps. 

For the purposes of our SLR's design, and to embed Open-Source Intelligence (OSINT) and Social Media Intelligence (SOCMINT) principles, we define \textit{social media} data as any online medium where users can interactively communicate, exchange or influence others. We accept external data sources, such as manifesto or news sites if interactive---such as via comment sections. Furthermore, we propose and use a novel quality assessment criteria to filter irrelevant or ambiguous studies.

\subsection{Trends and Shortfalls in Prior SLRs}
\label{section:related-work}

Searching for \textit{Extremism}, \textit{Radicalisation}, \textit{Hate speech} (ERH) and related terms, resulted in ten literature reviews ranging from January 2011 to April 2021~\cite{Aldera21_onlineextremism_textual_review, Gaikwad21_online_extremism_detection_review_tools, gaikwad2020bibliometric, Adek21_systematics_radical_extremist_review, Rini20SysLitReviewOfHateTextMining, Istaiteh20_randshatespeechlitreview, agarwal2015_onlineradical_civilunrest_review, Mullah21_ml_hate_areview, stephens21socialsciencelitreview, borum2011radicalizationsocialsciencelitreview}. Aldera et al. observed only one survey before 2011 (covering 2003-2011) and another in 2013, indicating the limited, exclusionary, but developing nature of reviews in this ERH detection area~\cite{Aldera21_onlineextremism_textual_review}.

Prior SLRs seldom delineated or elaborated on \textit{Extremism}, \textit{Hate Speech} and \textit{Radicalisation}. Neither “extremism”, "radicalism"~\cite{Gaikwad21_online_extremism_detection_review_tools, Adek21_systematics_radical_extremist_review, Aldera21_onlineextremism_textual_review, stephens21socialsciencelitreview, agarwal2015_onlineradical_civilunrest_review, gaikwad2020bibliometric, borum2011radicalizationsocialsciencelitreview} or “hate speech” oriented SLRs~\cite{Rini20SysLitReviewOfHateTextMining, Istaiteh20_randshatespeechlitreview, Mullah21_ml_hate_areview} cross-reference each other despite 26.3\% of the data reviewed in the “hate speech” oriented review by Adek et al. encompassing hate speech in a political context~\cite{Adek21_systematics_radical_extremist_review}. This lack of overlap presents an industry-wide challenge for social media companies who may oversee developments in \textquote{hate speech} detection which could transfer to a \textquote{extremism/radicalisation detection} model.

SLRs prior to 2015 found that deep learning approaches (DLAs), such as Convolutional Neural Networks (CNN) and Long Short-Term Memory (LSTM), resulted in 5-20\% lower F1-scores than non-deep approaches (e.g., Naïve Bayes, Support Vector Machines, and Random Forest classifiers)~\cite{Gaikwad21_online_extremism_detection_review_tools, Adek21_systematics_radical_extremist_review, Aldera21_onlineextremism_textual_review, agarwal2015_onlineradical_civilunrest_review, Rini20SysLitReviewOfHateTextMining, Istaiteh20_randshatespeechlitreview}. DLAs post-2015 indicated a pivotal change towards higher-performing language transformers such as Bidirectional Encoder Representations from Transformers (BERT) models~\cite{bert_original_study}.

\subsubsection{Domains and Criteria}\hfill\\
No review delineated or removed studies that did not use English social media data. This presents three areas of concern for researchers when attempting to compare the performance of models:
\begin{enumerate}
\item \textbf{Results may not be comparable}, if they use culture-specific lexical datasets, or language models trained on other languages.
\item \textbf{Linguistic differences and conveyance in language}---as what may be culturally appropriate for the majority class may appear offensive to minority groups and vice-versa.
\item \textbf{The choice of language(s) influences the distribution of target groups}---with a bias towards Islamic extremism given its global reach in both Western (predominantly ISIS) and Eastern countries (e.g., with studied online movements in the Russian Caucasus Region~\cite{mashechkin19_russia_caucasus_jihad_ml}). It is worth investigating whether Gaikwad et al.'s finding that 64\% of studies solely target \textquote{Jihadism} corroborates with our study, which targets only English data~\cite[p. 17]{Gaikwad21_online_extremism_detection_review_tools}.
\end{enumerate}

Our SLR incorporates the key approaches of dataset evaluation (including their accessibility, labels, source and target group(s)), data collection and scraping approaches, Machine Learning and Deep Learning algorithms, pre-processing techniques, research biases, and socio-legal contexts. Unlike prior SLRs, our SLR conceptualises all elements for ERH context mining---consisting of a user's ideological \textit{radicalisation} to an \textit{extremist} position, \textit{and} then projected via \textit{hate speech}.

\subsection{Research Questions}
Our Research Questions (RQ) investigate the full process of \textit{ERH Context Mining}---incl. data collection, annotation, pre-processing, model generation and evaluation. These RQs consist of:
\begin{enumerate}
\item What are the working definitions for classifying online \textit{Extremism}, \textit{Radicalisation}, and \textit{Hate Speech}?
\item What are the methodologies for collecting, processing and annotating datasets?
\item What are the computational classification methods for ERH detection?
\item What are the highest performing models, and what challenges exist in cross-examining them?
\end{enumerate}

Given the overlap of studies across prior SLRs targeting extremism \textit{or} radicalisation \textit{or} hate speech, RQ1 addresses the similarities and differences between researchers' definitions of ERH concepts and their computational classification approach. We dissect the ERH component of ERH context mining and propose consensus-driven working definitions.

RQ2 addresses the vital context for ERH models---the data used and features extracted or filtered out from it. Furthermore, identifying frequently used \textit{benchmark} datasets provides the basis for critical appraisal of the state-of-the-art algorithmic approaches in the \textit{community detection}, \textit{multimedia}, and \textit{NLP} spheres in RQ3/4.
Covering algorithmic approaches is not in itself novel. However, we consider novel, niche, and overlooked features relevant for an ERH model to make accurate classifications---namely, bot/troll detection, transfer learning, the role of bias, and a hybridised evaluation of NLP \textit{and} non-textual community detection models. We also consider critical challenges, choice of metrics, and performance considerations not observed in prior SLRs.

\subsection{Databases}
Given the cross-disciplinary, global and socio-technical concepts for ERH detection, we queried the following range of software engineering, computer science, crime and security studies databases.

\begin{itemize}
\item ProQuest (with the \textquote*{peer-reviewed} filter, including queries to the below databases)
\item Association for Computing Machinery (ACM) Digital Library
\item SpringerLink
\item ResearchGate
\item Wiley
\item Institute of Electrical and Electronics Engineers (IEEE) Xplore
\item Association for Computational Linguistics Portal
\item Public Library of Science (PLOS) ONE Database
\item Google Scholar---as a last line to capture other journals missed in the above searched databases
\end{itemize}

\subsection{Search Strings}
The first round of study collection included automated database search strings. A second round included a targeted manual search strategy with dissected keyword combinations to expand coverage. All results were added to the Title and Abstract Screening list. The following database search query also included time filters (2015-2021) and peer-review-only filters:

\begin{verbatim*}
(“artificial intelligence” OR “machine learning” OR “data mining” OR “natural 
language processing” OR “multiclass classification” OR “model” OR “analysis” OR 
“intelligence” OR “modelling” OR “detection”) 
AND (“hate speech” OR “radicalisation” OR “radicalization” OR “extremism”) 
AND (“social media” OR “forums” OR “comments”OR “virtual networks” OR “virtual 
communities” OR “online communities” OR “posts” OR “tweets” OR “blogs”)
\end{verbatim*}

\subsection{Inclusion and Exclusion Criteria}
After attaining our 251 studies from our search strings, we read the journal metadata, title and abstract to screen studies. Our ranked criteria requires that all studies must:

\begin{enumerate}
\item Originate from a peer-reviewed journal, conference proceeding, reports, or book chapters.
\item Be written in English.
\item Involve a computational model (network relationship, textual and/or visual machine learning model) for identifying and classifying radicalisation, hate speech or extreme affiliation. 
\item Utilise social media platform(s) for generating their model.
\item Computationally identify ERH via binary, multiclass, clustering or score-based algorithms.
\item Focus on politicised discourse to exclude cyber-bullying or irrelevant benign discussions.
\item Published after the 1st January 2015---until the 1st July 2021.
\item Utilise English social media data if evaluating semantics and grammatical structure.
\end{enumerate}

In addition to those that do not abide to any of the above, we exclude studies that:
\begin{enumerate}
\item Are duplicates of existing studies.
\item Do not specify their target affiliation to exclude broad observational studies.
\end{enumerate}

We outline our further in-depth critical Quality Assessment (QA) criteria to filter irrelevant or ambiguous studies in our supplementary material’s \textit{Quality Assessment Criteria} subsection.

After the \textit{Title and Abstract Screen}, we read the full text of the 57 studies for the screening stages displayed in Table~\ref{tab:studies_found_and_filtered}. With 42 studies passing the \textit{Full Text Screen}, we then randomly selected studies from the bibliographies from this \textit{'snowball sample'} of the 42 studies until 5 studies fail QA.
\begin{table}[!ht]
  \caption{Studies found and filtered}
  \label{tab:studies_found_and_filtered}
  \small
  \begin{tabular}{lll}
    \toprule
    Screen Type&Study Count\\
    \midrule
    Search Strings&251\\
    Title and Abstract Screen&57\\
    Full Text Screening&42\\
    After Snowball Sampling&51\\
  \bottomrule
\end{tabular}
\end{table}

\subsubsection{Threats to Validity}\hfill\\
While we consider a concerted range of search strings, we recognise that ERH concepts is a wide spectrum. To focus on manifestly hateful, politicised, and violent datasets/studies, we excluded cyber-bullying or emotion-detection studies. The potential overlap and alternate terms for ERH (e.g., sexism as \textquote{misogyny classification}~\cite{coria20_metric_bert_misogyny}) could evade our search strings. Our pilot study, subsequent tweaks to our search method, and snowball sampling minimise this lost paper dilemma.

This study does not involve external funding, and all researchers declare no conflicts of interest.

\section{Key Research Question (RQ) Findings}
\label{section:rq-summaries}
Across the 51 studies between 2015-2021, ERH research is gaining popularity---with 4 studies from 2015-2016 increasing to 25 between 2019-2020 (and 4 studies from January to July 2021). We present our SLR's core findings in this section, with in-depth RQ analysis in Sections~\ref{section:rq1},~\ref{section:rq2},~\ref{section:rq3}, and~\ref{section:rq4}.

\subsection{Summary of the Social ERH Definitions Used by Researchers}
\textit{RQ1: What are the working definitions for computationally classifying online Extremism, Radicalisation, and Hate Speech?}\\
Across the 51 studies, there are seldom delineations between the researchers choice of \textit{Extremism}, \textit{Radicalisation}, and \textit{Hate Speech} as the study’s focus--–with the consensus that hate speech is equivalent to extremist or radical views. Hence, researchers approach extremism and radicalisation as an organisationally affiliated form of hate speech.

The consensus on hate speech’s working definition is any form of subjective and derogatory speech towards protected characteristics expressed directly or indirectly in textual form–--predominantly via racism or sexism. Benchmark datasets utilise human-annotated labels on single-post instances of racially or gender-motivated straw man arguments, stereotyping, or post causing offensive towards the majority of annotators (via inter-annotator agreement). Only 20\% of studies consider explicit rules or legal frameworks for defining hate speech~\cite{mashechkin19_russia_caucasus_jihad_ml, Nouh19_understanding_radical_mind, benigni2017online_extremism_sustain_it_isis, hall20_machines_unified_understanding_radicalizing, bilbao18_political_classification_cnn, johnston20identifying_extremism_dl, kapil20_deep_nn_multitask_learning_hs, waseem16hateful_symbols_hs_twitter, macavaney19hs_challenges_solutions, zampieri_etal19_OLID_baseline_tests}, with others relying on either an implicit ‘consensus’ on hate speech or utilise benchmark datasets. Benchmark datasets typically consider categorising their data into explicit categories of racism~\cite{waseem16_are_you_racist_dataset, waseem16hateful_symbols_hs_twitter, Davidson17_benchmark_hs_study, de-gibert18-white-supremacy-dataset}, sexism~\cite{waseem16_are_you_racist_dataset, waseem16hateful_symbols_hs_twitter, basile19_benchmark_mulitlingual_immigrant_women_dataset}, aggression~\cite{kumar_etal18_trac_benchmark_dataset, basile19_benchmark_mulitlingual_immigrant_women_dataset}, or offensiveness~\cite{Davidson17_benchmark_hs_study, zampieri_etal19_OLID_baseline_tests, basile19_benchmark_mulitlingual_immigrant_women_dataset}; including hate categorisation via visual memes and textual captions~\cite{kiela20_hateful_memes_original, multioff_meme_images, rudinac17_graph_conv_networks, Aggarwal21_hateful_meme}.

Extremism and radicalisation are equivalent terms in existing academia. Islamic extremism is the target group in 77\% of US-originating extremism studies. \textquote{Far-right extremism} and \textquote{white supremacy} are used interchangeably, a form of cultural bias given the variety of right-wing politics worldwide. Only one study considered radicalisation as an ideological movement \textit{over time}~\cite{bartal20_roles_trolls_affiliation}.

\subsection{Summary of the Data Collection, Processing, and Annotation Processes}
\begin{figure}[!ht]
    \centering
    \begin{minipage}[b]{0.33\linewidth}
        \includegraphics[width=\textwidth]{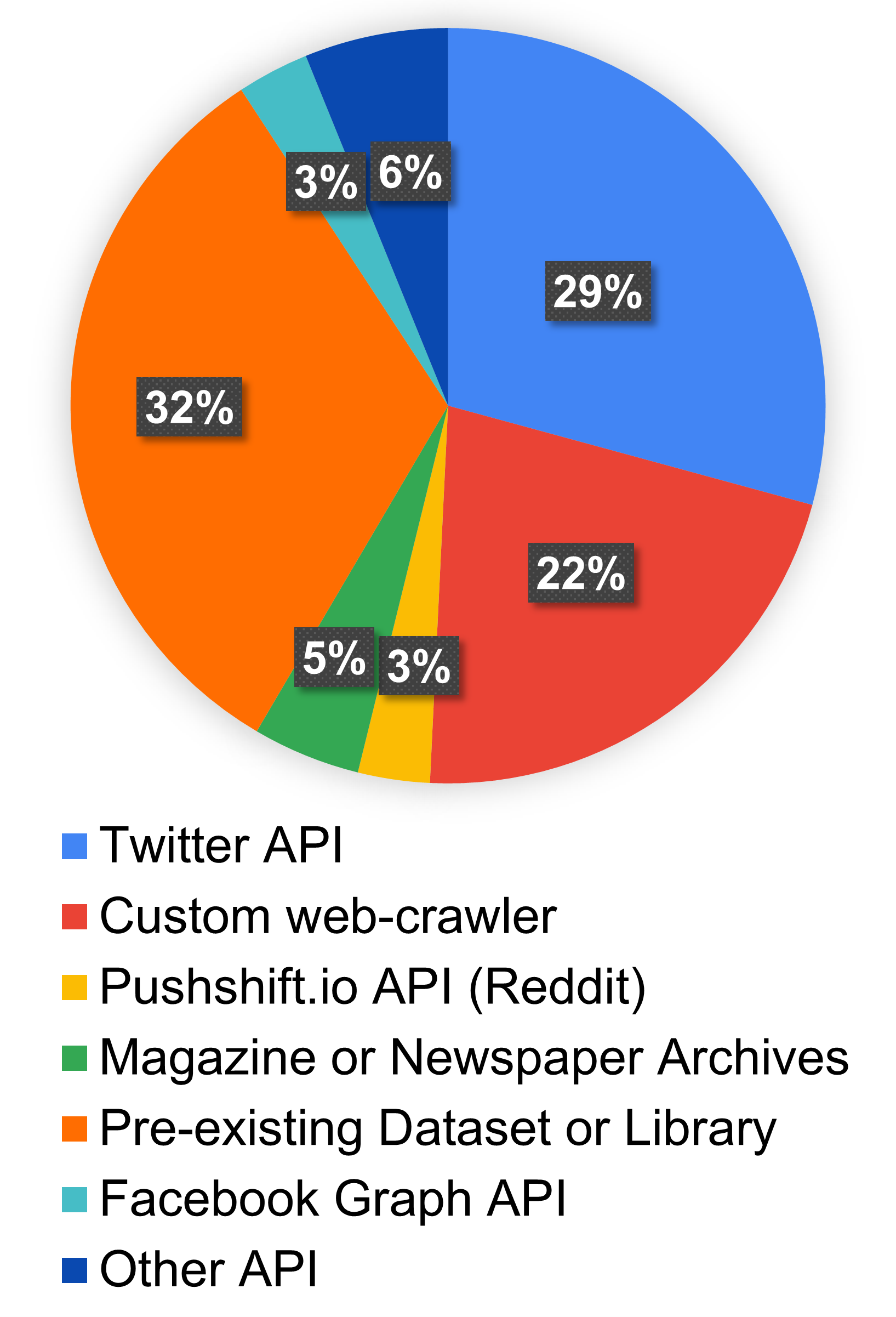}
        \caption{Method to collect data.}
        \label{RQ2-Method-of-Pulling-Data-Pie-Chart}
    \end{minipage}
    \begin{minipage}[b]{0.66\linewidth}
        \includegraphics[width=\textwidth]{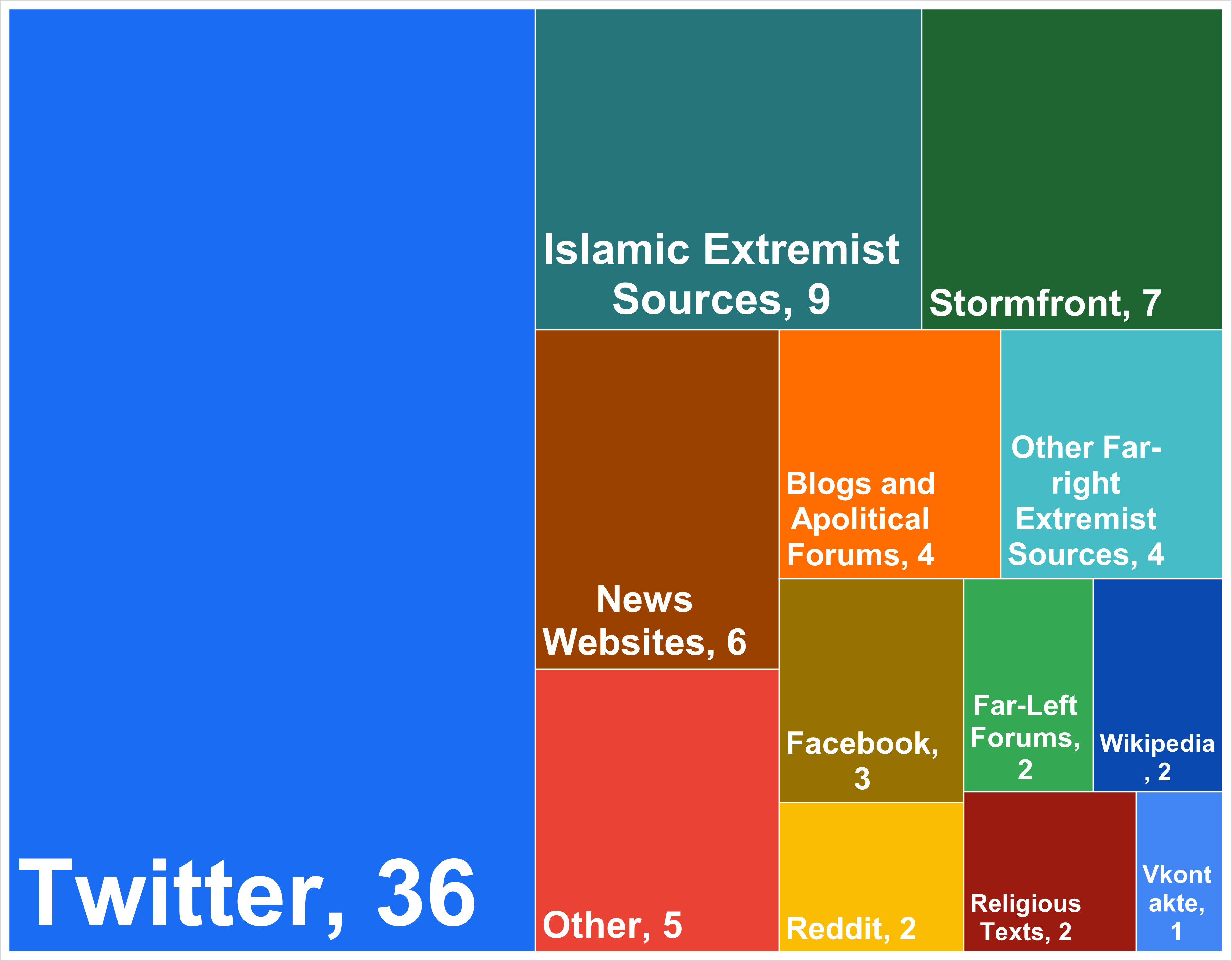}  
        \caption{Number of studies per social media platform studied.}
        \label{Platforms-studied-size-map}
    \end{minipage}
\end{figure}

\textit{RQ2: What are the methodologies for collecting, processing and annotating datasets?}\\
Collecting non-hateful and hateful ERH instances varies between supervised and unsupervised (clustering) tasks. Supervised learning typically utilises manual human annotation of textual posts extracted via tools presented in Figure~\ref{RQ2-Method-of-Pulling-Data-Pie-Chart}. Semi or unsupervised data collection can include grouping ideologies by platform, thread, or relation to a suspended extremist account. Islamic extremism studies frequently used manifestos and official Islamic State magazines as a ‘ground truth’ for textual similarity-based approaches for extremism detection. We found a direct correlation between the availability of open and official research tools, and the platform of choice by researchers. Biases extend geographically, with no studies utilising data or groups from Oceanic countries.

\begin{figure}[!ht]
    \centering
    \begin{minipage}[b]{0.49\linewidth}
        \includegraphics[width=\textwidth]{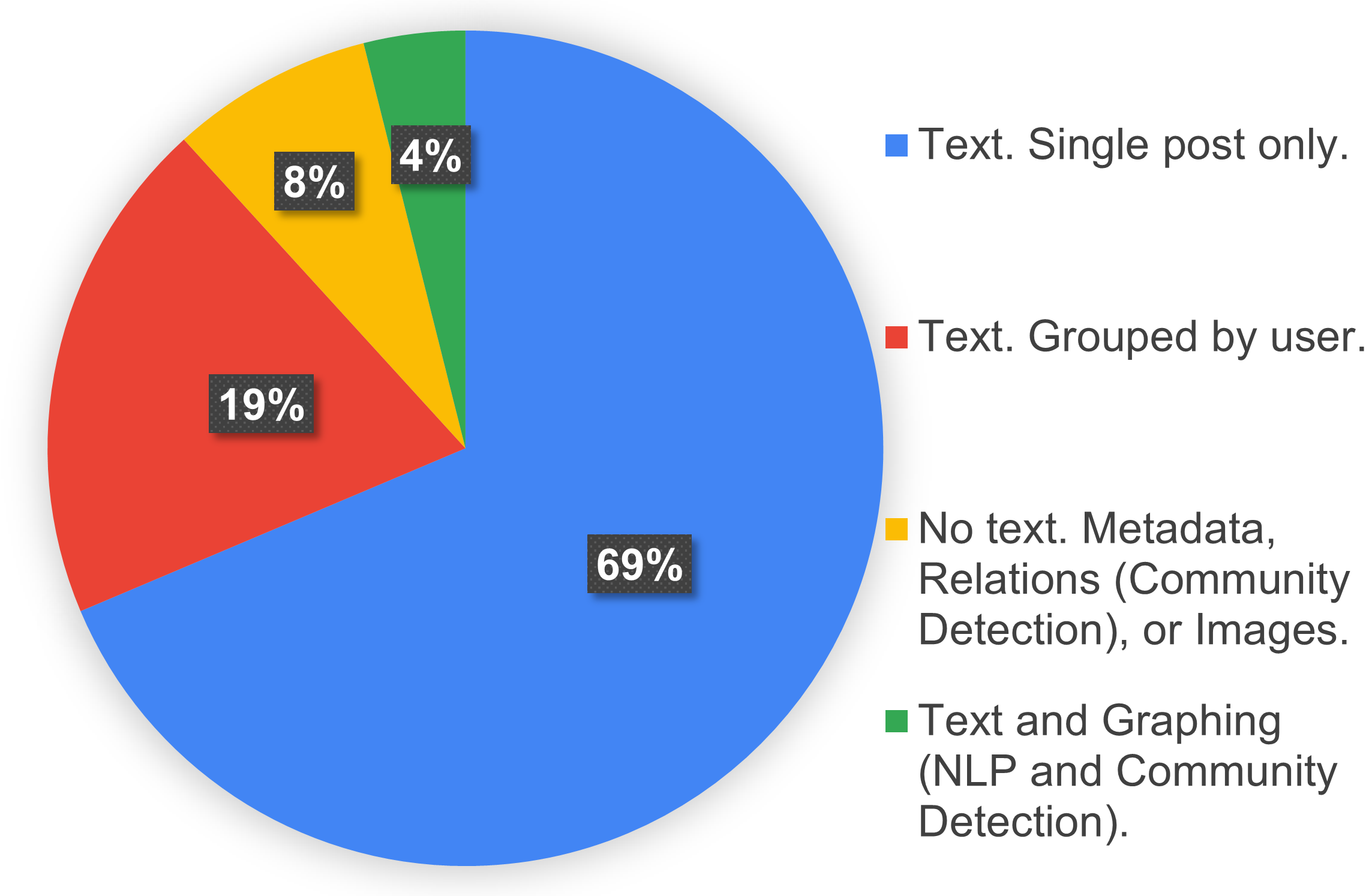}
        \caption{Type of data used for an ERH classification.}
        \label{RQ2-3-Type-of-Data-Used-Pie-Chart}
    \end{minipage}
    \hfill
    \begin{minipage}[b]{0.49\linewidth}
        \includegraphics[width=\textwidth]{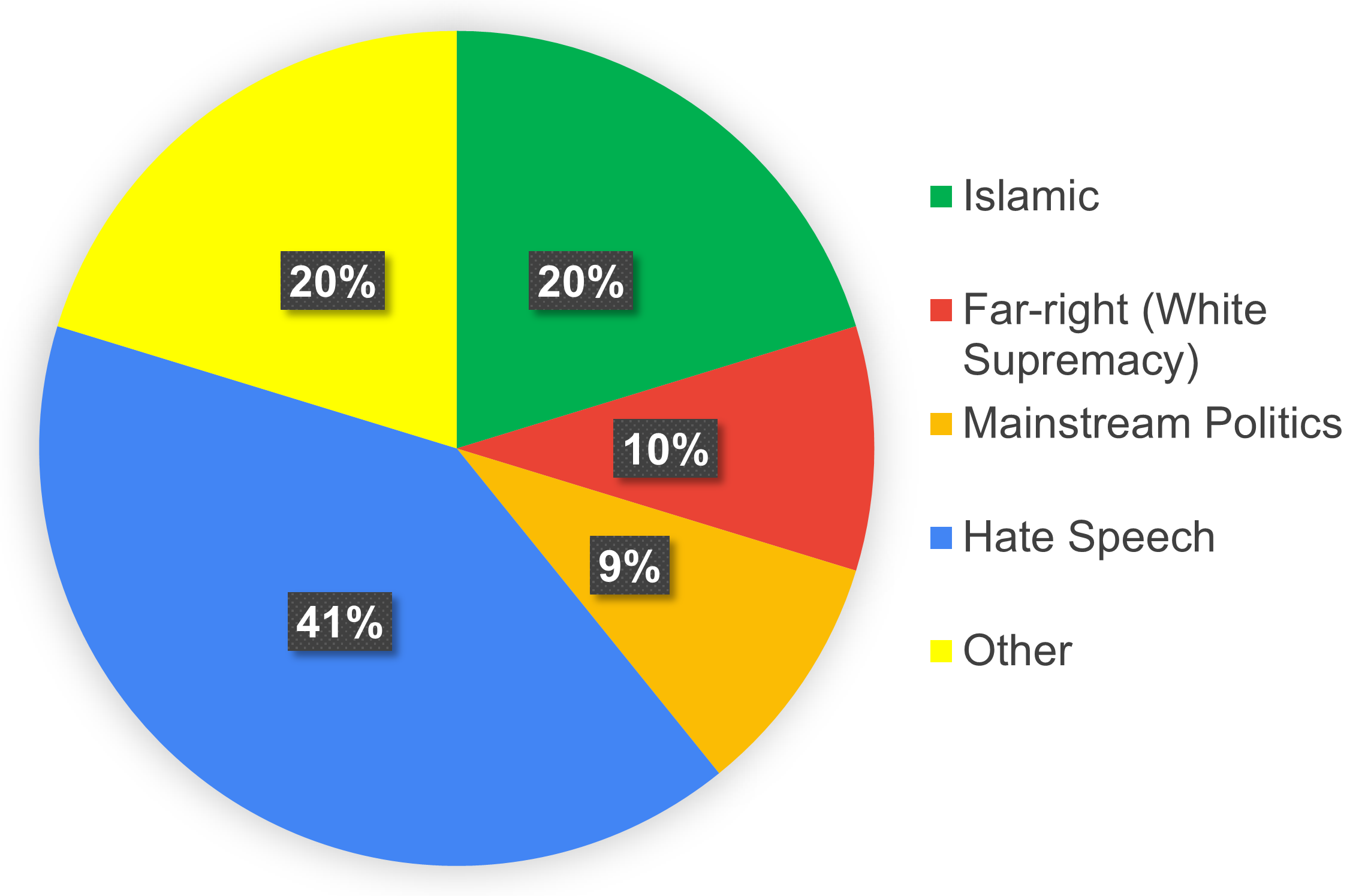}
        \caption{Distribution of target groups.}
        \label{RQ2-Target-Groups-Pie-Chart}
    \end{minipage}
\end{figure}

Figure~\ref{Platforms-studied-size-map} displays the skew for Twitter as the dominant platform for hate speech research. Despite the nuance of conversations, 69\% of studies classify hate on a single post per Figure~\ref{RQ2-3-Type-of-Data-Used-Pie-Chart}.

Data processing often utilises extracting statistically significant ERH features---such as hateful lexicons, emotional sentiment, psychological signals, ‘us vs them’ mentality (higher occurrence of first and third-person plural words~\cite{Grover19_alt-right_subreddits}), and references to political entities.

We categorise and frame the two approaches for dataset annotation: \textit{organisational} or \textit{experience-driven} annotation. Organisational annotation utilises non-governmental anti-hate organisations~\cite{hatebase} or ‘expert’ annotator panels---determined via custom tests or by tertiary degree. Organisational annotation relies on crowdsourced annotators, balanced by self-reported political affiliation. Inter-rater agreement or Kappa coefficient are the sole metrics for measuring annotator agreement.

\subsection{Summary of the State-of-the-art Computational ERH Detection Models}
\textit{RQ3: What are the computational classification methods for ERH detection?}\\
ERH detection includes text-only Natural Language Processing (NLP), network-related community detection, and hybrid image-text models. Between 2015-2021, there was a notable shift from traditional Machine Learning Algorithms (MLAs) towards contextual Deep Learning Algorithms (DLAs) due to higher classification performance---typically measured by macro F1-score.

Notably, only 3 of the 21 community detection studies utilised Deep Learning Algorithms (DLAs)~\cite{mashechkin19_russia_caucasus_jihad_ml, Nouh19_understanding_radical_mind, rudinac17_graph_conv_networks}. Instead, community detection researchers tended to opt for graph-based models such as heterogeneous graph models converting follower/following, reply/mention, and URL networks with numeric representations for logistic regression or decision trees~\cite{bartal20_roles_trolls_affiliation, benigni18communitymining_UNSUPERVISED, Nouh19_understanding_radical_mind, Moussaoui19_twitter_terrorism_communities, buchanan17_ethics_critique_twitter_study, hung16_fbi_radical}. Community detection Machine Learning Algorithms (MLAs) performance varied by \textasciitilde0.3 F1-score (mean between studies) dependent on the selection of features. Statistically significant features for performant MLA models include gender, topics extracted from a post's URL(s), location, and emotion via separate sentiment algorithms such as ExtremeSentiLex~\cite{pais20_unsupervised_extreme_sentiments} and SentiStrength~\cite{sentistrength_original_study}.

For textual non-deep NLP studies, researchers classified text via converting the input into word embeddings via Word2Vec, GloVe, or frequent words via Term Frequency-Inverse Document Frequency (TF-IDF), and parsing it into Support Vector Machines, decision trees, or logistic regression models. As these embeddings do not account for word order, context and nuance is often lost---leading to higher false positives on controversial political threads. Conversely, DLAs utilise \textit{positional} and \textit{contextual} word embeddings for context-\textit{sensitivity} using Long Short Term Memory (LSTM) Convolutional Neural Networks and Bidirectional Encoder Representations from Transformers (BERT) leading to their higher performance as outlined in RQ4.

\subsection{Summary of ERH Models' Classification Performance}
\textit{RQ4: What are the highest performing models, and what challenges exist in cross-examining them?}\\
By 2021, Support Vector Machines on emotional, sentiment, and derogatory lexicon features were the last non-deep MLA to attain competitive F1-scores for NLP tasks compared to DLAs such as Convolutional Neural Networks (CNNs) and neural language transformers. As of 2021, BERT-base attained the highest macro F1-score average across the seven benchmark datasets. However, cross-examining models \textit{between} datasets present various challenges due to varying criteria, social media domain, and choice of metrics. Likewise, non-textual community detection and traditional MLA studies resulted in lower classification F1-score by \char`\~0.15 and \char`\~0.2 respectively. While BERT, attention-layered Long Short Term Memory (BiLSTM), and other ensemble DLAs attain the highest F1-scores, no studies consider their performance trade-offs with their high computational complexity. Our recommendations propose further research in prompt engineering, distilled models, and hybrid multimedia-text studies---as we only identified one hybrid image-text study.

While textual DLAs outperform community detection models, grouping unknown instances enable network models to identify bot networks and emergent terror cells. Hence, there is a growing area of research for hybrid semi-supervised NLP and community detection models to identify new groups and radical individuals in a domain we frame as \textit{meso-level} and \textit{micro-level} classification.

\subsection{Geographic Trends---Islamophobia and Exclusion in the Academic Community}

\begin{figure}[!ht]
  \centering
  \includegraphics[width=\linewidth]{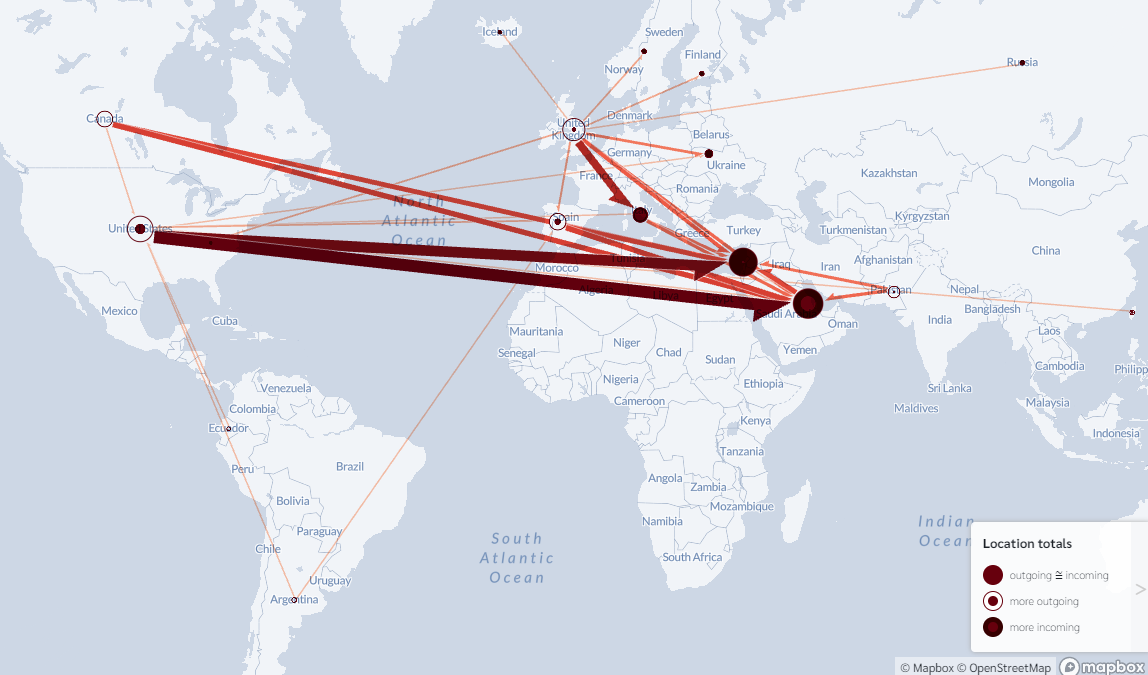}
  \caption{Relations between researchers country of origin and their dataset's country of focus (global/indiscriminate studies excluded). Created via Flowmap.blue~\cite{flowmap_blue} with Mapbox~\cite{mapbox} and OpenStreetMap~\cite{openstreetmap}}
  \label{country-of-origin-vs-country-of-focus-flow-map}
\end{figure}

To identify ERH hot spots in research, we present the first cross-researcher examination of their institution's location compared to their dataset(s) geolocational origin in Figure~\ref{country-of-origin-vs-country-of-focus-flow-map}. For clarity, we filter out the 29 indiscriminate global studies.

Despite the decline of the Islamic State as a conventional state-actor post-2016, western academic research remains skewed towards researching Islamic extremist organisations operating from the Middle East. 24\% of US-originating ERH studies targeted Islamic extremism, compared to 19\% focusing on violent far-right groups and 19\% for left vs right polarised speech (in discussions containing hate speech). Despite more Islamic extremist studies from US-oriented research, over 90\% of terrorist attacks and plots in the US were from far-right extremists in 2020~\cite{csis20_terrorism_in_us_proportions}.

European-origin studies have a reduced bias, where 25\% target far-right white supremacy and 29\% on Islamic extremism. Islamic extremism's popularity is a global trend for 20\% of all studies, shown in Figure~\ref{RQ2-Target-Groups-Pie-Chart}. Hence, there is a clear Islamophobic trend in academia---given the aversion of far-right groups, and the lack of a change in the distribution of targeted groups between 2015-2021.

Researcher ethics and socio-legal considerations present a critical \textit{international} research gap, as only 13\% of US and 28\% of European studies included discussions on annotation ethics, expression laws, or regulation. This US vs. Europe discrepancy likely emanates from the data collection and autonomous decision-making rights guaranteed under the EU's GDPR~\cite{GDPR}.

\subsubsection{The Case for Oceania and the Global South}\hfill\\
Despite developments post-2015, such as two New Zealand terrorist attacks~\cite{nzsis_chch_report, lynn_mall_nz_terror_attack21} and five in Australia~\cite{aussie_asio_terror_report20}, the rise of racial violence in South Africa~\cite{collins19}, and the 2019-ongoing COVID-related radicalisation~\cite{covid_extremism_study}; \textit{no studies considered targeting Oceania or English-recognised countries in the global south}. Likewise, applying these datasets across intersectional ethnic, sexual, and cultural domains presents a threat to validity as terms considered mundane or inoffensive to one group may be considered inflammatory to another. Datasets are also biased towards racism towards a minority group~\cite{Davidson17_benchmark_hs_study, waseem16hateful_symbols_hs_twitter}, which may bias English hate speech in a white minority country such as South Africa. Investigating language trends and model performance on Mela-,  Micro- and Polynesian groups could also offer insights in the role of religion, values (such as \textit{tikanga} values in New Zealand's M\=aori population), taboos, lexicons, and social structures unique to indigenous cultures.

\section{Socio-technical Context in Research---Consensus-driven Definitions}
\label{section:rq1}
\textit{What are the Working Definitions for Classifying Online Extremism, Radicalisation, and Hate Speech?}

Empirically, studies often provide a generalised social definition in their introduction or background and utilise technical criteria to annotate instances for (semi)supervised learning tasks.

Hence, this research question consists of two parts: the \textit{socio-legal} ERH definitions, and the \textit{technical} implementation and classification thereof outlined in the existing literature.

We identified an unexpected overlap between the definitions and models between \textit{extremism} and \textit{radicalisation} studies, whereby researchers frame these concepts as synonymous with hate speech \textit{with a political/organisational affiliation}. Hate speech studies focus on protected groups as binary \textquote{hate or not}~\cite{mashechkin19_russia_caucasus_jihad_ml, ahmad19_lstm_cnn_extrem_sentiment, Nouh19_understanding_radical_mind, benigni2017online_extremism_sustain_it_isis, saif17_semantic_graph_radicalisation, Soler-Company_auto_classification_linguistical_analysis_extrem, waseem16hateful_symbols_hs_twitter, chandrasekharan17bag_of_com, de-gibert18-white-supremacy-dataset, kiela20_hateful_memes_original, Aggarwal21_hateful_meme, multioff_meme_images}, or multiclass \textquote{racism, sexism, offensive, or neither} text~\cite{Davidson17_benchmark_hs_study, waseem16hateful_symbols_hs_twitter, lozano17_racism_pol_tweets, gamback17_cnn_hs, badjatiya17_lstm_dl_hs, naseem19_lstm_abusive_hs_twitter, wullah21_deep_GPT_BERT, mozafari20racialbias_hs_study, basile19_benchmark_mulitlingual_immigrant_women_dataset, perez19_benchmark_mulitlingual_immigrant_women, pitsilis18_recurrent_nn_hs}, with a consensus that \textit{\textquote{Extremism = Radicalisation = Hate speech with an affiliation}}. Alternatively, we propose a novel computationally grounded framework and definitions to \textit{seperate} and \textit{expand} ERH in Section~\ref{section:futureworkERH} to underline the holistic stages of \textit{extremists} temporal \textit{radicalisation} through disseminating \textit{hateful} media.

\subsubsection{Socio-legal Context Provided in Existing Literature}\hfill\\
The largest discrepancy were between studies that discussed legal or ethical context to ERH, which constituted only 20\% of studies~\cite{mashechkin19_russia_caucasus_jihad_ml, Nouh19_understanding_radical_mind, benigni2017online_extremism_sustain_it_isis, hall20_machines_unified_understanding_radicalizing, bilbao18_political_classification_cnn, johnston20identifying_extremism_dl, kapil20_deep_nn_multitask_learning_hs, waseem16hateful_symbols_hs_twitter, macavaney19hs_challenges_solutions, zampieri_etal19_OLID_baseline_tests}. The remaining 80\% relied on an implicit consensus of ‘hate speech’ (often synonymous with toxic, threatening, and targeted speech), or 'extremism' (often UN designated terrorist organisations like ISIS~\cite{un_isis_designation}).

Waseem and Hovy~\cite{waseem16hateful_symbols_hs_twitter} outlined a unique eleven-piece criteria to identify offensive ‘hate speech’ including considerations for politicised \textit{extremist} speech via tweets that “promotes, but does not directly use, hate speech or violent crime” and “shows support of \textit{problematic} hash tags" (although "problematic" was not defined). Hate speech \textit{as a supervised learning task} resulted in two categories---sexism and racism. A \textit{sexist} post requires gender-oriented slurs, stereotypes, promoting gender-based violence, or straw man arguments with gender as a focus (defined as a logical fallacy aimed at grossly oversimplifying/altering an argument in bad faith~\cite{waseem16hateful_symbols_hs_twitter}). The ambiguity for sexism classification by \textit{human annotators} was responsible for 85\% of inter-annotator disagreements~\cite{waseem16hateful_symbols_hs_twitter}.

\subsection{Researchers' Consensus-driven Definitions for ERH Concepts}
We aggregate the trends in ERH based on the definitions used throughout the 51 studies, and observe that ERH concepts reflect their computational approach more than their social definitions. Despite radicalisation being a social process of ideological movement, existing work considers the term as synonymous to political hate speech/extremism.

\begin{Definition}{Hate speech (researchers' consensus)}{working_def_hate_speech}
  \textit{Hate speech is the subjective and derogatory speech towards protected characteristics expressed directly or indirectly to such groups in textual form.\textbf{*}}
\end{Definition}

\textbf{*}N.B: there is a significant bias in hate speech categorical classification in practice, whereby no studies considered categories outside of sexism (including gender-identity) or racism.

\begin{Definition}{Extremism (researchers' consensus)}{working_def_extremism}
  \textit{Organisational affiliation to an ideology that discriminates against protected inalienable characteristics or a violent political organisation. Affiliation does not always include manifestly hateful text and may include tacit or explicit organisational support. Extremist studies often classify organisational affiliation based on text (NLP) \textbf{and} community networks (follower, following, or friend relationships).}
\end{Definition}

The current academic consensus among researchers demonstrates a considerable overlap between \textquote{extremism} and \textquote{hate speech} definitions. In practice, extremism exclusively focused on racism detection, or in the specific context of Jihadism~\cite{ahmad19_lstm_cnn_extrem_sentiment, Nouh19_understanding_radical_mind, benigni2017online_extremism_sustain_it_isis}, white supremacy~\cite{rudinac17_graph_conv_networks,johnston20identifying_extremism_dl,Soler-Company_auto_classification_linguistical_analysis_extrem, owoeye19_extremist_farright_senti}, Ukrainian separatists~\cite{benigni18communitymining_UNSUPERVISED}, anti-fascism (Antifa)~\cite{johnston20identifying_extremism_dl}, and the sovereign citizen movement~\cite{johnston20identifying_extremism_dl}. Of the 13 studies targeting \textit{extremism}, only one considered extremism by the ADL's \textit{politically-fringe-but-not-violent} definition~\cite{abubakar19_lstm_gru_comparative_multiclass}. Tying extremism to the study of mainstream terrorist-affiliated groups neglects rising movements, ethical movements using unethical terror-tactics, and non-violent fragments of other \textit{terrorist} groups, such as a reversion to \textquote{fringe} activism. Hence, \textit{extremism's} working definition is similar to \textit{terrorism} when considering \textit{group affiliation detection}. If investigating \textit{extremist ideologies}, then the definition is synonymous with those in \textit{hate speech} studies.

\textit{Extremism's} working definition is exceptionally biased towards support for Islamic extremist movements (10 out of 13 studies~\cite{mashechkin19_russia_caucasus_jihad_ml,ahmad19_lstm_cnn_extrem_sentiment, benigni18communitymining_UNSUPERVISED,Nouh19_understanding_radical_mind,Moussaoui19_twitter_terrorism_communities,buchanan17_ethics_critique_twitter_study,johnston20identifying_extremism_dl,Soler-Company_auto_classification_linguistical_analysis_extrem,owoeye19_extremist_farright_senti,pais20_unsupervised_extreme_sentiments}), with far-right ideologies a distant second (5 out of 13 studies~\cite{benigni18communitymining_UNSUPERVISED,rudinac17_graph_conv_networks,johnston20identifying_extremism_dl, Soler-Company_auto_classification_linguistical_analysis_extrem, owoeye19_extremist_farright_senti}). These organisational and ideological biases is potentially a result of US security discourse and national interests (via the \textquote{War on Terror}). \textit{White supremacy} and \textit{far-right ideologies} are separate terms used interchangeably without distinction.

\begin{Definition}{Radicalisation (researchers' consensus)}{working_def_radicalisation}
  \textit{No discernible difference between extremism's definition with both terms used interchangeably.
  Radicalisation = extremism = politically affiliated hate invoking or supporting speech.}
\end{Definition}

Definitions and algorithmic approaches on \textit{radicalisation} detection relied on political \textit{hate speech}, or \textit{extremism} via ideological affiliation---with 5 of the 8 \textit{radicalisation} studies targeting textual or network affiliation to the Islamic State (IS)~\cite{rehman21_language_of_isis_twitter, Moussaoui19_twitter_terrorism_communities, hall20_machines_unified_understanding_radicalizing, araque20_radical_emotion_signals_similarity, saif17_semantic_graph_radicalisation}, and 2 on white supremacy~\cite{Grover19_alt-right_subreddits, scrivens20_right-wing_posting_online}.

The only other notable deviations from this \textit{extremism = radicalisation} dilemma is Bartal et al.'s~\cite{bartal20_roles_trolls_affiliation} focus on radicalisation as a \textit{temporal} movement with apolitical \textit{roles}. Their study investigated the temporal movement from a \textquote{Novice} (new poster) classification  towards an \textquote{influencer} role based on their network relations and reply/response networks. Chandrasekharan et al. defined \textquote{radicalisation} as the process of an entire subreddit’s patterns up to and including the time of its ban to map subreddit-wide radicalisation~\cite{chandrasekharan17bag_of_com}. Only two studies are the exception to the \textit{extremism = radicalisation = politicised hate speech} consensus per the remainder of the 51 studies~\cite{bartal20_roles_trolls_affiliation, chandrasekharan17bag_of_com}.

\begin{figure}[!ht]
  \centering
  \includegraphics[width=\textwidth]{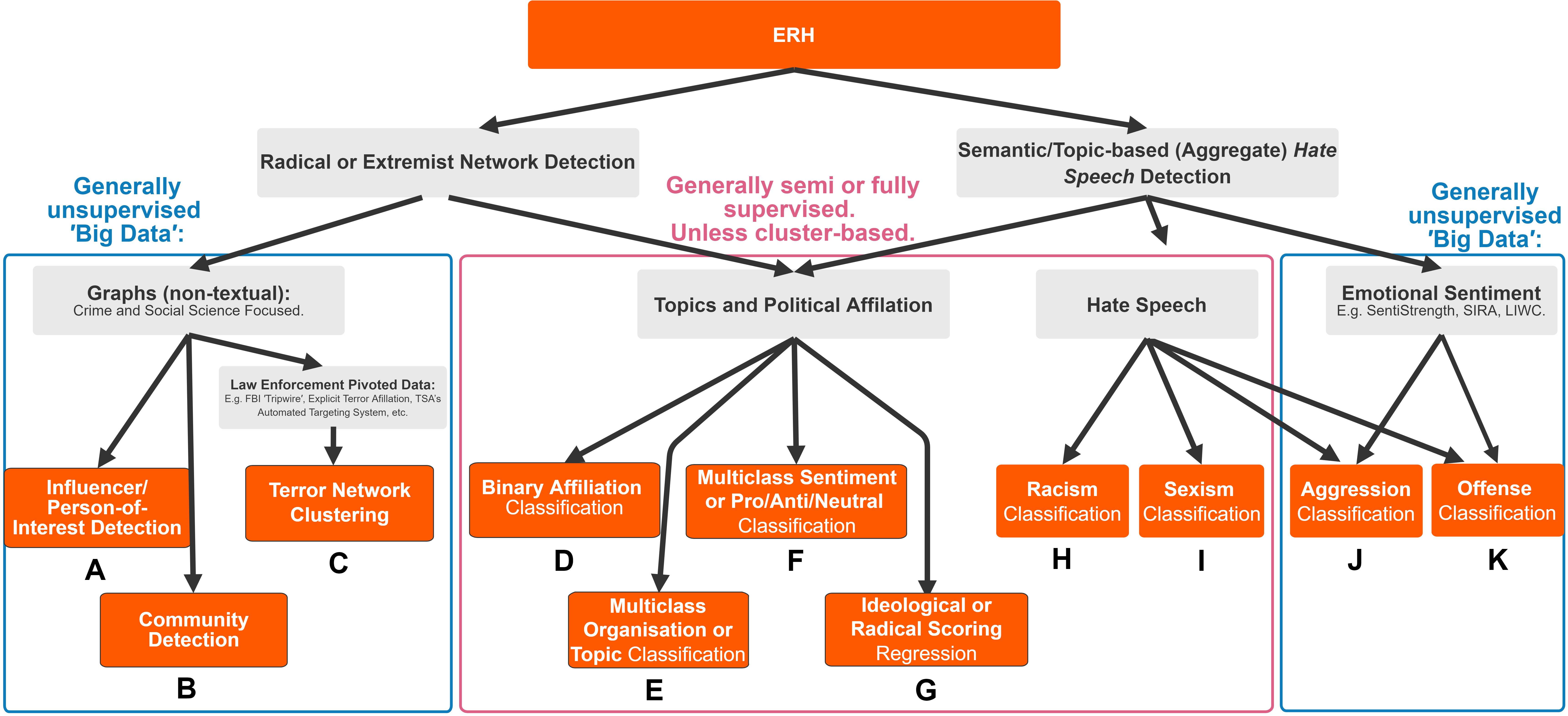}
  \caption{ERH Definition Tree---visualising how ERH definitions deviate based on their algorithmic approach.}
  \label{ERH-definition-tree}
\end{figure}

\subsection{Correlation Between Definitions and Algorithmic Approach}

Uniquely, 66\% of publications in the field of social-science or security studies utilised network-driven community detection models, with extremism defined in a law enforcement context by emphasising a user’s network-of-interactions between known annotated extremists. Hung et al. defined extremism in a semi-supervised OSINT and HUMINT surveillance manner---requiring links between an extremist virtual \textit{and} a physical/offline presence to extremist stimuli through incorporating the FBI’s tripwire program~\cite{hung16_fbi_radical}. Approaching extremism using relational properties via interactions, geographic proximity, profile metadata, and semantic or network similarity raises ethical dilemmas vis-à-vis individuals who have/had a solely virtual presence or those interested in opposing opinions~\cite{buchanan17_ethics_critique_twitter_study, marwick16_data_society_risky_researchers_practices, conway21_extremism_terrorism_research_ethics_study}. The relationship between definitions and algorithmic approaches indicate that \textit{radicalisation} studies skew towards community detection models, \textit{extremism} towards hybrid NLP and community detection models, and \textit{hate speech} to a text-only NLP endeavour.

\begin{table}[!ht]
  \caption{Table of references for studies in each category (A, B...K) in the above ERH definition tree diagram.}
  \label{ERH-tree-reference-table}
  \tiny
  \begin{tabular}{p{0.03\linewidth}p{0.24\linewidth}p{0.03\linewidth}p{0.24\linewidth}p{0.03\linewidth}p{0.24\linewidth}}
    \toprule
    Label&Studies&Label&Studies&Label&Studies\\
    \midrule
    A&\cite{bartal20_roles_trolls_affiliation, Moussaoui19_twitter_terrorism_communities, benigni2017online_extremism_sustain_it_isis, hung16_fbi_radical}&B&\cite{benigni18communitymining_UNSUPERVISED, Moussaoui19_twitter_terrorism_communities,  shi16_structural_similarity_networks, benigni2017online_extremism_sustain_it_isis, pais20_unsupervised_extreme_sentiments, rudinac17_graph_conv_networks}&C&\cite{shi16_structural_similarity_networks, hung16_fbi_radical}\\
    
    D&\cite{mashechkin19_russia_caucasus_jihad_ml, ahmad19_lstm_cnn_extrem_sentiment, Nouh19_understanding_radical_mind, benigni2017online_extremism_sustain_it_isis, saif17_semantic_graph_radicalisation, Soler-Company_auto_classification_linguistical_analysis_extrem, waseem16hateful_symbols_hs_twitter, chandrasekharan17bag_of_com, de-gibert18-white-supremacy-dataset, kiela20_hateful_memes_original, Aggarwal21_hateful_meme, multioff_meme_images}&E&\cite{Moussaoui19_twitter_terrorism_communities, rudinac17_graph_conv_networks, johnston20identifying_extremism_dl, Preotiuc-Pietro_pol_ideology_twitter, belcastro20_Italy_Polarisation, pais20_unsupervised_extreme_sentiments, rudinac17_graph_conv_networks}&F&\cite{abubakar19_lstm_gru_comparative_multiclass, rehman21_language_of_isis_twitter, sharma_indian_mass_media, weir_positing_text_analysis_extrem, araque20_radical_emotion_signals_similarity, owoeye19_extremist_farright_senti, ahmad19_lstm_cnn_extrem_sentiment, rustam21_covid_extremes, kapil20_deep_nn_multitask_learning_hs, pais20_unsupervised_extreme_sentiments, jia21_covid19_hs_ensemble, basile19_benchmark_mulitlingual_immigrant_women_dataset, zampieri_etal19_OLID_baseline_tests, zampieri_etal19_OLID_semeval_competitors_results, perez19_benchmark_mulitlingual_immigrant_women}\\
    
    G&\cite{Grover19_alt-right_subreddits, scrivens20_right-wing_posting_online, scrievens18_signs_of_extremism_SIRA, benigni2017online_extremism_sustain_it_isis, Tien_unite_the_right_polarization, hall20_machines_unified_understanding_radicalizing, bilbao18_political_classification_cnn}&H&\cite{waseem16hateful_symbols_hs_twitter, lozano17_racism_pol_tweets, gamback17_cnn_hs, badjatiya17_lstm_dl_hs, naseem19_lstm_abusive_hs_twitter, wullah21_deep_GPT_BERT, mozafari20racialbias_hs_study, pitsilis18_recurrent_nn_hs, basile19_benchmark_mulitlingual_immigrant_women_dataset, perez19_benchmark_mulitlingual_immigrant_women}&I&\cite{waseem16hateful_symbols_hs_twitter, gamback17_cnn_hs, badjatiya17_lstm_dl_hs, naseem19_lstm_abusive_hs_twitter, wullah21_deep_GPT_BERT, mozafari20racialbias_hs_study, pitsilis18_recurrent_nn_hs, basile19_benchmark_mulitlingual_immigrant_women_dataset, perez19_benchmark_mulitlingual_immigrant_women}\\
    
    J&\cite{karan1818_cross_domain, naseem19_lstm_abusive_hs_twitter, wullah21_deep_GPT_BERT, basile19_benchmark_mulitlingual_immigrant_women_dataset, zampieri_etal19_OLID_baseline_tests, zampieri_etal19_OLID_semeval_competitors_results, perez19_benchmark_mulitlingual_immigrant_women}&K&\cite{karan1818_cross_domain, macavaney19hs_challenges_solutions, zhu19_offensive_tweets_bert, liu19_transfer_learning_BERT, naseem19_lstm_abusive_hs_twitter, wullah21_deep_GPT_BERT, mozafari20racialbias_hs_study, zampieri_etal19_OLID_baseline_tests, zampieri_etal19_OLID_semeval_competitors_results, perez19_benchmark_mulitlingual_immigrant_women, kiela20_hateful_memes_original, Aggarwal21_hateful_meme, multioff_meme_images, rudinac17_graph_conv_networks}&-&-\\
   
  \bottomrule
\end{tabular}
\end{table}

\subsubsection{Privacy and Ethics-driven Regulation}\hfill\\
No studies integrated or mentioned existing AI ethics regulation or standards, such as those emerging from the EU~\cite{eu_guidelines_ai}, or private-sector self-regulation such as the IEEE P700x Series of Ethical AI Design~\cite{koene18_ieee_p70xx}. Researchers should consider the application and use cases for their proposed models---as autonomous legal decision making, injurious use of data (outside of a reasonable purpose), or erasure (a challenge for persistent open-source datasets), may violate regulations such as the EU's General Data Protection Regulation (GDPR)'s Article 22, 4, and 17 respectively~\cite{GDPR}.

Prominently, Mozafari et al. evaluated hate speech with a ethno-linguistic context, recognising that certain racist slurs were dependent on the culture and demographic using them~\cite{mozafari20racialbias_hs_study}.

\section{Building ERH Datasets---Collection, Processing, and Annotation}
\label{section:rq2}
\textit{What are the methodologies for collecting, processing and annotating datasets?}

This RQ outlines the dominant platforms of choice for ERH research, the APIs and methods for pulling data and its underlying ethical considerations. Geographic mapping demonstrates the marginalisation of Oceania and the global south in academia. We critically evaluate the sentimental, relationship, and contextual feature extraction and filtering techniques in community detection and NLP studies. We conclude with the key recommendations for future data collection research.

\subsection{Prominent Platforms, Pulling, and Populations}
This subsection outlines the common social media platforms, the method for sampling and extracting (\textquote{pulling}) textual and network/relationship data, and the type of data used in ERH datasets.

40\% of studies relied on Twitter tweets for ERH detection, with Twitter being the dominant platform for research per Figure~\ref{Platforms-studied-size-map}. Twitter's mainstream global reach paired with its data-scraping Twitter API enabled researchers to target specific hashtags (topics or groups), real-time tweet streams and reach of tweets and their community networks. Hence, the Twitter API is also the most used method for scraping data, with other methods outlined in Figure~\ref{RQ2-Method-of-Pulling-Data-Pie-Chart}. Unfortunately, revised 2021 Twitter Academic API regulations removed access to tweets from suspended
accounts~\cite{twitter_api_suspended21}, limiting datasets to those pre-archived. Currently, the Waseem and Hovy datasets are not available due to relying on the Twitter API and suspended tweets~\cite{waseem16_are_you_racist_dataset, waseem16hateful_symbols_hs_twitter}.

For far-right ERH detection, researchers used custom web-scrapers to pull from the global white supremacist forum \textit{Stormfront}---containing topics ranging from political discussions, radicalising "Strategy and Tactics", and "Ideology and Philosophy" sections, and regional multilingual chapters~\cite{scrivens20_right-wing_posting_online, kapil20_deep_nn_multitask_learning_hs,mozafari20racialbias_hs_study, Soler-Company_auto_classification_linguistical_analysis_extrem, macavaney19hs_challenges_solutions, wullah21_deep_GPT_BERT, weir_positing_text_analysis_extrem}. As a supplement or alternate to searching and collecting hateful posts en masse, five studies considered extremist \textquote{ground truth} instances by comparing textual similarity from Tor-based terror-supporting anonymous forums~\cite{ahmad19_lstm_cnn_extrem_sentiment} and websites~\cite{weir_positing_text_analysis_extrem}, radical Islamist magazines and manifestos~\cite{rehman21_language_of_isis_twitter,Nouh19_understanding_radical_mind,araque20_radical_emotion_signals_similarity}. Interestingly, no studies considered extracting ground truths from far-right manifestos or media. Likewise, no studies considered recent low-moderation anonymised forums such as 8Chan (now 8kun) or Kiwifarms, which were extensive hubs for propaganda dissemination from the Christchurch shooter~\cite{nzsis_chch_report}; Parler, notable for its organisational influence during the 2021 Capitol Hill riots~\cite{parler_capitol_riot}; Telegram, TikTok, or Discord, despite reports on its use for sharing suicides, mass shootings, and group-lead harassment of minority groups~\cite{isd21_hatescape_extreme_right_tiktok, isd21_extreme_right_discord}). Hence, there is a prevalent and concerning trend towards NLP studies on mainstream platforms, which may overlook the role of emergent, pseudo-anonymised or multimedia-oriented platforms.

\subsubsection{Data Collected}\hfill\\
62\% of all studies evaluate ERH on a single post-by-post basis, with NLP the dominant approach per Figure~\ref{RQ2-3-Type-of-Data-Used-Pie-Chart}. Conversely, grouping posts on a \textit{per user basis} frequently included annotations from cyber-vigilante groups such as the Anonymous affiliated OSINT \textquote{Controlling Section} (CtrlSec) group's annotations of ISIS-supporting accounts~\cite{hall20_machines_unified_understanding_radicalizing}. However, Twitter claims that CntrlSec annotations were "wildly inaccurate and full of academics and journalists"~\cite{cameron15_twitter_isis_cntrlsec_statement}. Hence, researchers should avoid unvetted cyber-vigilante data, and consider anonymising datasets to further benefit user privacy, researcher ethics, and model performance by reducing false positives (i.e., censorship).

While NLP text detection is the dominant detection approach, 23 of the 51 studies investigated data sources outside of textual posts per Figure~\ref{RQ2-Special-Type-of-Data-Used-Bar-Chart}. Research gaps include the lack of multimedia and law enforcement studies, with only three hybrid text-image detection~\cite{rudinac17_graph_conv_networks,kiela20_hateful_memes_original, multioff_meme_images} and one study utilising FBI anonymous tips, Automated Targeting System and Secure Flight data~\cite{hung16_fbi_radical,dhs_auto_targeting_system_program}.

\subsubsection{Data Collection and Annotation Bias}\hfill\\
Due to the varying fiscal costs, biases, and time trade-offs, there is no consensus for selecting or excluding annotators for supervised learning datasets. Hence, we frame that annotator selection falls within two varying groups: \textit{experience-driven} selection and \textit{organisation-driven} selection. 

For the former, experience-driven selection includes studies that utilised self-proclaimed \textquote{expert} panels as determined by their location and relevant degrees~\cite{waseem16hateful_symbols_hs_twitter}, are anti-racism and feminist activists~\cite{waseem16_are_you_racist_dataset}, or work on behalf of a private anti-discrimination organisation~\cite{hatebase}. However, assembling annotators by specific characteristics may be time-consuming or costly, such as crowdsourcing \textit{tertiary} annotators via Amazon Mechanical Turk, or Figure Eight~\cite{gamback17_cnn_hs, badjatiya17_lstm_dl_hs, macavaney19hs_challenges_solutions, Preotiuc-Pietro_pol_ideology_twitter, mozafari20racialbias_hs_study}.

Conversely, an organisation-driven selection approach focuses on agreement by a crowdsourced consensus. Instead of relying on specific experience, researchers utilised custom-made tests for knowledge of hate speech criteria based on the researchers own labelled subset~\cite{waseem16_are_you_racist_dataset, zampieri_etal19_OLID_baseline_tests}. Likewise, organising annotator pools can also include balancing annotators self-reported political affiliation to reduce political bias~\cite{Preotiuc-Pietro_pol_ideology_twitter}. Researchers use \textit{Inter-rater agreement}, and \textit{Kappa Coefficient} to determine a post's ERH classification. For racism, sexism, and neither classifications, annotation Fleiss' Kappa values ranged between 0.607~\cite{de-gibert18-white-supremacy-dataset} to 0.83~\cite{zampieri_etal19_OLID_baseline_tests}, indicating moderate to strong agreement~\cite{data_mining_book_eibe}.

Thirdly, unsupervised clustering enables mass data collection without time-consuming annotation via Louvain grouping (to automatically group text/networks to identify \textit{emergent} groups)~\cite{benigni18communitymining_UNSUPERVISED, benigni2017online_extremism_sustain_it_isis, shi16_structural_similarity_networks}, or grouping based on a thread's affiliation (e.g., the now-banned r/altright~\cite{Grover19_alt-right_subreddits} and v/[n-word]~\cite{chandrasekharan17bag_of_com}). Although not all posts from an extremist platform may be manifestly hateful, as evident in the 9507 post \textquote{non-hate} class in the Stormfront benchmark dataset from de Gibert et al.~\cite{de-gibert18-white-supremacy-dataset}.

Research continues to skew towards radical Islamic extremism per Figure~\ref{RQ2-Target-Groups-Pie-Chart}, while the plurality (41\%) target generic \textquote{hate speech} in \textquote{hate or not}, or delineations for racism, sexism, and/or offence.

\subsubsection{Benchmark Datasets}\hfill\\
We define a benchmark dataset as any dataset evaluated by three or more studies. The majority of studies used custom web-scrapped datasets or Tweets pulled via the Twitter API per Figure~\ref{RQ2-Method-of-Pulling-Data-Pie-Chart}.

\begin{table}[!ht]
  \caption{Datasets used by three or more studies.}
  \label{RQ2-benchmark-dataset-table}
  \small
  \begin{tabular}{p{0.1\linewidth}p{0.05\linewidth}p{0.23\linewidth}p{0.1\linewidth}p{0.23\linewidth}p{0.13\linewidth}}
    \toprule
    Dataset&Year&Categories&Platform of origin&Collection strategy&Used By\\
    \midrule
    
    Waseem and Hovy~\cite{waseem16hateful_symbols_hs_twitter}&2016&16914 tweets: 3383 \textit{Sexist} , 1972 \textit{Racist}, 11559 \textit{Neutral}&Twitter&11-point Hate Speech Criteria&\cite{karan1818_cross_domain, badjatiya17_lstm_dl_hs, waseem16hateful_symbols_hs_twitter, wullah21_deep_GPT_BERT, pitsilis18_recurrent_nn_hs, mozafari20racialbias_hs_study, naseem19_lstm_abusive_hs_twitter, macavaney19hs_challenges_solutions, kapil20_deep_nn_multitask_learning_hs, jia21_covid19_hs_ensemble}\\
    
    FifthTribe \cite{kaggle_isis_dataset}&2016&17350 \textit{pro-ISIS} accounts&Twitter&Annotated pro-ISIS accounts& \cite{Nouh19_understanding_radical_mind, araque20_radical_emotion_signals_similarity, rehman21_language_of_isis_twitter, sanchez2019detection}\\
    
    de Gibert \cite{de-gibert18-white-supremacy-dataset}&2018&1196 \textit{Hate}, 9507 \textit{Non-hate}, 74 \textit{Skip} (other) posts&Stormfront&3 annotators considering prior posts as context& \cite{macavaney19hs_challenges_solutions, kapil20_deep_nn_multitask_learning_hs, wullah21_deep_GPT_BERT, de-gibert18-white-supremacy-dataset}\\
    
    OffenseEval (OLID)~\cite{zampieri_etal19_OLID_baseline_tests}&2019&14100 tweets. (30\%) \textit{Offensive} or Not; \textit{Targeted} or \textit{Untargeted} insult; towards an \textit{Individual}, \textit{Group}, or \textit{Other}&Twitter&Three-level hierarchical schema, by 6 annotators& \cite{zampieri_etal19_OLID_baseline_tests, zhu19_offensive_tweets_bert, kapil20_deep_nn_multitask_learning_hs, liu19_transfer_learning_BERT}\\
    
    HatEval~\cite{basile19_benchmark_mulitlingual_immigrant_women_dataset}&2019&10000 tweets distributed with \textit{Hateful} or Not, \textit{Aggressive} or Not, \textit{Individual} targeted or \textit{Generic}&Twitter&Crowdsourced via Figure Eight, with 3 judgements per Tweet& \cite{macavaney19hs_challenges_solutions, wullah21_deep_GPT_BERT, kapil20_deep_nn_multitask_learning_hs, basile19_benchmark_mulitlingual_immigrant_women_dataset, perez19_benchmark_mulitlingual_immigrant_women}\\
    
    Hatebase-Twitter~\cite{Davidson17_benchmark_hs_study}&2019&25000 tweets: \textit{Hate speech}, \textit{Offensive}, \textit{Neither}&Twitter& 3 or more CrowdFlower annotators per tweet&\cite{Davidson17_benchmark_hs_study, macavaney19hs_challenges_solutions, kapil20_deep_nn_multitask_learning_hs, wullah21_deep_GPT_BERT, jia21_covid19_hs_ensemble, naseem19_lstm_abusive_hs_twitter, mozafari20racialbias_hs_study}\\
    
    TRAC~\cite{kumar_etal18_trac_benchmark_dataset}&2018&15000 English and Hindi posts; Overtly, Covertly, or Not Aggressive&Facebook&Kumar et al.~\cite{kumar18_trac_original_dataset} subset, 3 annotators per post, comment or unit of discourse& \cite{karan1818_cross_domain, kapil20_deep_nn_multitask_learning_hs, macavaney19hs_challenges_solutions}\\
  \bottomrule
\end{tabular}
\end{table}

\subsection{Feature Extraction Techniques}
\begin{figure}[!ht]
  \centering
  \includegraphics[width=0.8\linewidth]{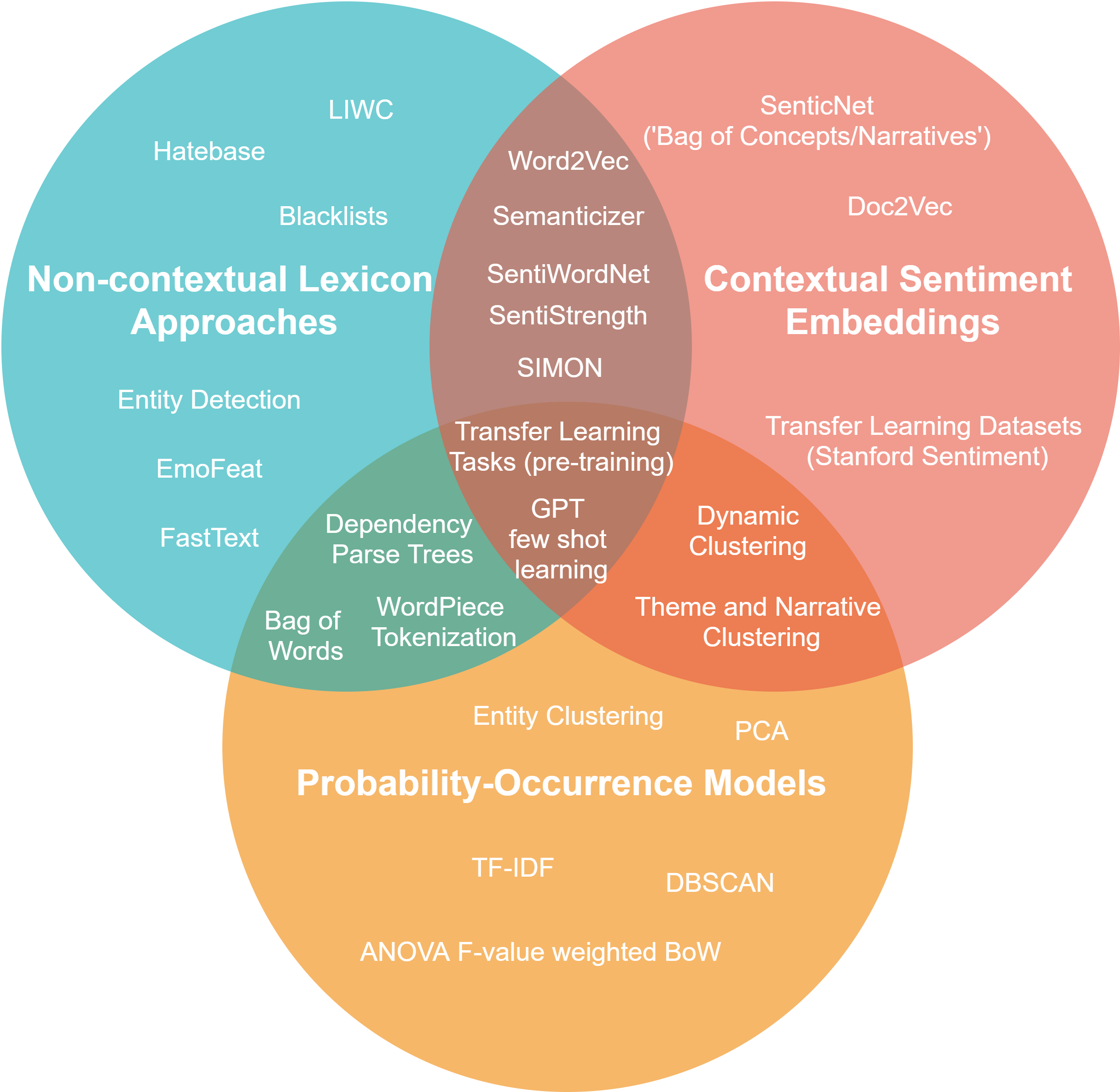}
  \caption{Types of feature extraction techniques.}
  \label{RQ2-feature-extraction-approaches-venn-diagram}
\end{figure}

Figure~\ref{RQ2-feature-extraction-approaches-venn-diagram} outlines the three types of feature extraction techniques. \textit{Non-contextual} lexicon approaches relate to word embeddings for entities, slurs, and emotional features. However, non-contextual blacklists and \textit{Bag of Words} (BoW) lexicon approaches cannot identify context, concepts, emergent, or dual-use words (see the \textit{Supplementary Material's Algorithm Handbook} section for comprehensive definitions)~\cite{waseem16_are_you_racist_dataset, naseem19_lstm_abusive_hs_twitter, perez19_benchmark_mulitlingual_immigrant_women, de-gibert18-white-supremacy-dataset, mozafari20racialbias_hs_study}. \textit{Contextual} sentiment embeddings expand on lexicons by embedding a form of context via positional tokens to establish an order to sentences.

We group unsupervised term clustering and dimensionality reduction methods under the \textit{Probability-Occurrence Models} category. The two dominant approaches include weighted ANOVA-based BoW approaches and Term Frequency-Inverse Document Frequency (defined in the \textit{Supplementary Material}), which weigh the importance of each word in the overall document and class corpus. Contextual sentiment embeddings result in higher F1-scoring models (per RQ4) due to their context-sensitivity and compatibility as input embeddings for deep learning models~\cite{kapil20_deep_nn_multitask_learning_hs, naseem19_lstm_abusive_hs_twitter, macavaney19hs_challenges_solutions}.

Community detection features require mapping following, friend, followee, and mention dynamics. Furthermore, other statistically significant metadata includes profile name, geolocation (to investigate ERH \textit{as a disease}), gender, and URLs~\cite{waseem16hateful_symbols_hs_twitter, wullah21_deep_GPT_BERT, mozafari20racialbias_hs_study, benigni18communitymining_UNSUPERVISED, Nouh19_understanding_radical_mind}. URLs can identify rabbit holes for misinformation or alternate forums via PageRank~\cite{page99pagerank} and Hyperlink-induced Topic Search (HITS)~\cite{Kleinberg_HITS_Algo}---which extracts keywords, themes and topical relations across the web~\cite{mashechkin19_russia_caucasus_jihad_ml, page99pagerank}.

\subsection{Data Filtering}
\begin{figure}[!ht]
  \centering
  \includegraphics[width=0.94\linewidth]{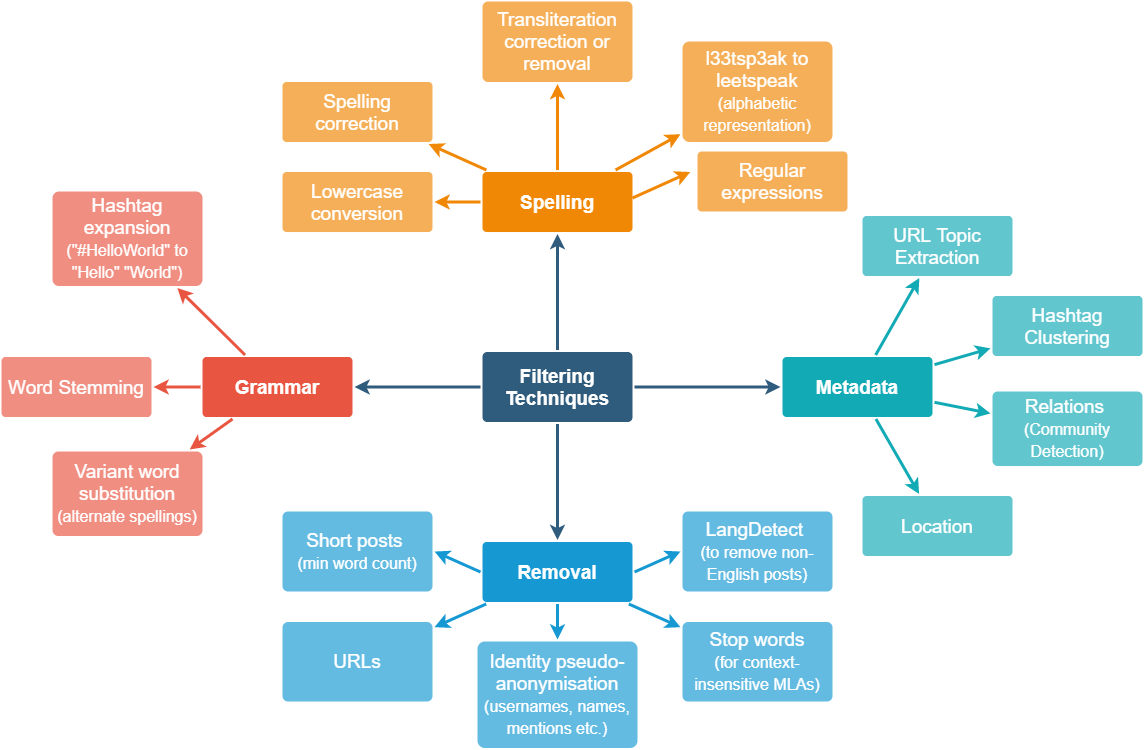}
  \caption{Types of data filtering techniques across NLP and community detection studies.}
  \label{RQ2-data-filtering-approaches}
\end{figure}

For context-insensitive BoW and non-deep models, stop words (e.g. \textit{the}, \textit{a}, etc.), misspellings, and web data are often filtered out via regular expressions and parsing libraries~\cite{hall20_machines_unified_understanding_radicalizing, ahmad19_lstm_cnn_extrem_sentiment, Grover19_alt-right_subreddits, Nouh19_understanding_radical_mind, lozano17_racism_pol_tweets, macavaney19hs_challenges_solutions, chandrasekharan17bag_of_com, mozafari20racialbias_hs_study, jia21_covid19_hs_ensemble}. Compared to semantic or reply networks, community detection models tend to extract \textit{metadata} for separate clustering for entity and concept relationships. All data filtering techniques are thereby aggregated and branched in Figure~\ref{RQ2-data-filtering-approaches}.

\begin{figure}[!ht]
    \centering
    \begin{minipage}[b]{0.49\linewidth}
        \includegraphics[width=\textwidth]{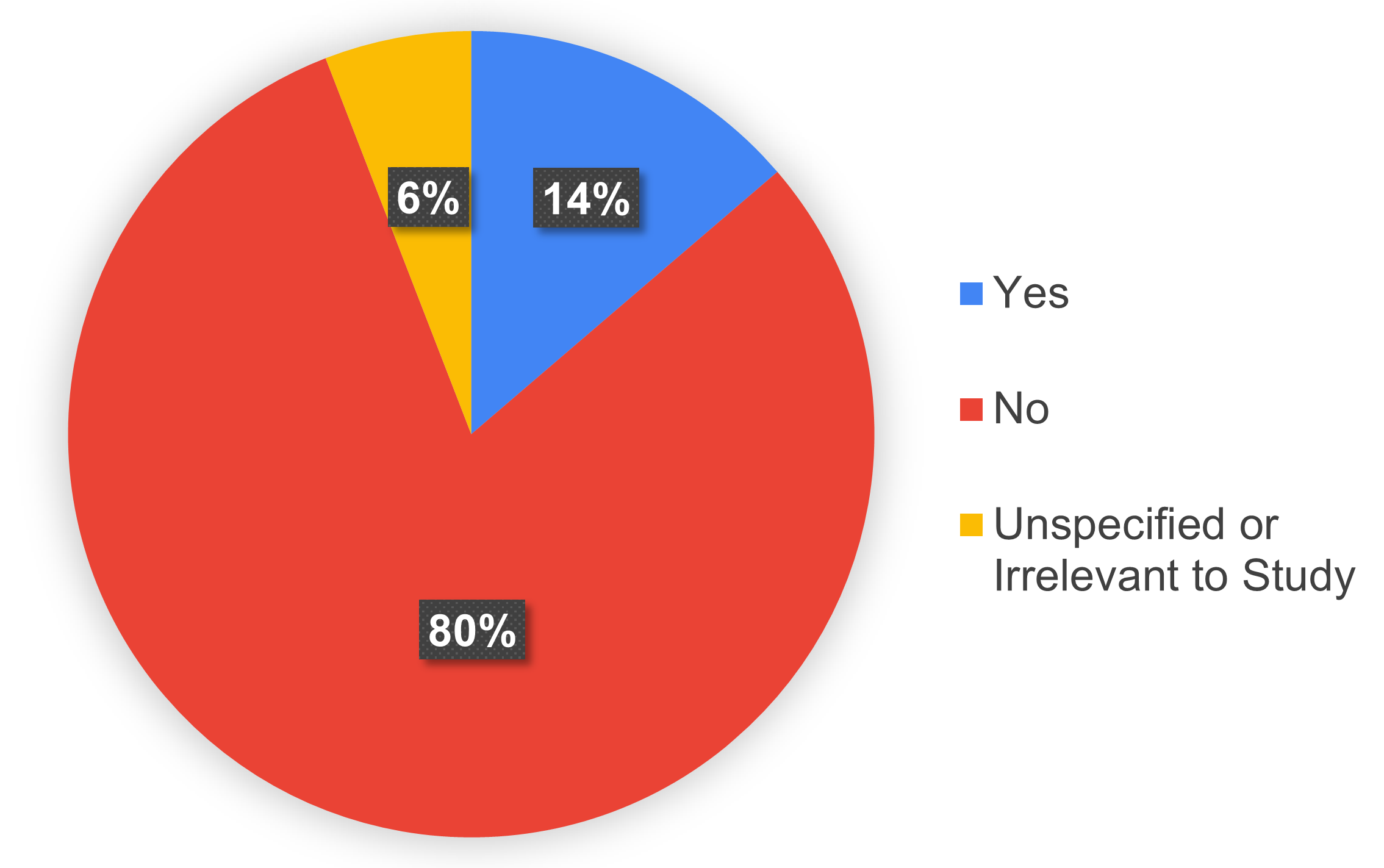}
        \caption{Studies with Bot or Troll Filtering.}
        \label{RQ2-Bots-or-Trolls-Pie-Chart}
    \end{minipage}
    \hfill
    \begin{minipage}[b]{0.49\linewidth}
        \includegraphics[width=\textwidth]{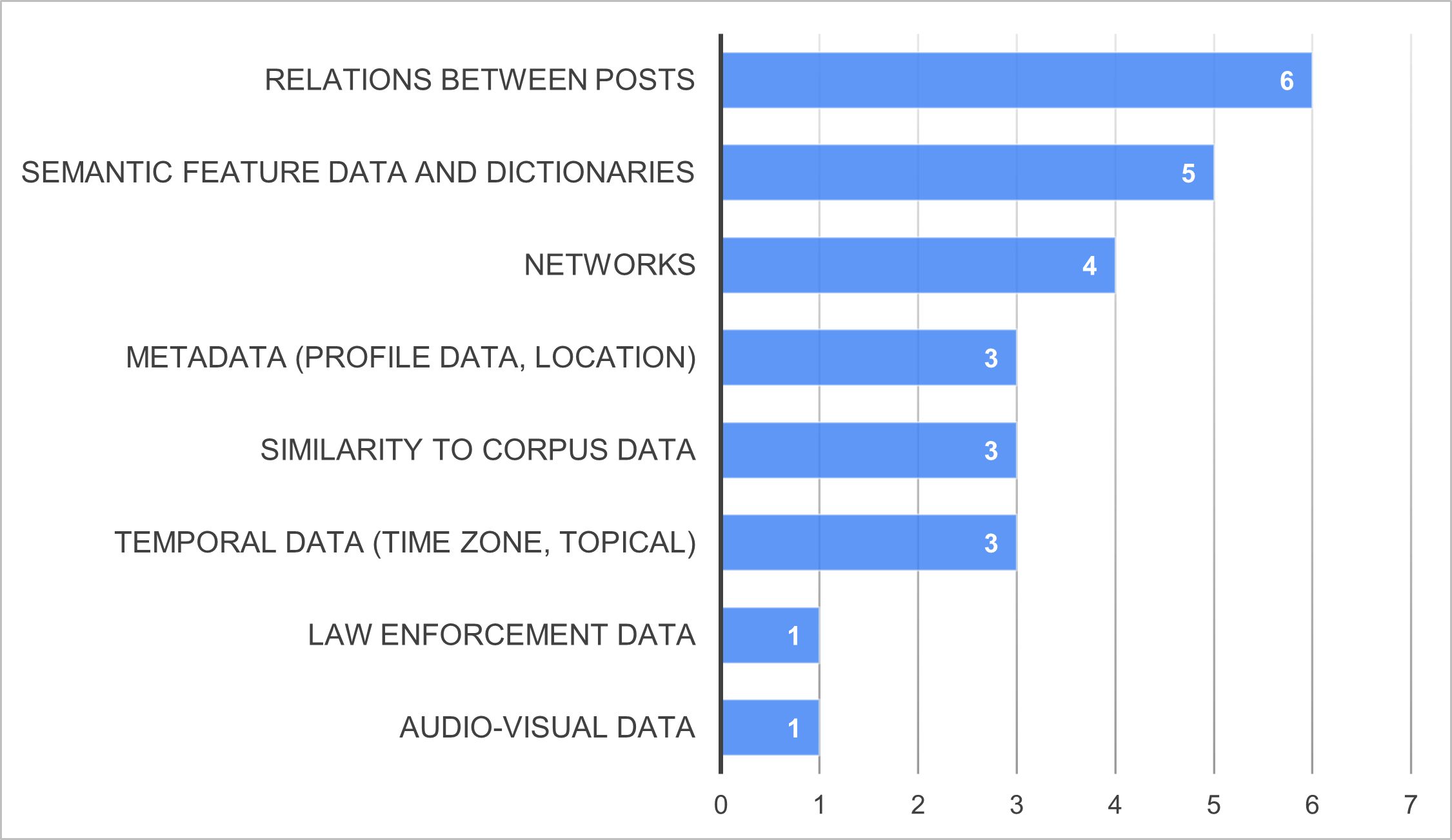}
        \caption{Special Type of Data Used.}
        \label{RQ2-Special-Type-of-Data-Used-Bar-Chart}
    \end{minipage}
\end{figure}

No studies considered \textit{satire}, \textit{comedy}, or \textit{irony} to delineate genuine extremism and online culture. Researchers' implicit consensus is to treat \textit{all} posts as part of the ERH category if it violates their criteria, regardless of intent. Conversely, Figure~\ref{RQ2-Bots-or-Trolls-Pie-Chart} displays the 14\% of the studies filtered bots by removing \textit{but not classifying} bot accounts from the ERH datasets~\cite{bartal20_roles_trolls_affiliation, benigni18communitymining_UNSUPERVISED, Grover19_alt-right_subreddits, benigni2017online_extremism_sustain_it_isis,Soler-Company_auto_classification_linguistical_analysis_extrem, lozano17_racism_pol_tweets, Preotiuc-Pietro_pol_ideology_twitter}. Strategies include removing duplicate spam text, filtering Reddit bots by name, and setting minimum account statistics for verification---such as accounts with that share hashtags to at least five other users to combat spam~\cite{benigni18communitymining_UNSUPERVISED}. Likewise, Lozano et al. limited eligible users for their dataset to have at least 100 followers, with more than 200 tweets, and at least 20 favourites to other tweets~\cite{lozano17_racism_pol_tweets}. This operates on the assumption that bots are short-lived, experience high in-degree between similar (bot) accounts, and seldom have real-world friends or followers---as discovered by Bartal et al.~\cite{bartal20_roles_trolls_affiliation}

Outside of removing suspicious bot accounts via human annotation in dataset generation, computational means to \textit{explicitly categorise bots or trolls} remains an area for future research.

\subsection{Key Takeaways for Dataset Domain, Pre-processing, and Annotation}
Twitter's accessible API, popularity and potential for relationship modelling via reply and hashtag networks, makes it the platform of choice for research (Figure~\ref{RQ2-Method-of-Pulling-Data-Pie-Chart}). Despite the rise of far-right extremism post-2015, Islamic extremism in the US and Europe remains the target group for the majority of organisation-based studies, with no studies considering far-right/left manifestos. The marginalisation of Oceania and the global south by datasets predominantly containing US white hetero males indicates a structural bias in academia. For feature extraction, we recommend using:

\begin{enumerate}
\item \textit{Contextual sentimental embeddings}---due to their compatibility with deep learning models and highest performance, per Table~\ref{f1-score-benchmark-table}.
\item \textit{Pre-defined lexicons}---assuming they remain up-to-date with online culture.
\item \textit{Probability-occurrence models}---ideal for large-scale clustering of emergent groups~\cite{chandrasekharan17bag_of_com, wullah21_deep_GPT_BERT, rudinac17_graph_conv_networks}.
\end{enumerate}

We do not recommend pre-defined lexicons on non-English text, new groups or ideologies---as these lexicons may not translate to different concepts and slurs. We recommend adaptive semi or unsupervised learning via contextual embeddings and entity/concept clustering for edge cases.

Research currently lacks multidomain datasets, pseudo-anonymous platforms, multimedia (i.e., images, videos, audio and livestreams), and extraction of comedy, satire, or bot features.

\section{Community Detection, Text, and Image ERH Detection Algorithms}
\label{section:rq3}
\textit{What are the computational classification methods for ERH detection?}

Between 2015-2021, non-deep Machine Learning Algorithms (MLAs) shifted towards Deep Learning Algorithms (DLAs) due to their superior performance and context-sensitivity (Table~\ref{f1-score-benchmark-table}). Support Vector Machines (SVM) and a case of a Random Forest (RF) model were the last remaining non-deep MLAs post-2018 to outperform DLAs. Studies seldom hybridise relationship network modelling and semantic textual analysis. Ongoing areas of research in MLAs consist of identifying psychological signals to compete with DLAs such as Bidirectional Encoder Representations from Transformers (BERT), Convolutional Neural Networks (CNN) and Bidirectional Long Short-Term Memory (BiLSTM) models (defined further in the \textit{Supplementary Material's Algorithm Handbook} section). DLAs are best oriented for text-only tasks and for hybrid image-caption models~\cite{rudinac17_graph_conv_networks, kiela20_hateful_memes_original, Aggarwal21_hateful_meme, multioff_meme_images}. Future NLP studies should consider higher-performing neural languages models over BERT-base---such as RoBERTa~\cite{roberta_original_study}, Sentence-BERT~\cite{sbert_original_study}, or multi-billion parameter transformers such as GPT-3~\cite{brown2020language}, GPT-Neo~\cite{gpt_neo_original_software}, or Jurassic-1~\cite{jurassic1_original_white_paper}.

\subsection{Observed Non-deep Machine Learning Algorithms (MLAs)}
\begin{figure}[!ht]
  \centering
  \includegraphics[width=0.85\linewidth]{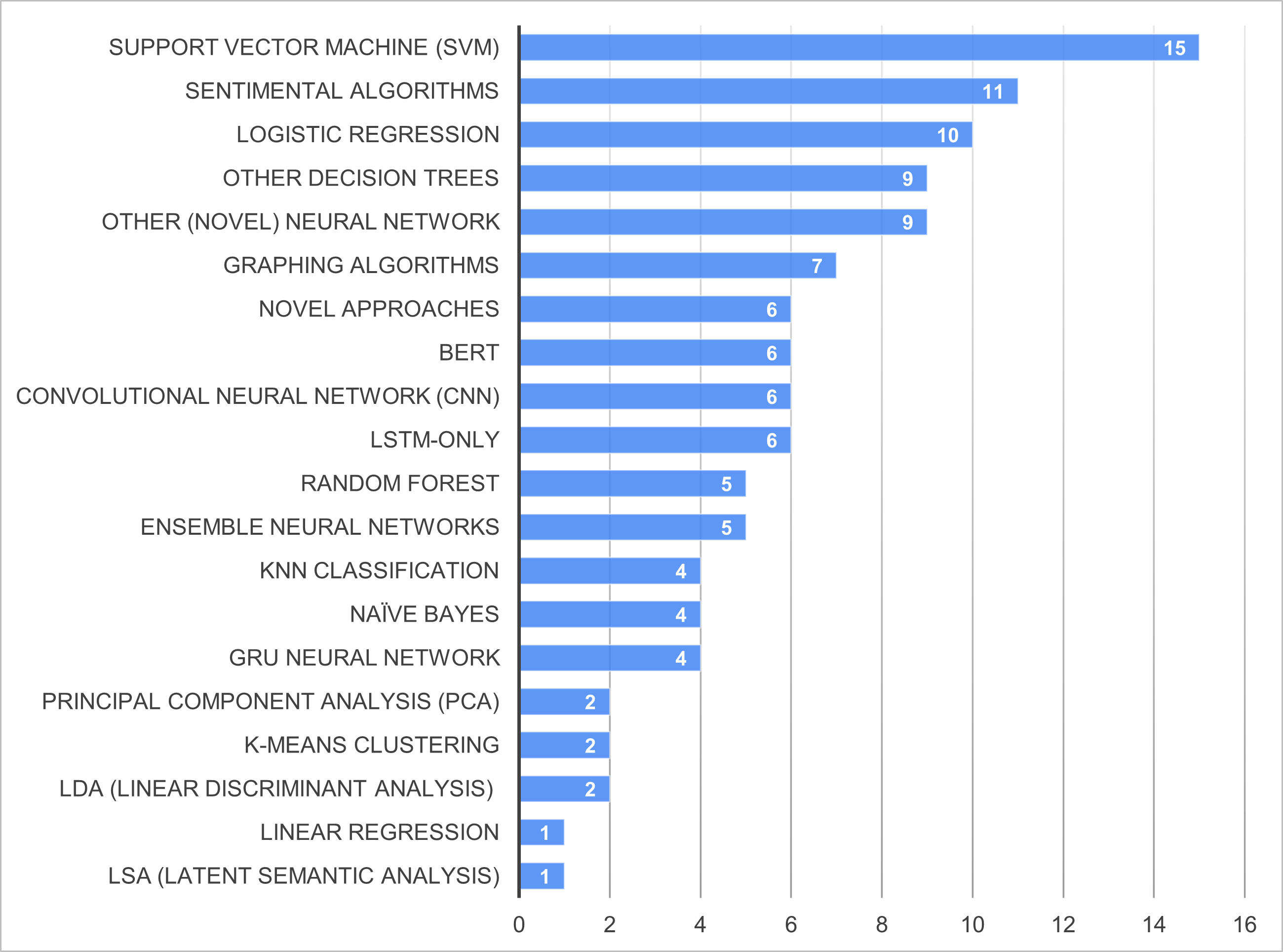}
  \caption{Number of instances of Machine Learning Algorithms (MLAs) used for ERH detection.}
        \label{RQ3-Target-Algorithm-Used-Bar-Chart}
\end{figure}

Studies investigating non-deep MLAs tend to test multiple models, typically Support Vector Machines (SVMs), Random Forest (RF), and Logistic Regression. Figure~\ref{RQ3-Target-Algorithm-Used-Bar-Chart} outlines the distribution of both deep and non-deep approaches, with SVM again the most popular MLA in 15 of the 51 studies.

Non-deep MLAs consistently under-performed for multiclass classification, whereby Ahmad et al. identified that a prior Naïve Bayes model could not distinguish between \textquote{Racism} and \textquote{Extremism} classes due to a low F1-score of 69\%; while their LSTM and GRU model could detect such nuance with a 84\% F1-score~\cite{ahmad19_lstm_cnn_extrem_sentiment}. Likewise, application-specific sentimental algorithms paired with MLAs resulted in lower F1-scores compared to context-sensitive BERT models---\textit{which do not require manual feature extraction}~\cite{wullah21_deep_GPT_BERT, kapil20_deep_nn_multitask_learning_hs, naseem19_lstm_abusive_hs_twitter, macavaney19hs_challenges_solutions}. Sharma et al. claimed that SentiStrength was "...not robust to various sentence structures and hence fails to capture important semantics like sarcasm, negation, indirect words, etc. at the phrase level"~\cite[pg. 5]{sharma_indian_mass_media}---a critique shared in six other non-deep \textit{sentiment scoring} studies~\cite{scrivens20_right-wing_posting_online, Nouh19_understanding_radical_mind, hall20_machines_unified_understanding_radicalizing, lozano17_racism_pol_tweets, zhu19_offensive_tweets_bert, pais20_unsupervised_extreme_sentiments}. Consider the hypothetical case of "I am \textit{not} happy with \textit{those} people", whereby context-insensitive (orderless) embeddings will not detect the negation of \textit{happy} nor the implicit euphemism for \textquote{\textit{those} people}.

Hence, researchers have three options when designing ERH models: 

\begin{enumerate}
\item Avoid complex textual feature extraction and filtering by prioritising DLA development, or
\item Prioritise manual textual and metadata feature extraction, such as psychological signals, emotions, sarcasm, irony, temporal data, and/or
\item Consider community detection (relationship network or topic modelling) features.
\end{enumerate}

\subsubsection{Non-deep Machine Learning Algorithms in Community Detection Studies}\hfill\\
There is a discrepancy in the choice of algorithmic approach compared to NLP-oriented models where less than a third of the community detection studies considered Deep Learning (DL) models~\cite{mashechkin19_russia_caucasus_jihad_ml, Nouh19_understanding_radical_mind, rudinac17_graph_conv_networks}, while NLP-only studies were majority DL (15 of 29). A reason for this discrepancy would be the limited research in social media \textit{network} analysis \textit{without} investigating textual data, instead opting to cluster group affiliation via K-means~\cite{bartal20_roles_trolls_affiliation, benigni18communitymining_UNSUPERVISED, Moussaoui19_twitter_terrorism_communities}, NbClust~\cite{bartal20_roles_trolls_affiliation}, weighted bipartite graphing into Louvain groups~\cite{benigni18communitymining_UNSUPERVISED, benigni2017online_extremism_sustain_it_isis}, and fast greedy clustering algorithms~\cite{Moussaoui19_twitter_terrorism_communities}.

We observed that graphing relationship networks result in two types of classification categories:

\begin{enumerate}
\item \textit{Meso-level affiliation}---semi or unsupervised affiliation of a \textit{user to an extremist group or organisation}, with a bias towards Islamic extremist groups~\cite{benigni2017online_extremism_sustain_it_isis, Moussaoui19_twitter_terrorism_communities, benigni18communitymining_UNSUPERVISED, Nouh19_understanding_radical_mind, shi16_structural_similarity_networks, saif17_semantic_graph_radicalisation, rudinac17_graph_conv_networks}.
\item \textit{Micro-level affiliation}---(semi)supervised \textit{person-to-person} affiliation to an annotated extremist, such as radicalising \textit{influencers}~\cite{bartal20_roles_trolls_affiliation, mashechkin19_russia_caucasus_jihad_ml, buchanan17_ethics_critique_twitter_study}, and legal \textit{person-of-interest} models~\cite{hung16_fbi_radical}.
\end{enumerate}

For organisational affiliation, information for clustering included the use of hashtags shared by suspended extremist Twitter users and unknown (test) users~\cite{rehman21_language_of_isis_twitter, Nouh19_understanding_radical_mind, benigni2017online_extremism_sustain_it_isis, araque20_radical_emotion_signals_similarity}.

For identifying a user's affiliation to other \textit{individuals}, researchers preferred non-textual graph-based algorithms as they reduce memory complexity and avoid the perils in classifying ambiguous text~\cite{Moussaoui19_twitter_terrorism_communities, benigni18communitymining_UNSUPERVISED}. Furthermore, 2016-2019 demonstrated a move from investigative graph search and dynamic heterogeneous graphs via queries in SPARQL~\cite{hung16_fbi_radical, Moussaoui19_twitter_terrorism_communities} towards Louvain grouping on bipartite graphs as a higher-performing (by F1-score) classification method~\cite{benigni2017online_extremism_sustain_it_isis, shi16_structural_similarity_networks, benigni18communitymining_UNSUPERVISED}.

For hybrid NLP-community detection models, researchers mapped text \textit{and} friend, follower/ing, and mention networks via decision trees and kNN~\cite{mashechkin19_russia_caucasus_jihad_ml, Nouh19_understanding_radical_mind}, or used Principal Component Analysis on extracted Wikipedia articles to map the \textit{relationships} between discussed events and entities~\cite{rudinac17_graph_conv_networks}.

An emerging field of community detection for extremism consists of knowledge graphs. Knowledge graphs represents a network of real-world entities, such as events, people, ideologies, situations, or concepts~\cite{data_mining_book_eibe}. Such network representations can be stored within graph databases, word-embeddings, or link-state models~\cite{data_mining_book_eibe, hung16_fbi_radical, yamada20_wikipedia2vec}. Link-state knowledge models consist of undirected graphs where nodes represent entities and edges represent links between entities, such as linking Wikipedia article titles with related articles based on those referenced in the article, as used in Wikipedia2Vec~\cite{yamada20_wikipedia2vec}. Hung et al. consider a novel hybrid OSINT and law-enforcement database graph model-–-which unifies textual n-grams from social media to shared relationships between other individuals and law enforcement events over time~\cite{hung16_fbi_radical}.

Four studies consider model relationships to \textit{individual} extremist affiliates~\cite{bartal20_roles_trolls_affiliation, mashechkin19_russia_caucasus_jihad_ml, buchanan17_ethics_critique_twitter_study, hung16_fbi_radical}. In a direct comparison between text and relationship detection models, Saif et al. observed that text-only semantic analysis outperformed their graph-based network model by a +6\% higher F1-score~\cite{saif17_semantic_graph_radicalisation}.

\subsection{Deep Learning Algorithms (DLAs)}
\begin{figure}[!ht]
  \centering
  \includegraphics[width=\textwidth]{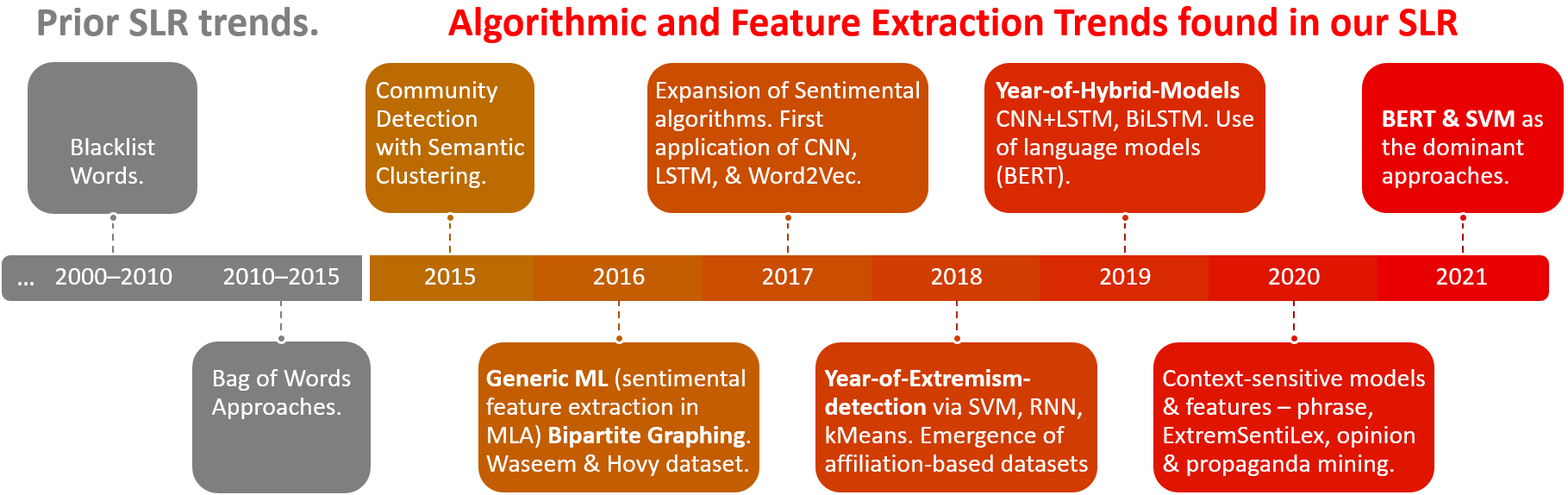}  
  \caption{Patterns of adoption for ERH detection algorithms over time. Colour change ordered by F1-score trend (low to high). Brown = \textasciitilde0.75 F1-score on benchmark datasets, Red = \textasciitilde0.9 F1-score, Grey = No Data.}
  \label{RQ3-algorithms-over-time-chart}
\end{figure}

DL studies are rising, with less than a third of studies including DLAs pre-2019~\cite{bilbao18_political_classification_cnn, rudinac17_graph_conv_networks, gamback17_cnn_hs, badjatiya17_lstm_dl_hs, chandrasekharan17bag_of_com, pitsilis18_recurrent_nn_hs, de-gibert18-white-supremacy-dataset}. The percentage of all studies which included a DLA per year was 0\% in 2016, 27.3\% in 2017~\cite{rudinac17_graph_conv_networks, gamback17_cnn_hs, badjatiya17_lstm_dl_hs, chandrasekharan17bag_of_com}, and 33.3\% in 2018~\cite{bilbao18_political_classification_cnn, pitsilis18_recurrent_nn_hs, de-gibert18-white-supremacy-dataset}, compared to being the majority post-2018 (81.8\% in 2019~\cite{mashechkin19_russia_caucasus_jihad_ml, abubakar19_lstm_gru_comparative_multiclass, ahmad19_lstm_cnn_extrem_sentiment, Nouh19_understanding_radical_mind, johnston20identifying_extremism_dl, macavaney19hs_challenges_solutions, zhu19_offensive_tweets_bert, liu19_transfer_learning_BERT, naseem19_lstm_abusive_hs_twitter, zampieri_etal19_OLID_baseline_tests, zampieri_etal19_OLID_semeval_competitors_results, perez19_benchmark_mulitlingual_immigrant_women}, 54.5\% in 2020~\cite{sharma_indian_mass_media, kapil20_deep_nn_multitask_learning_hs, belcastro20_Italy_Polarisation, mozafari20racialbias_hs_study, kiela20_hateful_memes_original, multioff_meme_images} and 80\% in 2021~\cite{wullah21_deep_GPT_BERT, jia21_covid19_hs_ensemble, Aggarwal21_hateful_meme})---with Figure~~\ref{RQ3-algorithms-over-time-chart} displaying the shift towards DLAs since 2015.

Between 2017-2018 Convolutional Neural Networks (CNN) using Long-Short Term Memory (LSTM), GRU, Recurrent Neural Networks (RNN), or graph-based layers were the sole DLAs~\cite{bilbao18_political_classification_cnn, rudinac17_graph_conv_networks, gamback17_cnn_hs, badjatiya17_lstm_dl_hs, pitsilis18_recurrent_nn_hs, de-gibert18-white-supremacy-dataset}. From 2019-2021, various new approaches such as SenticNet 5~\cite{pais20_unsupervised_extreme_sentiments}, ElMo (Embeddings from Language Model)~\cite{perez19_benchmark_mulitlingual_immigrant_women}, custom neural networks such as an Iterative Opinion Mining using Neural Networks (IOM-NN) model~\cite{belcastro20_Italy_Polarisation}, and attention-based models such as BiLSTM~\cite{zampieri_etal19_OLID_baseline_tests, naseem19_lstm_abusive_hs_twitter, zampieri_etal19_OLID_semeval_competitors_results, perez19_benchmark_mulitlingual_immigrant_women, mozafari20racialbias_hs_study}. Since 2019, there is an emerging consensus towards BERT~\cite{sharma_indian_mass_media, macavaney19hs_challenges_solutions, zhu19_offensive_tweets_bert, liu19_transfer_learning_BERT, wullah21_deep_GPT_BERT, mozafari20racialbias_hs_study, zampieri_etal19_OLID_semeval_competitors_results} due to its easy open-source models on the Hugging Face platform and high performance per Table~\ref{f1-score-benchmark-table}.

\subsubsection{Deep Learning for Community Detection}\hfill\\
Only 3 of the 21 DL studies considered relationship network models~\cite{mashechkin19_russia_caucasus_jihad_ml, Nouh19_understanding_radical_mind, rudinac17_graph_conv_networks}. Whereby, Mashechkin et al. grouped self-proclaimed Jihadist forums and VK users with Jihadist keywords as a "suspicious users" category~\cite{mashechkin19_russia_caucasus_jihad_ml}. Uniquely, the researchers implemented a \textit{Hyperlink-induced Topic Search} (HITS) approach to calculate spatial network proximity between annotated extremists and unknown instances. HITS identifies \textit{hubs}, which are influential web pages as they link to numerous other information sources/pages known as \textit{authorities}~\cite{Kleinberg_HITS_Algo}. The influence of an authority depends on the number of hubs that redirect to the authority. An example of HITS in-action would be an extremist KavazChat forum (a hub) with numerous links to extremist manifestos (authorities)~\cite{Kleinberg_HITS_Algo, mashechkin19_russia_caucasus_jihad_ml}. Evaluating influence in these graph networks requires measuring spatial proximity via \textit{betweenness centrality}~\cite{betweenness_centrality} and depth-first search shortest paths where proximity to a known extremist via following/reposting them constitutes an extremist classification. However, such relations do not accommodate for replies to deescalate, deradicalise, or \textit{oppose} extremist speech.

\subsubsection{Visual-detection Models for ERH Detection}\hfill\\
Despite the emergence of multimedia sources for radicalisation and ideological dissemination, only three studies considered multimodal image and image-text sources---utilising image memes with superimposed text from the Facebook hateful meme dataset~\cite{kiela20_hateful_memes_original} and the MultiOFF meme dataset~\cite{multioff_meme_images}. Only one study considered the post's text (i.e., text not displayed on the image itself) as context via the multimedia Stormfront post \textit{and} image data from Rudinac et al.~\cite{rudinac17_graph_conv_networks}.

For the Facebook hateful meme and MultiOFF datasets include images with superimposed captions~\cite{kiela20_hateful_memes_original, multioff_meme_images}. Both Kiela et al.~\cite{kiela20_hateful_memes_original} and Aggarwal et al.~\cite{Aggarwal21_hateful_meme} extract caption text via Optical Character Recognition (OCR) models–--a computer vision technique to convert images with printed/visual text into machine-encoded text~\cite{data_mining_book_eibe}. The three hateful meme studies utilised either both (multimodal) or one (unimodal) of the image and its caption~\cite{kiela20_hateful_memes_original, Aggarwal21_hateful_meme, multioff_meme_images}. The multimodal Visual BERT-COCO model attained the highest accuracy of 69.47\%, compared to 62.8\% for a caption text-only classifier or 52.73\% for image only, 64\% for the ResNet152 model~\cite{Aggarwal21_hateful_meme}; and 84.70\% for the baseline (human)~\cite{kiela20_hateful_memes_original}.

\begin{figure}[!ht]
  \centering
  \includegraphics[width=\textwidth]{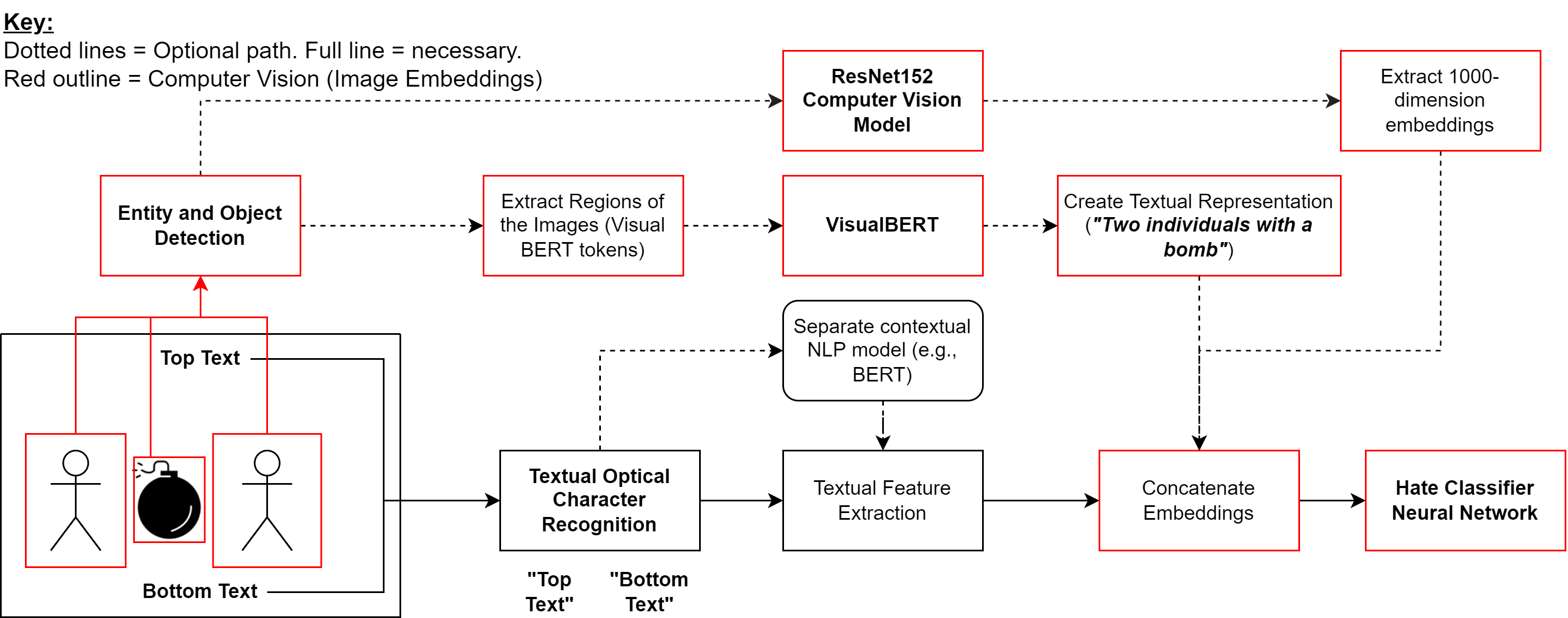}  
  \caption{Deep learning pipeline for visual-text ERH detection based on the hateful meme studies~\cite{kiela20_hateful_memes_original, Aggarwal21_hateful_meme,  multioff_meme_images}.}
  \label{fig:visual_models_architecture}
\end{figure}

The highest performing multimodal model relied on Visual BERT~\cite{kiela20_hateful_memes_original}. Visual BERT extends textual BERT by merging BERT’s self-attention mechanism to align input text with visual regions of an image~\cite{visualbert}. VisualBERT models are typically pretrained for object detection, segmentation, and captioning using the generic Common Objects in COntext (COCO) dataset~\cite{coco_dataset}, such that the model can segment and create a textual description of the objects behind an image such as “two \textit{terrorists} posing with a \textit{bomb}” (Figure~\ref{fig:visual_models_architecture}). Training otherwise acts the same as BERT-–-which involves masking certain words/tokens from a textual description of the image of what the image depicts, and VisualBERT predicting the masked token(s) based on the image regions. We aggregate and generalise all visual ERH detection studies architectural pipelines in Figure~\ref{fig:visual_models_architecture}~\cite{Aggarwal21_hateful_meme, kiela20_hateful_memes_original, multioff_meme_images, rudinac17_graph_conv_networks}.

No hateful meme dataset studies consider accompanying text from the original post. This raises concerns regarding posts satirising, reporting, or providing counter-speech on hateful memes.

Only one study investigated a contextual textual post \textit{and} accompanying images through a proposed Graph Convolutional Neural Network (GCNN) model~\cite{rudinac17_graph_conv_networks}. This GCNN approach extracted semantic concepts extracted from Wikipedia, such as identifying that an image was a \textit{KKK rally}---attaining a 0.2421 F1-score for detecting forum thread affiliation across 40 Stormfront threads~\cite{rudinac17_graph_conv_networks}.

\section{Model Performance Evaluation, Validation, and Challenges}
\label{section:rq4}
\textit{What are the highest performing models, and what challenges exist in cross-examining them?}

\begin{figure}[!ht]
  \centering
  \includegraphics[width=0.87\linewidth]{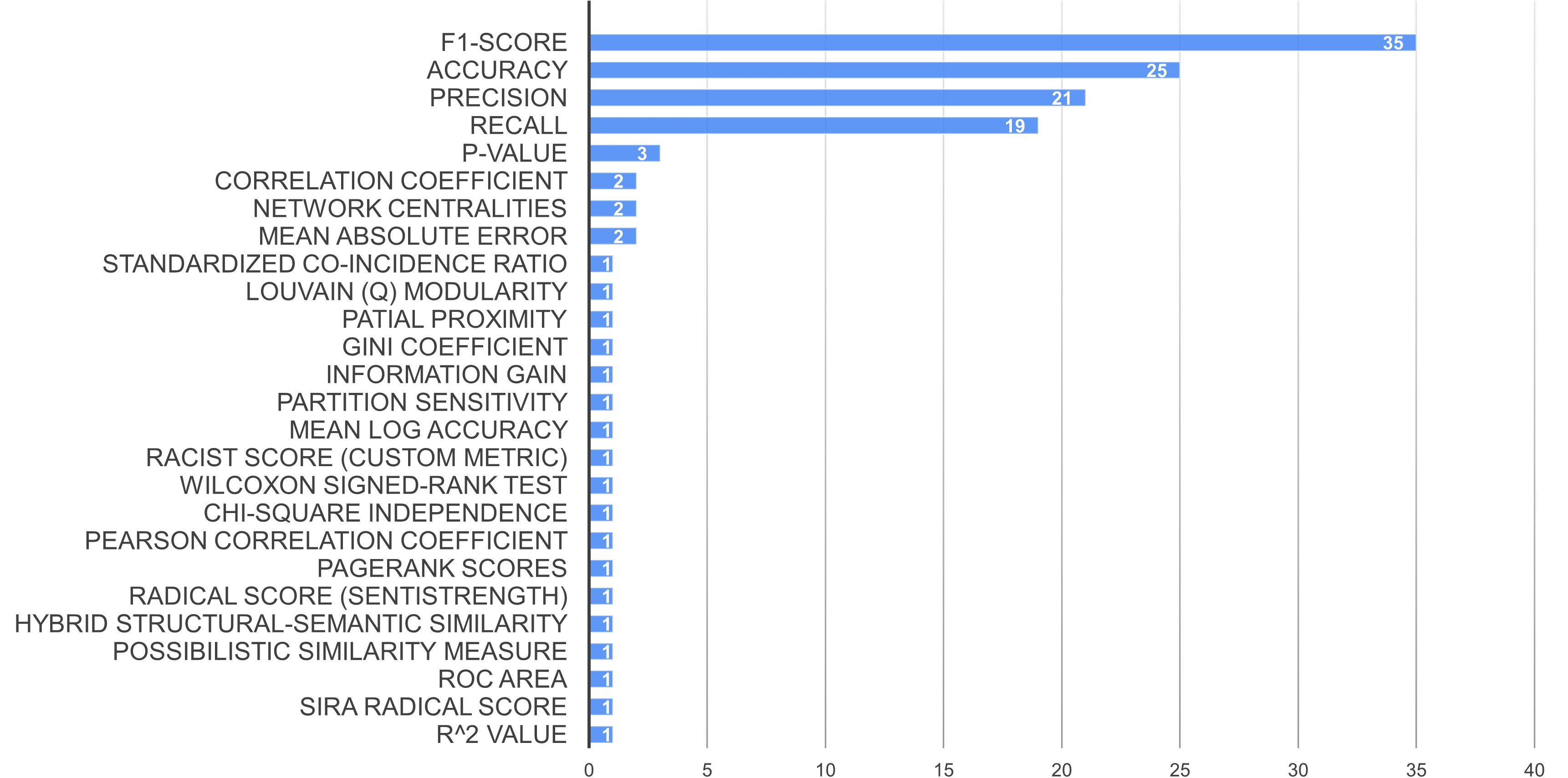}
  \caption{Distribution of metrics used across the 51 studies---demonstrating a lack of standardisation.}
        \label{RQ4-metrics-used-bar-chart}
\end{figure}

Evaluating model performance presents three core challenges for future researchers:

\begin{enumerate}
\item \textbf{Dataset domain differences}---which may include or exclude relevant features (e.g., gender, location, or sentiment) and may involve numerous languages or groups (e.g., Islamic extremists vs. white supremacists) who will express themselves with different lexicons~\cite{johnston20identifying_extremism_dl, borum2011radicalizationsocialsciencelitreview}.
\item \textbf{Criteria differences}---different standards for ERH definitions, criteria, filtering, and annotation threaten cross-dataset analysis between models~\cite{waseem16_are_you_racist_dataset, buchanan17_ethics_critique_twitter_study}. Binary classification can result in higher accuracy compared to classifying nuanced and non-trivial subsets of hate such as racism/sexism~\cite{waseem16_are_you_racist_dataset}, overt/covert aggression~\cite{kumar_etal18_trac_benchmark_dataset}, or hateful group affiliation~\cite{johnston20identifying_extremism_dl}.
\item \textbf{Varying and non-standard choice of metrics}---Figure~\ref{RQ4-metrics-used-bar-chart} displays the 28 metrics, which vary depending on whether the study investigates community detection via closeness, in-betweenness, and eigenvectors; or NLP, often via accuracy, precision, and F1-scores.
\end{enumerate}

\begin{table}[!b]
  \caption{Models ranked by macro F1-score for the benchmark datasets across studies (inter-study evaluation).}
  \label{f1-score-benchmark-table}
  \small
  \begin{tabular}{p{0.15\linewidth}p{0.25\linewidth}p{0.25\linewidth}p{0.25\linewidth}}
    \toprule
    Dataset&1st Highest&2nd Highest&3rd Highest\\
    \midrule
    Waseem and Hovy~\cite{waseem16hateful_symbols_hs_twitter}&0.966 (BERT with GPT-2 fine-tuned dataset~\cite{wullah21_deep_GPT_BERT})&0.932 (Ensemble RNN~\cite{pitsilis18_recurrent_nn_hs})&0.930 (LSTM + Random Embedding + GBDT~\cite{badjatiya17_lstm_dl_hs})\\
    
    FifthTribe~\cite{kaggle_isis_dataset}&1.0 (RF~\cite{Nouh19_understanding_radical_mind})&0.991-0.862 (SVM~\cite{araque20_radical_emotion_signals_similarity})&0.87 (SVM~\cite{rehman21_language_of_isis_twitter})\\
    
    de Gibert~\cite{de-gibert18-white-supremacy-dataset}&0.859 (SP-MTL LSTM, CNN and GRU Ensemble~\cite{kapil20_deep_nn_multitask_learning_hs})&0.82 (BERT~\cite{macavaney19hs_challenges_solutions})&0.73 (LSTM baseline metric~\cite{de-gibert18-white-supremacy-dataset})\\

    TRAC FB~\cite{kumar_etal18_trac_benchmark_dataset}&0.695 (CNN + GRU~\cite{kapil20_deep_nn_multitask_learning_hs})&0.64 (LSTM~\cite{kumar_etal18_trac_benchmark_dataset})&0.548 (FEDA SVM~\cite{karan1818_cross_domain})\\
    
    Hatebase Twitter~\cite{Davidson17_benchmark_hs_study}&0.923 (BiLSTM with Attention modeling~\cite{naseem19_lstm_abusive_hs_twitter})&0.92 (BERTbase+CNN / BiLSTM~\cite{mozafari20racialbias_hs_study}, 0.86 (with racial/sexual debiasing module)&0.912 (Neural Ensemble~\cite{macavaney19hs_challenges_solutions})\\
    
    HatEval~\cite{basile19_benchmark_mulitlingual_immigrant_women_dataset}&0.7481 (Neural Ensemble~\cite{macavaney19hs_challenges_solutions})&0.738 (LSTM-ELMo+BoW)~\cite{perez19_benchmark_mulitlingual_immigrant_women}&0.695 (BERT with GPT-2 fine-tuned dataset~\cite{wullah21_deep_GPT_BERT})\\
    
    OffensEval~\cite{zampieri_etal19_OLID_baseline_tests}&0.924 (SP-MTL CNN~\cite{kapil20_deep_nn_multitask_learning_hs})&0.839 (BERT~\cite{zhu19_offensive_tweets_bert})&0.829 (BERT 3-epochs~\cite{liu19_transfer_learning_BERT})\\
  \bottomrule
\end{tabular}
\end{table}

\subsection{Benchmark Dataset Performance (Inter-study Evaluation)}
We use macro F1-score as the target metric as it balances true and false positives among all classes, and is a shared metric across the benchmark datasets~\cite{waseem16hateful_symbols_hs_twitter, kaggle_isis_dataset, de-gibert18-white-supremacy-dataset, basile19_benchmark_mulitlingual_immigrant_women_dataset, Davidson17_benchmark_hs_study, zampieri_etal19_OLID_baseline_tests}. Table~\ref{f1-score-benchmark-table} outlines that the highest F1-scoring models reflect the move towards context-sensitive DLAs like BERT, as also displayed in Figure~\ref{RQ3-algorithms-over-time-chart}. SVMs and a single instance of a Random Forest classifier on sentimental features were the last standing non-deep MLAs~\cite{Nouh19_understanding_radical_mind, araque20_radical_emotion_signals_similarity, rehman21_language_of_isis_twitter, karan1818_cross_domain}. Given the variety of MLA and DLAs (Figure~\ref{RQ3-Target-Algorithm-Used-Bar-Chart}), approaches that frequently \textit{underperformed} included Word2Vec, non-ensemble neural networks such as CNN-only models, and baseline models~\cite{Davidson17_benchmark_hs_study, Grover19_alt-right_subreddits}. These baseline models include the HateSonar model by Davidson et al.~\cite{Davidson17_benchmark_hs_study}, Waseem and Hovy's n-grams and gender-based approach~\cite{waseem16hateful_symbols_hs_twitter}, LSTM model by de Gibert et al.~\cite{de-gibert18-white-supremacy-dataset}, and C-Support Vector Classification by Basile et al.~\cite{basile19_benchmark_mulitlingual_immigrant_women_dataset}. No studies discuss memory or computational complexity, an area worthy of future research as expanded in our \textit{Supplementary Material's} Section 3.

\subsection{Community Detection Performance}
While community detection models tend to produce F1-scores \textasciitilde0.15 lower than DLAs~\cite{bartal20_roles_trolls_affiliation, benigni18communitymining_UNSUPERVISED, Moussaoui19_twitter_terrorism_communities, benigni2017online_extremism_sustain_it_isis, shi16_structural_similarity_networks, hung16_fbi_radical}, these comparisons rely on \textit{different} datasets/metrics. Shi and Macy recommended using \textit{Standardised Cosine Ratio} as the standardised metric for structural similarity in network analysis, as it is not biased towards the majority class, unlike Jaccard or cosine similarity~\cite{shi16_structural_similarity_networks}. For community detection models on the same pro/anti-ISIS dataset~\cite{kaggle_isis_dataset}, F1-scores ranged from 0.74-0.93~\cite{benigni2017online_extremism_sustain_it_isis, araque20_radical_emotion_signals_similarity, rehman21_language_of_isis_twitter}.

Only one study cross-examined text and network features~\cite{saif17_semantic_graph_radicalisation}, with a hybrid dataset consisting of annotated anti/pro-ISIS users' posts \textit{and} number of followers/ing, hashtags, mentions, and location. \textit{Text-only semantic analysis} outperformed their network-only model (0.923 F1 vs. 0.866 respectively)~\cite{saif17_semantic_graph_radicalisation}. However, topic (hashtag) clustering and lexicon-based \textit{sentiment} detection via SentiStrength \textit{underperformed} compared to the network-only approach by a 0.07-0.1 lower F1~\cite{saif17_semantic_graph_radicalisation}. Thus, unsupervised clustering models are ideal for \textit{temporal radicalisation detection} and identification of \textit{emergent or unknown} groups or ideologies. There is insufficient evidence to conclude whether community detection is superior to NLP due to the lack of shared NLP-network datasets.

For supervised community detection tasks, researchers~\cite{saif17_semantic_graph_radicalisation, Moussaoui19_twitter_terrorism_communities, benigni2017online_extremism_sustain_it_isis} used network features via Naïve Bayes~\cite{Moussaoui19_twitter_terrorism_communities}, k-means~\cite{bartal20_roles_trolls_affiliation, benigni18communitymining_UNSUPERVISED, Moussaoui19_twitter_terrorism_communities}, SVM~\cite{Nouh19_understanding_radical_mind}, and decision trees~\cite{Nouh19_understanding_radical_mind, mashechkin19_russia_caucasus_jihad_ml}. The highest F1-score community detection model was a \textit{hybrid NLP and community detection model} using network features, keywords and metadata (i.e., language, time, location, tweet/retweet status, and whether the post contained links or media) with a Naïve Bayes classifier---attaining a 0.89 F1-score~\cite{Moussaoui19_twitter_terrorism_communities}.

\section{Future Research Directions}
\label{section:futureworkERH}
In this section, we offer an alternate to the \textit{radicalisation = extremism = political hate speech} consensus from RQ1 and models observed in RQ3/4 to present a new framework for \textit{delineating and expanding} ERH for future work. Overall, we propose an uptake roadmap for ERH context mining to expand the field into new research domains, deployments for industries, and elicit governance requirements.

\subsection{Ideological Isomorphism---a Novel Framework for \textit{Radicalisation Detection}}
\begin{Definition}{Ideological Isomorphism (Computational Definition for Radicalisation)}{def_ideological_isomorphism}
  \textit{The temporal movement of one's belief space and network of interactions from a point of normalcy towards an extremist belief space. It is an approach to detecting radicalisation with an emphasis on non-hateful sentiment as ringleaders and/or influencers pull and absorb others towards their hateful group's identity, relationships, and beliefs.}
\end{Definition}

As outlined in our novel tree-diagram dissection of ERH definitions to their computational approach in Figure~\ref{ERH-definition-tree}, there is considerable overlap in approaches between the otherwise unique fields of \textit{extremism}, \textit{radicalisation} and \textit{politicised hate speech}. Radicalisation’s working definition suffers from ambiguity in the majority of studies due to its interchangeability towards extremist affiliation and no considerations for temporal changes. Radicalisation's computational definition should reflect a behavioural, psychological, and ideological move towards extremism over time. While extremist ideologies and outwards discourse towards victim groups may be manifestly hateful, radicalisation towards target audiences may involve non-hateful uniting and persuasive speech~\cite{Kinnvall21_psychology_extremist_identification}. Hence, we propose that \textit{radicalisation detection should not be a single-post classification}. Rather, models should consider micro (individual), meso (group dynamics), and macro (global events and trends) relations. The roots for radicalisation result from an individual's perceptions of injustice, threat, and self-affecting fears on a micro-level. On a meso-level, this can include the rise of community clusters based on topics and relationships. Socially, a radicalised user draws on an extremist group's legitimacy, connections and group identity, trends, culture and memes~\cite{Kinnvall21_psychology_extremist_identification, ludemann18_polemics_4chan, meindl17_mass_shootings_imitation}.

Hence mapping ideological isomorphism requires temporal modelling to:

\begin{enumerate}
\item Detect the role of users or groups polarising or pulling others towards extreme belief spaces (i.e., \textit{ideological isomorphism}), akin to detecting online influencers~\cite{bartal20_roles_trolls_affiliation, Grover19_alt-right_subreddits, Moussaoui19_twitter_terrorism_communities, priyank21_ringleader_detection}. Studies should also consider the role of alienation as a radicalising factor via farthest-first clustering.
\item Further research into the role of friendship and persuasion by adapting sentimental approaches to consider positive reinforcement towards hateful ideologies akin to existing research in detecting psycho-behavioural signals~\cite{Nouh19_understanding_radical_mind}. Furthermore, there lacks research in computationally detecting social factors such as suicidal ideation or mental health.
\item Investigate the interactions between groups across social media platforms as radicalisers themselves, such as the promotion of extremist content by recommendation algorithms.
\item Utilise community detection metrics such as centrality, Jaccard similarity, and semantic similarity over time as measurements for classifying radicalisation for meso-level NLP (topic) and graph-based (relational) clustering, leaving content moderation as a separate task.
\item Consider the role of satire, journalism, and martyrs as areas for radicalisation clustering.
\end{enumerate}

\subsection{Morphological Mapping and Consensus-building---a novel computationally-grounded framework for extremism detection}
\begin{Definition}{Morphological Mapping and Consensus-building (Extremism)}{def_mapping_extremism}
  \textit{The congregation of users into collective identities (\textquote{in-groups}) in support of manifestly unlawful actions or ideas.}
\end{Definition}
While ideological isomorphism focuses on micro-level inter-personal relations, morphological mapping pertains to clustering meso-level beliefs and community networks to extremist ideologies. While we discovered various affiliation-based clustering approaches, no studies identified novel or emergent movements. Establishing a ground truth for a novel extremist organisation is challenging if such groups are decentralised or volatile. Hence, we recommend using manifestos, particularly unconsidered far-right sources, and influential offline and online extremists as a benchmark for identifying martyrdom networks and new organisations. Areas for future research include investigating the role of trolls, physical world attacks, or misinformation in narrative-building.

Our morphological mapping framework proposes to delineate \textit{Extremism} by considering the role of group identity and ideological themes behind hate speech by considering \textit{affiliation} across users and posts. When targeting extremism, pledging \textquote{support} to a terrorist organisation may not violate context-insensitive BoW hate speech classifiers–--hence it is not appropriate to categorise extremist affiliation under the same guise as post-by-post hate speech. Currently, extremism detection constitutes a binary \textquote{pro vs anti group} classification, which fails to capture the inner trends of radicalisation from peaceful, to fringe beliefs, to committing to violent-inducing beliefs online, and potentially to offline extremism. Investigating semi or unsupervised clustering (mapping) of groups will also aid Facebook's commitment to moderating militarised social movements, violence-inducing conspiracy networks, terrorist organisations, or hate speech inducing communities~\cite{facebookhate_stnd}.

Thus, we propose four prerequisites for studies to fall under the \textit{extremism detection} category:

\begin{enumerate}
\item Investigate the interactions and similarities between groups on mainstream and anonymous platforms to map group dynamics and extremist networks. For privacy, we recommend group-level (non-individualistic) network and semantic clustering.
\item Map affiliation and group dynamics. Given the lack of definitions for extreme \textit{affiliation}, we recommend using Facebook's definition of affiliation as a basis---being the positive praise of a designated entity or event, substantive (financial) support, or representation on behalf of a group (i.e., membership/pledges)~\cite{facebookhate_stnd}.
\item Investigate hateful \textit{and} non-hateful community interactions, memes and trends, that reinforce group cohesion.
\item Map affiliation as a clustering task, akin to our proposed radicalisation framework but without the temporal component.
\end{enumerate}

\subsection{Outwards Dissemination--–\textquote{traditional} hate speech detection updated}
\begin{Definition}{Outwards Dissemination (Hate Speech)}{def_outwards_dissemination_hate_speech}
  \textit{Targeted, harassing, or violence-inducing speech towards other members or groups based on protected characteristics.}
\end{Definition}
Hence, the projection and mainstreaming of hateful ideologies through speech, text, images, and videos requires an \textit{outwards dissemination} of views shared by extremists, such as racism. The outwards dissemination of hate is a strictly NLP (text) and computer vision (entity and object) classification problem. We delineate hate speech with affiliation to violent extremist groups as such misappropriation could have devastating effects on one's image, well-being, and safety~\cite{bacchini19_african_face_recognition, buchanan17_ethics_critique_twitter_study, conway21_extremism_terrorism_research_ethics_study}.

All researchers should be aware that malicious actors may exploit existing ERH models for injurious surveillance and censorship. Future work should also consider the impacts of labels on society at large, whereby terms such as \textquote{far-right} as an alias for white supremacy is both misleading, infers a \textquote{right vs wrong} left-to-right spectrum, and ambiguous. We recommended decoupling religious contexts in favour of technical terms such as \textquote{radical Islamic extremism} or \textquote{terror-supporting martyrdom} to avoid grouping religiosity to a political ideology and terrorism.

Thus, we propose three key prerequisites for a study to be in the hate speech category:

\begin{enumerate}
\item Investigates textual or multimedia interactions only, whereby detecting cyber-bullying or extremist community networks should be separate tasks.
\item Decouple affiliation where possible. For instance, white supremacy instead of far-right (an ambiguous term) or organisational affiliation.
\item Consider models which include latent information, such as news, entities, or implied hate. Datasets should explain each classification with categories for disinformation and fallacies.
\end{enumerate}

Future work in outlining \textit{hate speech} would be a systematic socio-legal cross-examination of hate speech laws from governments and policies from social media platforms---including the emerging consensus vis-à-vis the harmonised EU Code of Conduct for countering illegal hate speech~\cite{eu_code_of_conduct_16}.

\subsection{Uptake Roadmap for Researchers, Industry, and Government}
\begin{figure}[!htbp]
  \centering
  \includegraphics[width=0.75\linewidth]{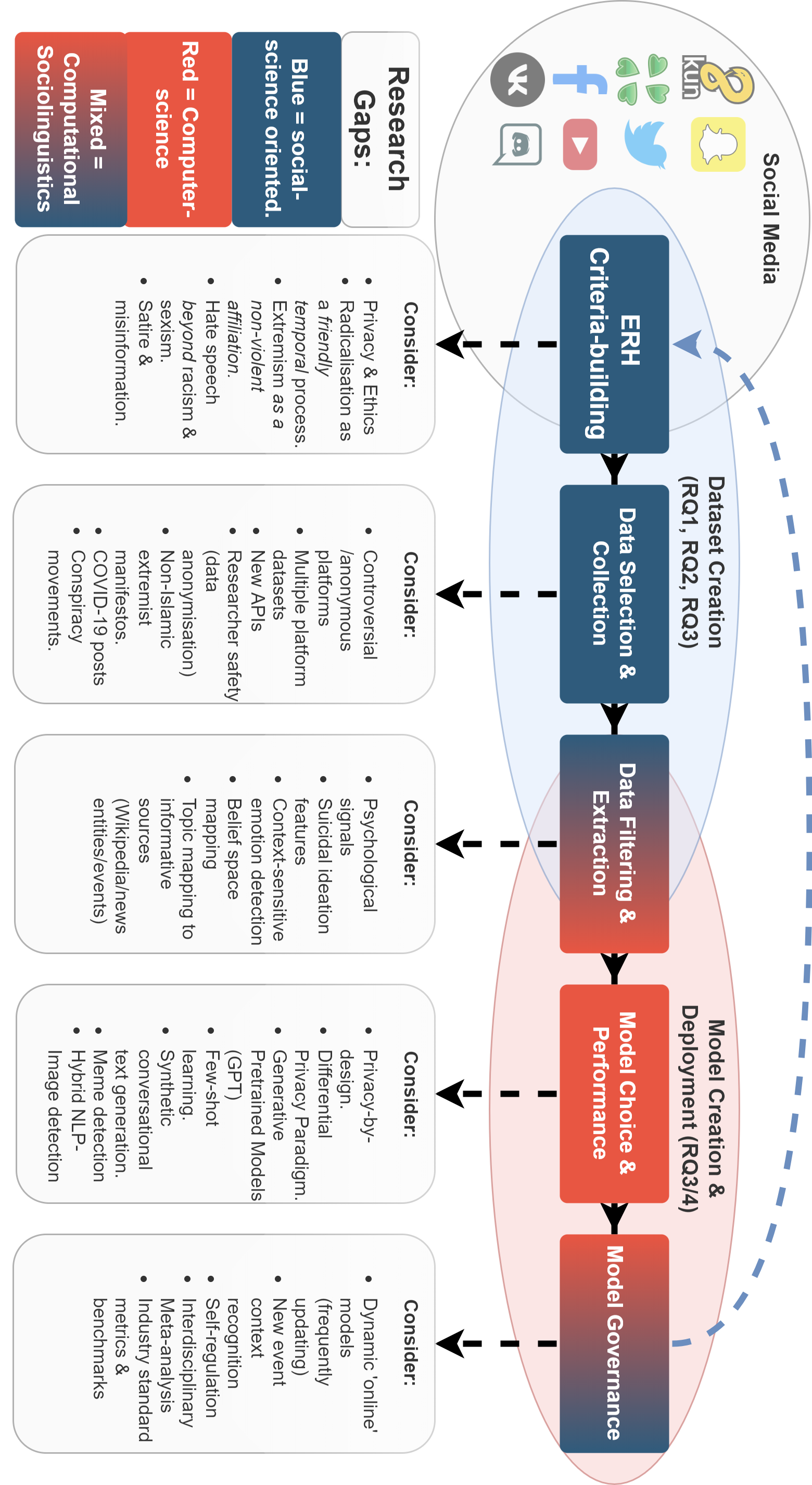}
  \caption{ERH Context Mining pipeline---with key identified research gaps.}
        \label{ERH_Context_Mining_Pipeline_v3}
\end{figure}

We present a pipeline for researchers, industries, and government analysts to approach ERH context mining per Figure~\ref{ERH_Context_Mining_Pipeline_v3}. In addition to this summary visualisation of our key dataset and model recommendations, we expand on our actionable recommendations for immediate next steps and long-term software requirements for ERH detection in our supplementary material.

Future SLRs should consider a mixture of academic studies, grey material, and technical reports to further encompass our proposed ERH context mining field's socio-legal component and explore \textit{industry} approaches. We recommend transforming our ethical recommendations for responsible research outlined in our SLR design into formalised interdisciplinary guidelines to protect privacy and researcher safety. ERH is never a singular end-goal, post, or unexpected event. Hence, detecting erroneous behaviour emanating from mental health crises can both avoid ERH online and offline, and present avenues for cooperation with third-parties such as suicide prevention and counselling groups. Finally, we recommend searching for \textit{multimedia-only} studies including for livestreams.

\section{Conclusion}
\label{section:conclusion}

ERH context mining is a novel and wide field that funnels to one fundamental aim---the pursuit to computationally identify hateful content to enact accurate content moderation. In our work, we harmonised \textit{Extremist} affiliation, \textit{Radicalisation}, and \textit{Hate Speech in politicised discussions} from 2015-2021 in a socio-technical context to deconstruct and decouple all three fields of our proposed \textit{ERH context mining} framework. Hence, we propose a novel framework consisting of \textit{ideological isomorphism} (radicalisation), \textit{morphological mapping} (extremism), and \textit{outwards dissemination} (politicised hate speech) based on our findings in RQ1. While hate speech included racism and sexism, other forms of discrimination were seldom considered. Extremism and radicalisation frequently targeted Islamic groups, particularly from US and European researchers. Binary post-by-post classification remain the dominant approach despite the complexity of online discourse.

There is a clear and present danger in current academia emanating from the unresolved biases in dataset collection, annotation, and algorithmic approaches per RQ2. We observed a recurring lack of consideration for satire/comedic posts, misinformation, or multimedia sources. Likewise, data lacked nuance without contextual replies or conversational dynamics, and were skewed towards the US and Europe---with the global south, indigenous peoples, and Oceania all marginalised.

Computationally, we identified that deep learning algorithms result in higher F1-scores at the expense of algorithmic complexity via RQ3/4. Context-sensitive neural language DLAs and SVM with sentimental, semantic, and network-based features outperformed models found in prior SLRs. However, state-of-the-art models still lack a contextual understanding of emergent entities, conversational dynamics, events, entities and ethno-linguistic differences. To combat injurious censorship and vigilantism, we recommended several areas for future work in context-sensitive models, researcher ethics, and a novel approach to framing ERH in SLRs and computational studies.

The poor design and abuse of social media threatens the fabric of society and democracy. Researchers, industries, and governments must consider the full start-to-finish ecosystem to \textit{ERH context mining} to understand the data, their criteria, and model performance. Without a holistic approach to delineating and evaluating \textit{Extremism}, \textit{Radicalisation}, and \textit{Hate Speech}, threat actors (extremists, bots, trolls, (non-)state actors) will continue to exploit and undermine content moderation systems. Hence, informed, accurate and ethical content moderation are core to responsible platform governance \textit{while} averting injurious censorship from biased models.

\bibliographystyle{ACM-Reference-Format}
\bibliography{Main}

\appendix

\end{document}


\title{Supplementary Material for:\\Down the Rabbit Hole: Detecting Online Extremism, Radicalisation, and Politicised Hate Speech}


\author{Jarod Govers}
\orcid{0000-0002-7648-318X}
\email{jg199@students.waikato.ac.nz}
\affiliation{
  \institution{ORKA Lab, Department of Software Engineering, University of Waikato}
  \streetaddress{Gate 1, Knighton Road}
  \city{Hamilton}
  \state{Waikato}
  \country{NZ}
  \postcode{3216}
}

\author{Philip Feldman}
\orcid{0000-0001-6164-6620}
\email{philip.feldman@asrcfederal.com}
\affiliation{
  \institution{ASRC Federal}
  \city{Beltsville}
  \state{Maryland}
  \country{US}
}

\author{Aaron Dant}
\orcid{0000-0001-5852-5262}
\email{aaron.dant@asrcfederal.com}
\affiliation{
  \institution{ASRC Federal}
  \city{Beltsville}
  \state{Maryland}
  \country{US}
}

\author{Panos Patros}
\orcid{0000-0002-1366-9411}
\email{panos.patros@waikato.ac.nz}
\affiliation{%
  \institution{ORKA Lab, Department of Software Engineering, University of Waikato}
  \streetaddress{Gate 1, Knighton Road}
  \city{Hamilton}
  \state{Waikato}
  \country{NZ}
  \postcode{3216}
}

\renewcommand{\shortauthors}{Govers et al.}

\settopmatter{printacmref=false} 
\maketitle

\noindent\makebox[\linewidth]{\rule{\linewidth}{0.6pt}}

\tableofcontents
\section{Definitions---The Algorithm Handbook}
This supplementary material document includes the supplementary material referenced in the main \textit{Down the Rabbit Hole: Detecting Online Extremism, Radicalisation, and Politicised Hate Speech} Systematic Literature Review. This document offers a \textquote{dictionary/look-up table} for the core algorithmic architectures for the non-deep machine learning and deep learning models mentioned throughout the SLR, alongside other side findings and design considerations. We contextualise the relevant strengths and weaknesses of the various algorithmic approaches for text and visual models for ERH detection. No new findings are in this handbook/\textquote{look-up table}. Hence, those familiar with the models listed in the contents above need not read this section.

\subsection{Definitions for Traditional (non-deep) Machine Learning Algorithms}
We aggregate common and historic non-deep machine learning algorithms into the \textquote{traditional} MLA category. Hence, this section defines each of the baseline models used for textual or community detection models---consisting of:

\begin{enumerate}
\item Sentimental Bag of Words approaches,
\item Naïve Bayes,
\item Decision Trees,
\item Support Vector Machines,
\item Clustering Models.
\end{enumerate}

\subsubsection{Bag of Words (BoW)}\hfill\\
BoW approaches simplify complex contextual sentences into a multiset (\textquote{bag}) of individual words by assigning a value or probability to each word in its relation to a specific document class. For instance, a BoW approach would deconstruct the contiguous sentence, “The Eldian people are the spawn of the devil” (where Eldian is a fictitious race), into an unordered bag of individual words. While \textquote{are}, \textquote{the} are unlikely to have a considerable influence on whether a sentence is hate speech or not, the use of \textquote{devil} and \textquote{Eldian [race]} is more frequently paired in hate speech than for non-hateful/off-topic text. The disregard of word order and the relationship of BoW approaches, and MLA models at large, constitute context-insensitive models. For instance, a BoW model does not know that  \textquote{I love the Eldian people but hate their food} is paring love -> Eldian, and hate -> food, and thus would consider \textquote{I hate the Eldian people but love their food} as identical. Likewise, BoW approaches do not consider alternate word meanings/uses (e.g., \textquote{I ran for government} vs. \textquote{I ran away}). Nonetheless, BoW approaches are core to word-specific \textquote{blacklists} in content moderation, such as banning users who use slurs in a post. However, for nuanced and often politicised discussions on controversial topics, simple blacklists can lead to injurious censorship---due to the context and use of such words.

\begin{figure}[ht]
  \centering
  \includegraphics[width=\linewidth]{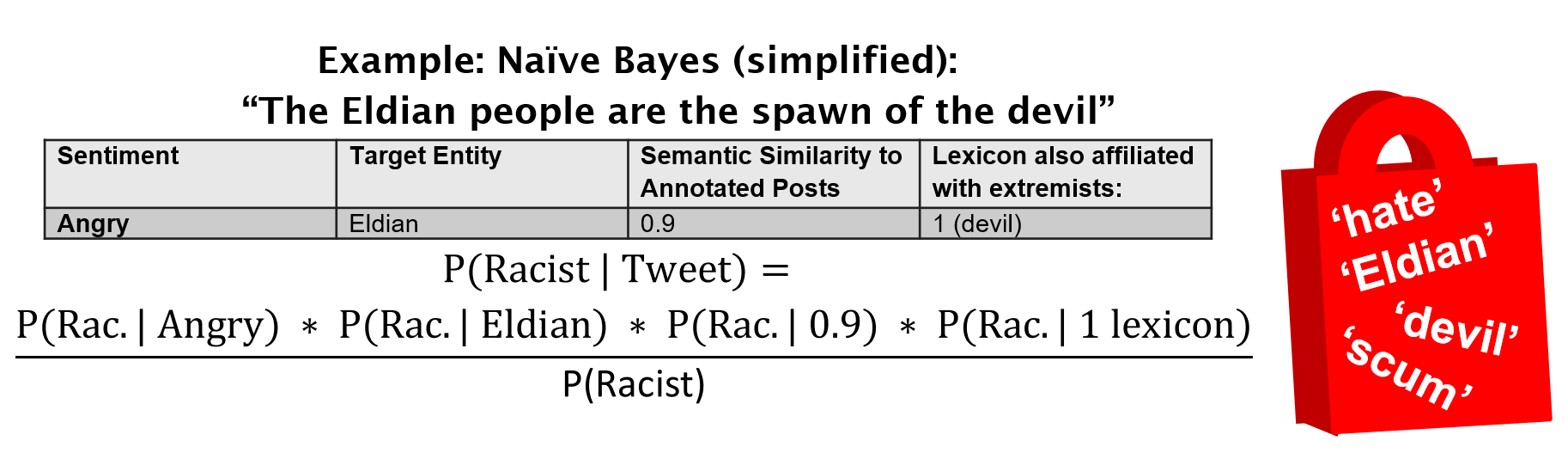}
  \caption{An abstracted example of Bag of Words approach within a Naïve Bayes classifier---demonstrating its lack of context sensitivity and the focus on key \textquote{racist} words for ERH detection tasks.}
  \label{naive-bayes-figure}
\end{figure}

Sentimental algorithms, such as SentiStrength~\cite{sentistrength_original_study} aggregate individual words into individual emotions---whereby \textquote{love} indicates a positive sentiment, while \textquote{hate} generally appears in vitriolic speech. Figure~\ref{naive-bayes-figure} outlines an abstracted representation of the sentiment classification based on the average sentiment score of a sentence. However, the context-insensitive BoW models again fails for nuanced cases, whereby Sharma et al.~\cite{sharma_indian_mass_media} identified that SentiStrength cannot detect negations (e.g., “I am NOT happy” where happy skews the final sentiment scores).

\subsubsection{Naïve Bayes}\hfill\\
Naïve Bayes classifiers represent types of probabilistic classifiers utilising Bayes theorem with the assumption that the influence of each variable for classification is independent of each other (i.e., \textit{naïve})~\cite{data_mining_book_eibe}. For document classification, notable features are assigned a probability for their occurrence given a specific class. For instance, a hate speech post that has an angry sentiment may have a P(0.8) (Probability of 80\%) of being hateful, given that a test hate speech dataset may be 80\% angry speech. Bayes rule represents these chains of (assumed) independent/unrelated probabilities to form a final probability for a test instance.

\textbf{Notable features for probability models include:}
\begin{itemize}
\item \textit{Textual features}---(e.g., sentimental scores, appearance of certain slurs/terms), 
\item \textit{Network data}---(e.g., probability that someone who is friends with a supremacist is also a supremacist, retweet relationships),
\item \textit{Metadata}---(e.g., length of a post, readability via a Flesch Reading Ease score, number of posts).
\end{itemize}

In the example of Figure~\ref{naive-bayes-figure}, the probability that the tweet is racist depends on the probability that the racist tweet is angry, contains racial terms (\textquote{Eldian}), the semantic similarity between known hate speech posts, and the appearance of a negative lexicon. Naïve Bayes can be a final classifier for aggregating context-sensitive embeddings (e.g., deep learning models) and multiple \textquote{ensembles} of approaches/models---via chaining their probabilities together with this Bayes rule. 

\subsubsection{Decision Trees}\hfill\\
\begin{figure}[ht]
  \centering
  \includegraphics[width=0.6\linewidth]{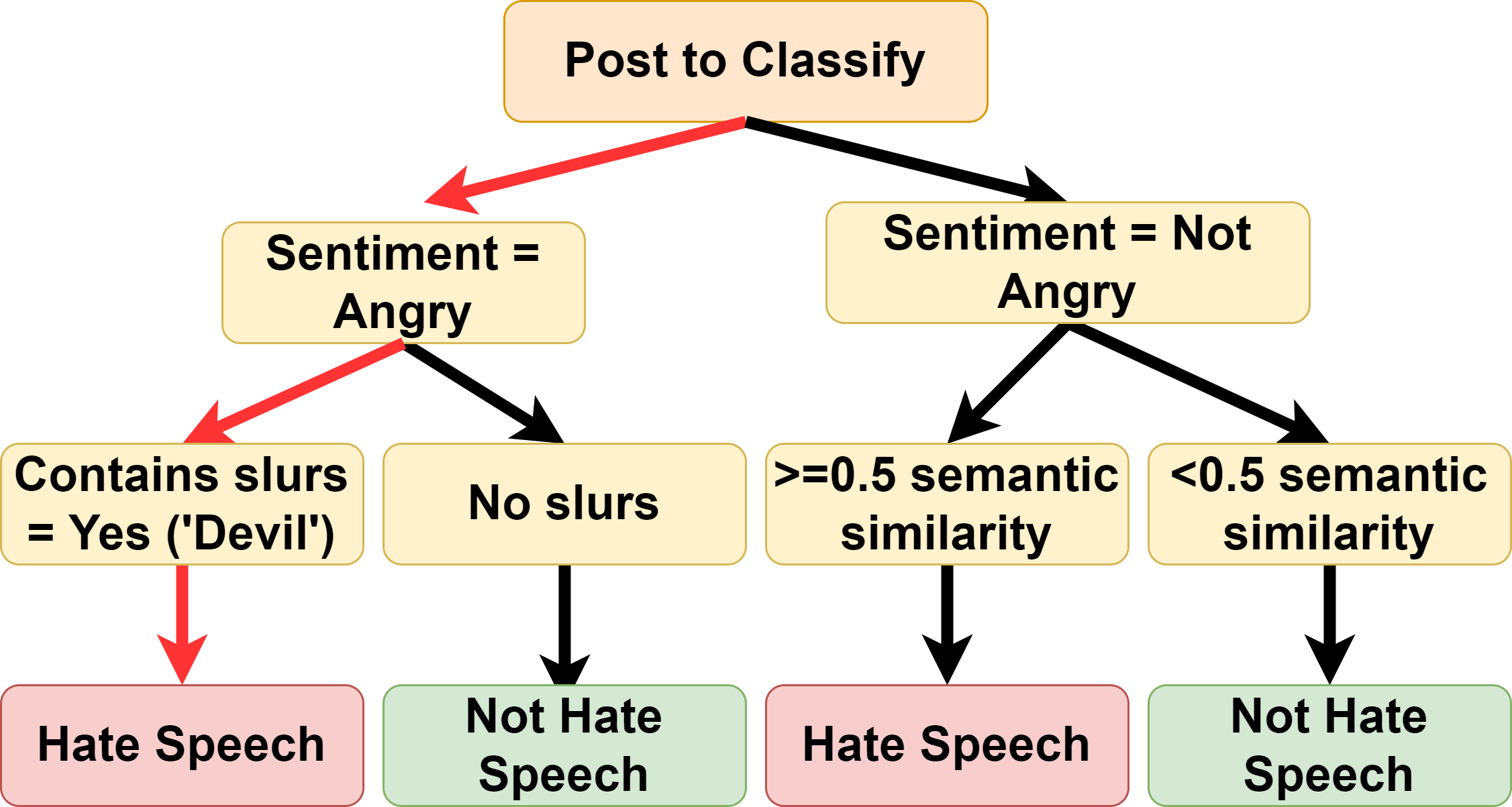}
  \caption{An example of a decision tree, with the leaf nodes constituting the classification. In the Eldian hate speech example, this would require traversing the left branches recursively for the final \textquote{Hate Speech} classification leaf (shown via the red arrows).}
  \label{decision-tree-example-figure}
\end{figure}

Chaining the correlations between features and their class likelihood can also span a tree of scenarios. If an annotated dataset indicates that a post is 80\% likely to be racist if a sentiment-scoring algorithm detects anger, then a binary decision emerges---if post contains angry words, then likely hate speech; if not, then not hate speech. These rules construct decision trees, where the root constitutes the instance (text, network, metadata, or image), and each node is a decision, with the leaves (final node) being the expected class value (i.e., the classification)~\cite{data_mining_book_eibe}. Hence, decision trees are not naïve as they rely on specific values of other features when traversing a tree’s branches for a prediction.

Creating an optimal tree that maximises accuracy and precision is not trivial due to the feature explosion of possible rules and tree nodes. Hence, Random Forest classifiers rely on a divide-and-conquer algorithm for generalising feature pairings into class classifications with a random initialisation~\cite{random_forest_original_study}. This recursive process requires finding optimal splits to maximise the separation of classes for a final leaf, with an example tree presented in Figure~\ref{decision-tree-example-figure}---where a random forest would consists of multiple trees as a \textit{forest}. Ideally, a leaf node should encapsulate instances of one class.

\textit{Random forests} generate multiple decision trees and select the final prediction based on the predictions from the majority of decision trees. Utilising multiple trees with a random initial tree state increases the range of features and values selected during the training step. Utilising multiple trees and testing the models on untrained \textquote{test} data minimises the risk of over-fitting to the training (i.e., a classifier which performs reliably on the training dataset but not on real-world data).

Random forests strengths include its ability to tie dependent and complex features while reducing over-fitting through pruning (i.e., reducing tree size to generalise the model). Hence, decision trees capture related concepts in hate speech where naïve BoW approaches do not---such as the appearance of anger/negative sentiment invoking the use of charged terms (e.g., racism as an emotional outlet) or frequency of posts and sentiment.

\subsubsection{Support Vector Machines (SVM)}\hfill\\
\begin{figure}[ht]
  \centering
  \includegraphics[width=0.6\linewidth]{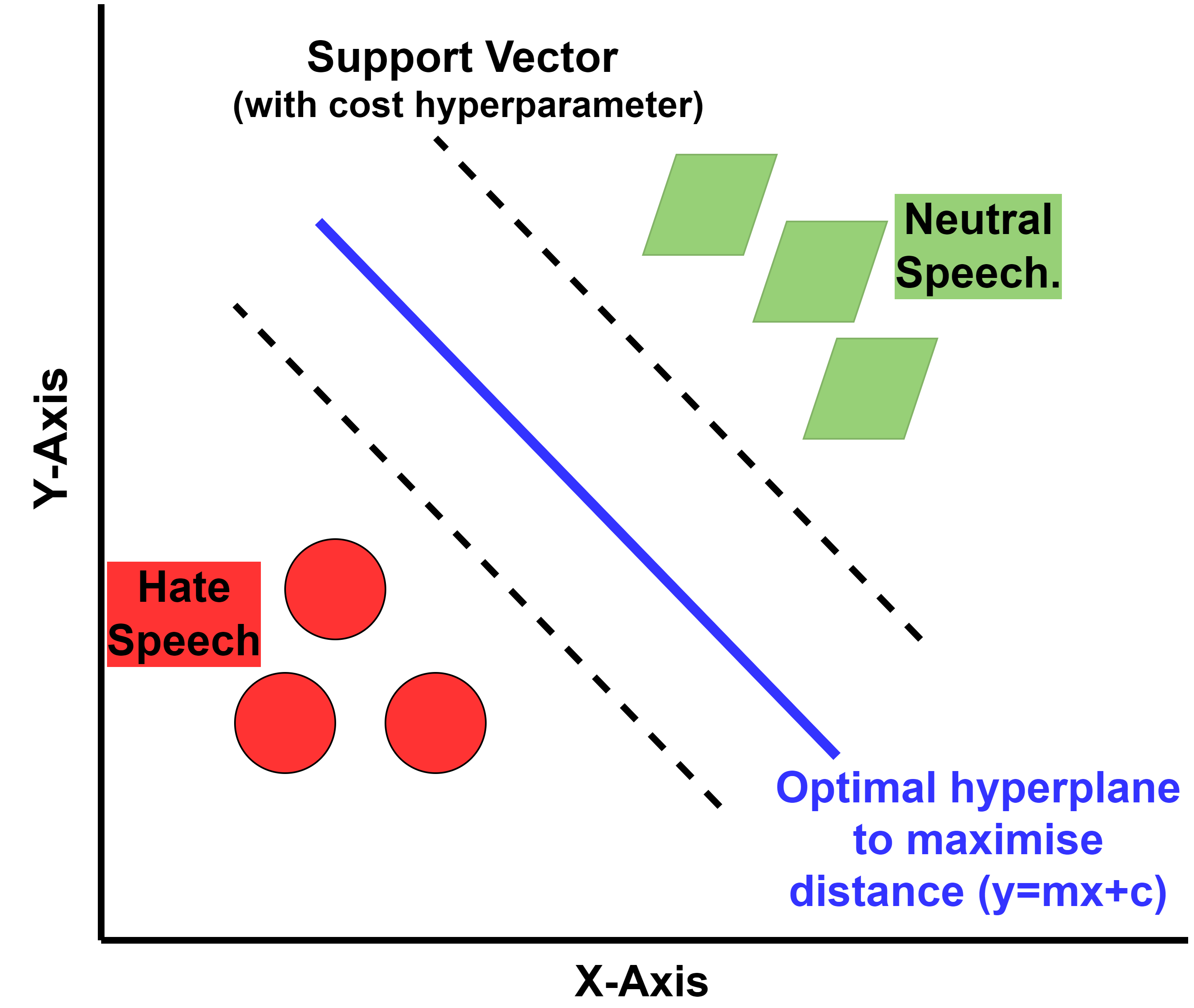}
  \caption{Support Vector Machine where instances beyond the boundaries (support vectors) are automatically assigned to the class.}
  \label{svm-example-figure}
\end{figure}

SVMs are another supervised learning model for classification and regression tasks, seeking to map instances in vector spaces to maximise the distance between classes~\cite{svm_original_study}, visualised in Figure~\ref{svm-example-figure}. Mapping features to multidimensional vectors can exponentially increase dimensions (an issue shared in deep-learning models). Thus, SVMs reduce irrelevant features through specific kernels---typically a linear, polynomial, Gaussian or sigmoid function. These kernels reduce the feature set to draw boundaries between two classes, similar to logistic regression. These boundaries are either hard (i.e., a binary classification) or soft---allowing outliers near the boundary for edge cases, like niche controversial and offensive, but not ostensibly targeting protected characteristics. SVM models are computationally faster and reduce memory compared to deep learning models~\cite{wullah21_deep_GPT_BERT, araque20_radical_emotion_signals_similarity}, while achieving comparative performance outlined in RQ4. Dimensionality reduction techniques can also reduce runtime by reducing the complexity of large feature spaces from textual or network data, such as via Principle Component Analysis~\cite{kmeans_pca}.

SVMs are the consistently highest performing MLAs per RQ4, while lowest complexity, with \(O(m * n)\) complexity for a Linear Kernel SVC---where m = feature count, and n = number of instances.

\subsubsection{Clustering and Nearest Neighbour Classifiers}\hfill\\
Instead of annotated hate speech datasets, clustering methods group by textual similarity via \textit{Natural Language Processing} (NLP), and network relations via \textit{Community detection}. Hence, clustering can work in cases of fully annotated datasets as supervised learning, semi-annotated datasets as semi-supervised learning, or unlabelled raw web scrapped data for unsupervised learning.

For supervised learning, \textit{K-Nearest Neighbour} (KNN) classifiers work via evaluating the nearest neighbours' likeliness when projecting the textual, network, or metadata features onto a multidimensional space~\cite{knn_original_study}. The \textquote{distance} between feature spaces typically rely on Euclidean, Manhattan, or Minkowski distance---where the latter two are suited for non-linear feature spaces. Non-euclidean distances are ideal where dimensions are not comparable, as Manhatten distance reduces noise/errors from outliers since the gradient has a constant magnitude.

Clustering examples for hate speech detection includes K-Means, which partitions n observations into k clusters~\cite{kmeans_pca}. K-Means automatically generates clusters, thus does not require annotated datasets. Hence, K-Means can detect novel groups, including emergent extremist organisations, or influential individuals~\cite{bartal20_roles_trolls_affiliation}. Unsupervised clustering’s strength for ERH detection is how it circumvents the definition issues for annotating data and can cluster large movements without costly annotation. However, K-Means may not identify manifestly hateful posts, as it does not abide by any standard imbued within strict annotation criteria. Evidently, in the cross examination of a naïve approach vs their proposed K-Means derived model by Moussaouri et al.~\cite{Moussaoui19_twitter_terrorism_communities}, the naïve approach outperformed the possibilistic clustering by 0.07-0.14 for accuracy 0.04-0.05 for precision.

\subsection{Definitions for Deep Learning Approaches}
\begin{figure}[!ht]
  \centering
  \includegraphics[width=0.74\linewidth]{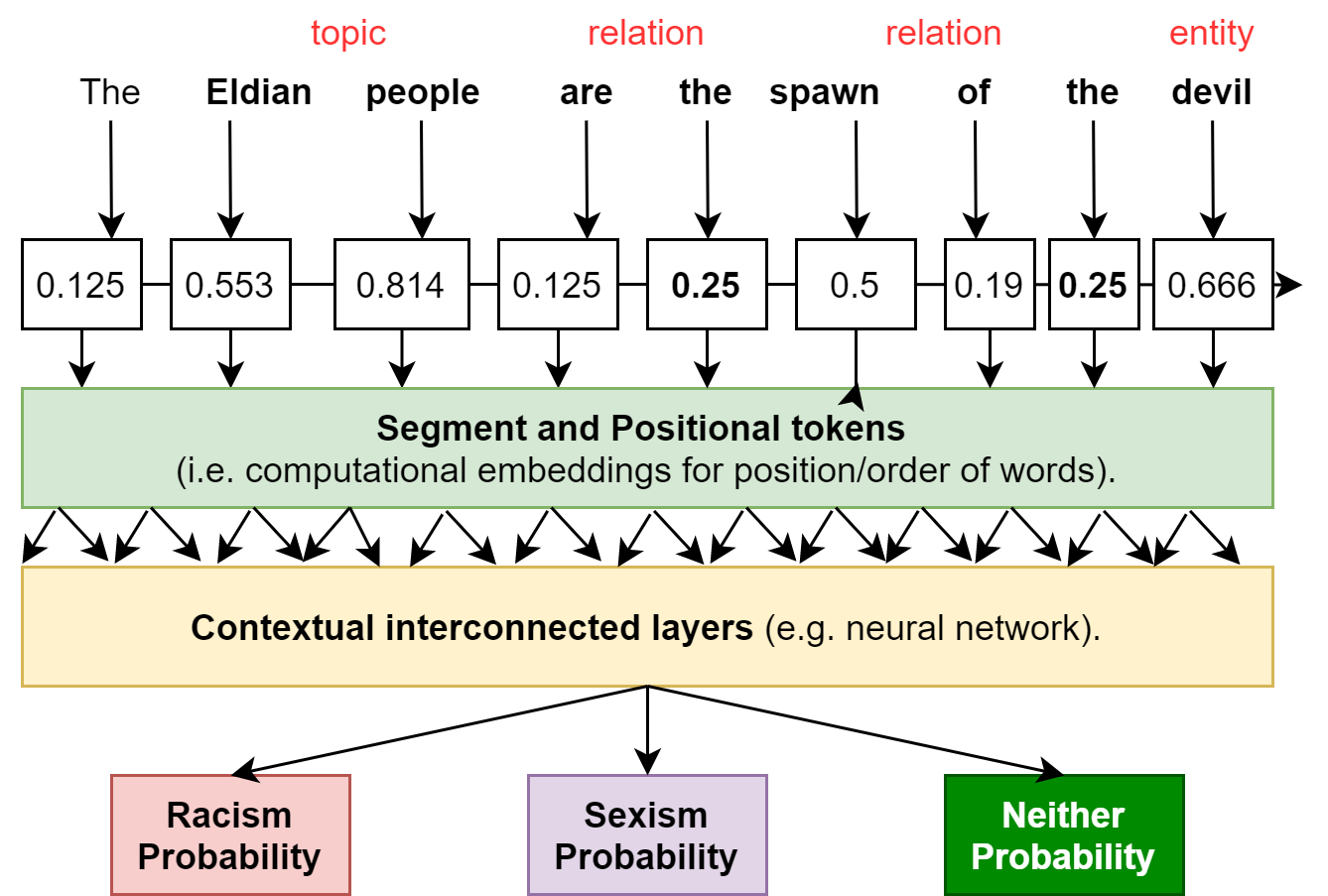}
  \caption{An abstracted example of a neural network for text. The top text represents its raw syntactic form, with its converted numeric embedding representation. These embeddings are responsible for altering the weights to increase token prediction or generation (for transformers) via backpropagation. The final output layer for this example would offer the probability that the given text is racist, sexist, or benign.}
  \label{context-sensitive-dnn-diagram}
\end{figure}

Deep learning represents a family of machine learning algorithms with multiple layers and complexity, typically via neural network architectures. Neural networks rely on training a network with a set of weights at each layer, known as \textit{neurons}. The first layer of a neural network utilises numeric representation of an instance (e.g., hateful text) in numeric \textquote{tokenised} form, which is adjusted throughout the hidden lower layers towards a final output (typically) classification layer. Each downwards training step results in readjusting the weights of the upper layers for the neurons---known as \textit{backpropagation}~\cite{backprop_original_study}. Figure~\ref{context-sensitive-dnn-diagram} displays this architecture for neural networks per our example. The benefit of DLAs in ERH detection is the preservation of word order and meaning (e.g., “I ran” vs “I ran for president”), thus displaying context-\textit{sensitivity}. Given dual-use words such as \textquote{queer}, or racially motivated slurs, understanding the surrounding contextual words is essential to reduce bias via misclassifications~\cite{mozafari20racialbias_hs_study}. DLAs dominant the benchmark dataset leader-board in RQ4.

\subsubsection{Convolutional Neural Networks (CNN)}\hfill\\
Convolutional Neural Networks (CNNs) expand on the neural network model through a convolutional layer---which acts as a learnable filter for textual or image embeddings~\cite{data_mining_book_eibe}. Moreover, CNNs include a pooling layer(s) to reduce the spatial complexity of the network's features. Reducing spatial size helps reduce the number of parameters and thus training time and memory footprint, while reducing over-fitting by generalising patterns in the training data.

\subsubsection{Long Short-Term Memory (LSTM) and Gated Recurrent Unit (GRU)}\hfill\\
LSTM and GRU aim to increase contextual awareness to process data sequences with long-term gradients to retain information on prior tokens~\cite{lstm_original_study, gru_original_study}. LSTM and GRU seek to reduce the vanishing gradients caused during backpropagation steps, which reduces classification performance as older trained instances are effectively \textquote{forgotten} due to later weight changes. Similarly, GRU’s are a gating approach with fewer parameters and thus higher runtime, enabling larger neural networks overall. CNN models with LSTM and GRU connections outperform CNNs on their own for hate speech detection~\cite{naseem19_lstm_abusive_hs_twitter, kapil20_deep_nn_multitask_learning_hs, kumar_etal18_trac_benchmark_dataset}. The highest performing BiLSTM model expands LSTM for bidirectional input, via two LSTMs---where tokens in the network utilise information from past (backwards) tokens/data and future (forwards) data~\cite{naseem19_lstm_abusive_hs_twitter}. The ability to uphold the temporal memory of prior tokens (attention) constitutes a Recurrent Neural Network (RNN).

\subsection{Language Transformer Models}
The state-of-the-art transformer architecture relies on \textit{self-attention}---the memory retention of neural networks where each token of a sequence is differentially weighted~\cite{bert_original_study, brown2020language}. Unlike Recurrent Neural Networks (i.e., neural networks where nodes follow a temporal sequence), a transformer's attention mechanism utilises context for any position for the token sequence. Hence, transformers can handle words out of order to increase understanding. Transformers offer greater classification performance (see RQ4) at the expense of memory and computational overhead. A considerable ethical threat of transformer models is their capability to predict future tokens (i.e., text generation). For instance, a malicious actor could create realistic automated trolls or radicalising synthetic agents as bots. Language models also risk data leakage of their trained data through predicting tokens found in the original trained dataset, such as names or addresses~\cite{brown2020language}.

\subsubsection{Cross-encoders (e.g., BERT)}\hfill\\
Bidirectional Encoding Representations from Transformers (BERT) is the most common cross-encoder observed for ERH detection~\cite{bert_original_study}, with the highest performance of all NLP models. Cross-encoders offer higher performance for classification tasks, through retaining information over a given sequence with a label (i.e., self-attention). BERT’s core strength is its memory retention of all tokens in a sentence, thus upholding full context-sensitivity of every word in the post it seeks to classify. However, cross-encoders are computationally expensive due to high parameter counts (110 million parameters for BERT-base, 365 million for BERT-large), an issue further outlined in RQ4. Hence, an area of ongoing research includes model distillation (optimising and reducing parameter count to reduce memory requirements and training time), specialised training datasets, and alternate layers~\cite{roberta_original_study, sbert_original_study, bigbird_original_study}. BERT is pre-trained on entries from English Wikipedia (2.5 million words) and the English BookCorpus (800 million words)~\cite{ bert_original_study}. Hence, such pre-trained models are then fine-tuned on a smaller dataset (typically 1000+ instances, per RQ2’s benchmark datasets) to optimise the BERT weights to detect hate speech with the context of its pre-trained corpus.

\subsubsection{Generative Pre-trained Models (GPT)}\hfill\\
Similarly, the state-of-the-art GPT transformer architecture expands on the encoder blocks (shared with BERT) to include decoder blocks~\cite{ brown2020language}. Hence, GPT works on a single token (i.e., word vector) and produces estimates for the sequence’s next token---ideal for tasks such as text generation, summarising, question answering, and information retrieval.

GPT models differ from BERT-based models via \textit{masked self-attention}---an alternate form of context-sensitivity where the model only knows the context of the prior words in the sentence. GPT-2/3~\cite{brown2020language}, GPT-Neo~\cite{gpt_neo_original_software}, and Jurassic-1~\cite{jurassic1_original_white_paper}, are notable 2019-2021 era multi-billion parameter models---where larger datasets and parameter count result in more human-like text generation and higher performance in information retrieval tasks~\cite{brown2020language}.

GPT’s core strength in ERH detection synthetic hate speech generation via a GPT model fine-tuned on a hateful corpus---as investigated by Wullah et al. (see RQ3)~\cite{ wullah21_deep_GPT_BERT}. However, state-of-the-art GPT models utilise up to 178B parameters, whereby memory and computational requirements scale linearly. Hence, future GPT work in synthetic text generation should consider inference tasks over fine-tuning. Specifically, inference utilises a pre-trained model’s on-demand text generation capability through prompts rather than altering each of the billions of weights. Using the auto-complete-like inference capabilities for generating realistic synthetic hate speech posts constitutes a novel case of prompt engineering in ERH detection and thus is a potential future research project.

\subsection{Definitions for Prominent Feature Extraction Techniques}
This subsection outlines the three most common feature extraction techniques used for textual ERH detection---as outlined in RQ2 in the SLR. These models seek to identify hateful lexicons from text, or create numerical representations for word or sentence meaning via embeddings. We deconstruct the six most common feature extraction techniques observed in our SLR.

\subsubsection{Word2Vec}\hfill\\
Word2Vec is a model to convert words into vector embeddings, which compares synonymous words (e.g., \textquote{hate} and \textquote{disgust}) via numerical vectors~\cite{mikolov2013_word2vec}. Word2Vec compares these word-to-vector embeddings via \textit{semantic similarity} by evaluating their cosine similarity between their vectors (e.g., comparing word vectors of an unknown class instance to words from a known \textquote{hate speech} instance(s) to make a \textquote{hate or not} classification). On a word-level basis, the vector value for \textquote{king} - value for \textit{man} + value for \textit{woman} would result in a vector similar to \textit{queen}~\cite{mikolov2013_word2vec}. In our case, a \textquote{Islamist extremist} and \textquote{ISIS} are semantically similar akin to \textquote{White Supremacy} and \textquote{Nazism}.

\subsubsection{Doc2Vec}\hfill\\
Similar to Word2Vec, Doc2Vec aggregates vector embeddings for \textit{paragraphs} in addition to individual words~\cite{doc2vec}. Thereby offering memory of the current context and paragraph's topic---useful for understanding a whole post's sentiment and meaning.

\subsubsection{N-grams}\hfill\\
N-grams represent contiguous sequences of n-number of characters for frequency analysis given their non-linear distribution in English, as well as when comparing a radical vs non-radical corpus~\cite{data_mining_book_eibe}. This linguistic model is often paired with methods such as TF-IDF or BoW.

\subsubsection{Term Frequency-Inverse Document Frequency (TF-IDF)}\hfill\\
TF-IDF determines the relevance of a word in a document by comparing its frequency \textit{in the document} compared to its inverse number for the frequency of that word \textit{across all documents}~\cite{data_mining_book_eibe}. Thereby, assigning each word a weight to signify its semantic importance compared to the wider corpus. For instance, radical Islamist \textit{dog-whistle terms} (i.e., coded or suggestive political messages intended to support a group) appeared disproportionately in extremist text compared to a neutral religious corpus~\cite{rehman21_language_of_isis_twitter}.

\subsubsection{SenticNet}\hfill\\
SenticNet embeds pattern matching, parser trees, and LSTM-CNN models for sentiment analysis, with the aim to replace a naïve BoW approach within a proclaimed bag of concepts and narratives~\cite{senticnet18}. Specifically, it includes feature extraction methods of concept parsing (i.e., understanding linguistic patterns in natural language into conceptual pairs), subjectivity and polarity inference, alongside personality and emotion extraction.

\subsubsection{Global Vectors for Word Representation (GloVe)}\hfill\\
GloVe offers an unsupervised learning algorithm for context-independent word-to-vector embeddings~\cite{pennington-etal-2014-glove}. While similar in creating vectors akin to Word2Vec, GloVe instead establishes word co-occurrences using matrix factorization (i.e., co-occurrence matrix of word [row] and context [usage of the word in the document]) and dimensionality reduction techniques.

\section{SLR Design Considerations}
This supplementary material section outlines the additional criteria and considerations for selecting papers and ensuring privacy-protections for users, groups and collected data. In essence, this section offers a meta-analysis of the ethics and selection process used throughout the SLR.

\subsection{Quality Assessment Criteria}
The following includes our paper inclusion quality check criteria---with a score of 13 or higher required for inclusion in the final paper selection (i.e., final 51 papers included).

We propose a critical criteria for quality assessment to filter irrelevant or ambiguous studies. Specifically, for a study that passed a title and abstract screen, we assess the study's clarity for ERH definitions and annotations (for objective and legible classifications), methodical clarity (i.e. outlining each study's algorithmic model, methods, data collection processes, and statistical analysis/evaluation methods), and socio-technical considerations. We weighted each quality assessment section to prioritise their research methodology and clarity in their technical methods over their \textit{Conceptual Quality} for studies encompassing broader socio-technical issues such as ethics, legality, or ERH clarity. After a ten paper pilot study, we selected a score threshold of 65\% to exclude irrelevant or ambiguous studies. Our supplementary material document includes the criteria and scoring for our quality assessment rubric.

\subsubsection{Computational Quality (0 = None, 1 = Partial, 2 = Full)}
\begin{enumerate}
\item Is the radicalisation/affiliation detection model clearly defined?
\item Is the radicalisation/affiliation detection model’s algorithm clearly defined?
\item Is the training data reputable?
\item Are the models results compared to similar state-of-the-art methods?
\item Is the methodology for designing and conducting their experiment clearly defined?
\item Are patterns and trends discussed and presented clearly?
\end{enumerate}
\subsubsection{Epistemological Quality (0 = None, 1 = Partial, 2 = Full)}
\begin{enumerate}
\item Does the source(s) (data or researchers) avoid any conflict of interests or expressed biases? (i.e., explicit support/funding from a political think tank or state agency).
\item Does the study provide a cited or evidence-based definition for “radicalisation”, “hate speech” or "extremist" affiliation?
\item Are the dataset annotations vetted by more than one annotator to reduce bias?
\end{enumerate}
\subsubsection{Conceptual Quality (0 or 0.5 value, as not critical but useful)}
\begin{enumerate}
\item Does the study discuss social or ethical issues in ERH detection (e.g. censorship)?
\item Do the authors discuss the legality of  their model or definitions?
\item Does the study evaluate its model across multiple social media platforms?
\item Does the study discuss regulatory frameworks or recommendations for social media platforms based on their findings? 
\end{enumerate}

\subsubsection{Researcher Ethics}\hfill\\
We focus on key terms and compositions of ERH examples to protect the privacy of the individuals exposed, as recommended by meta-studies on extremism research ethics~\cite{buchanan17_ethics_critique_twitter_study, conway21_extremism_terrorism_research_ethics_study, marwick16_data_society_risky_researchers_practices}. When linking ERH detection to real-world groups and events, we solely focus on events and organisations which resulted in media attention or criminal convictions. In no part during this SLR did we attempt to track users, groups, or correlate online users to any personally identifiable information (name, location, username etc.) given the ease of composing online data into a traceable online fingerprint.

Similar to the social norms in New Zealand in the aftermath of the Christchurch shooting, no extremists, terrorists, and/or criminals are referred by name to minimise publicity. We recognise the potential for political or cultural bias in this charged field by citing international non-partisan Non-governmental Organisations when framing ERH concepts, and avoid searching any party or ideology in our search strategy. Moreover, we encourage that our findings and recommendations invoke an open debate among social media platforms, governments, and the wider public. However, we do not condone the use of ERH detection in social media as a form of autonomous law. We recommend human-in-the-loop processes when handling or classifying data via independent reviews, privacy protections, and complaint and redress mechanisms for deployed models.

Our recommendations thereby focus on Open Source Intelligence (OSINT) oriented studies that do not consider governmental or private-conversation surveillance (with the exception of one hybridised study that appeared in our search~\cite{hung16_fbi_radical}). We thereby consider ERH detection as a \textit{computational} method aimed at garnering community-insights, trends, and flagging for \textit{social media platforms themselves} to use. Whether ERH detection \textit{policies} should encourage deplatforming, deranking, demonetisation, fact-checking, or targeted counter-speech/prevention programs require further research. We encourage open interdisciplinary research in public and private-communications---particularly ethical and legal discussions.

\section{The Case for Performance Engineering when Evaluating Models}
While high F1-scores help enforce community guidelines via accurate predictions and reduce injurious censorship from false positives, runtime performance trade-offs are seldom discussed. DLAs may perform within 1\% (F1-score) of their MLA counterparts in NLP studies but require significantly higher computational resources. For instance, fine-tuning a BERT-large model for NLP tasks requires Graphics or Tensor Processing Units (GPU or TPU), restricting researchers from testing large language models~\cite{zhu19_offensive_tweets_bert, wullah21_deep_GPT_BERT}. For community detection, uncompressed network models can include up to 27.4 million links~\cite{benigni2017online_extremism_sustain_it_isis}, which significantly increases computational and memory requirements for a minimal 1-5\% performance gain. Specifically, using a Possibilistic Approach (PA) with dimensionality reduction reduced subgraph mining runtime by up to 67\% (1500 seconds to ~500 seconds on an 8-core 3.2GHz system), while reducing accuracy by only ~4\%~\cite{Moussaoui19_twitter_terrorism_communities}. Furthermore, \textit{community-level insights} on topics with millions of tweets, relations, and discussions can lead to a network \textit{explosion} with a non-deterministic polynomial runtime~\cite{Moussaoui19_twitter_terrorism_communities, benigni18communitymining_UNSUPERVISED}. In graph-detection approaches, performance engineering and optimisation for mining frequent subgraphs and graph-traversal is an active area of research~\cite{Moussaoui19_twitter_terrorism_communities}). No NLP studies consider performance engineering for DLAs despite developments in model distillation and sentence-level embeddings~\cite{sbert_original_study}.

Thus, we recommend that researchers consider performance trade-offs in future work and investigate a possible standardised performance-complexity metric (e.g. parameter count vs. F1-score ratio) to build scalable, energy-efficient and fiscally-viable models. Moreover, fine-tuning or retraining DLAs, or regenerating frequent subgraphs for community detection, should be a frequent endeavour to adapt to the rapidly evolving topics, entities, and events throughout online discourse. Due to the computational costs of fine-tuning or training multi-billion parameter models, we recommend approaches that do not require expensive training, such as few-shot learning (i.e., giving several known instances of ERH and a unseen \textquote{test} instance) and prompt engineering~\cite{brown2020language}.

\section{Uptake Roadmap Expanded}
This supplementary section expands on the dataset and model research gaps highlighted in Figure 16 of the main \textit{Down the Rabbit Hole} SLR document. We categorise these research recommendations into eight core components for our proposed \textit{ERH Context Mining} research field.
\begin{figure}[b]
  \centering
  \includegraphics[width=0.8\linewidth]{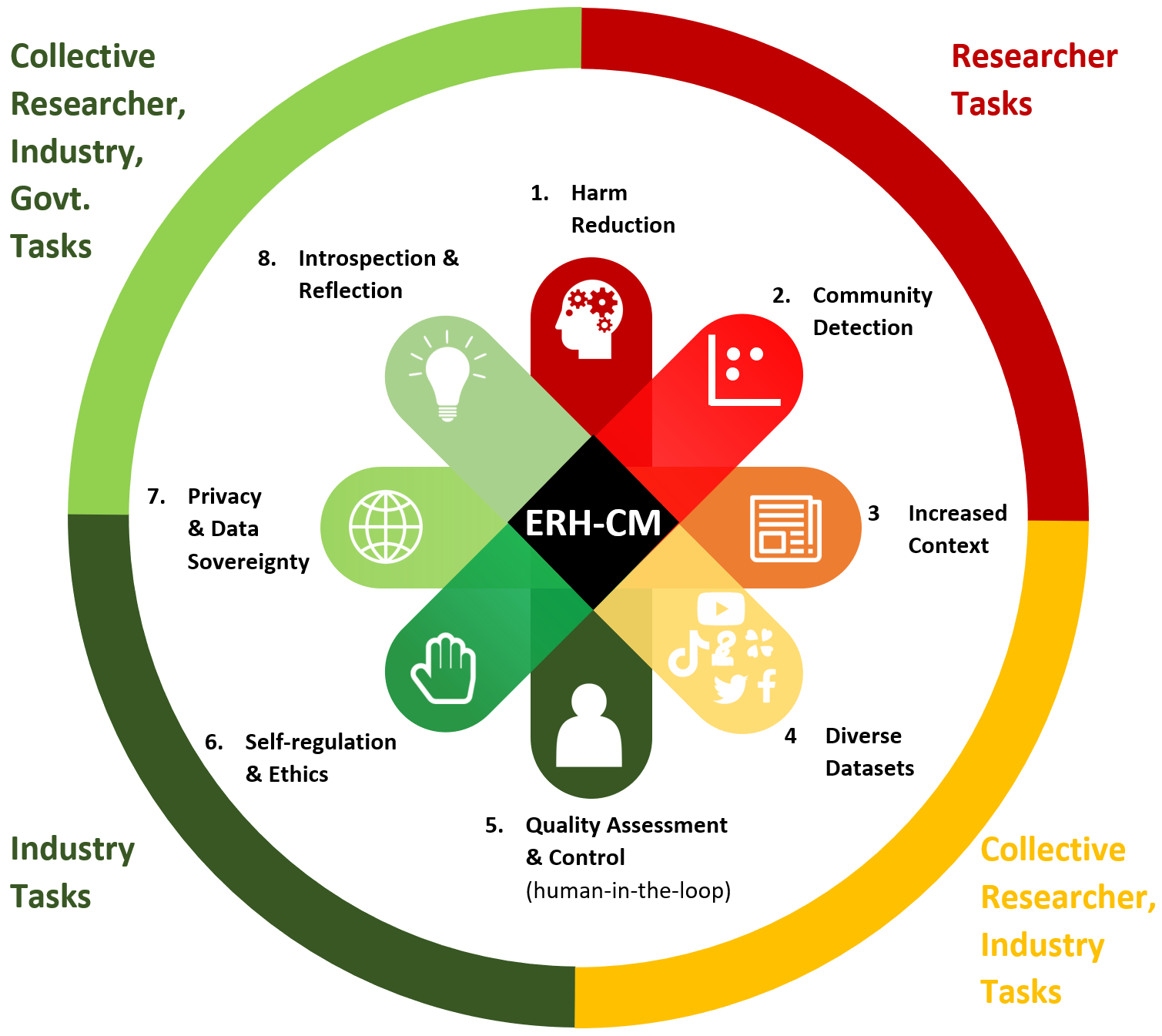}
  \caption{ERH Context Mining (ERH-CM) eight core components for Research, Industry, and Government.}
        \label{ERH_Context_Mining_Components_Figure}
\end{figure}

\subsection{Model Recommendations}
The two predominant recommendations for future work are investigating the role of \textit{changes in hateful affiliation or speech over time} to satisfy the temporal requirement for \textit{Radicalisation} detection, and to train models on multiclass datasets from multiple platforms. We note that only one study considered temporal data on both meso and macro (changes within and between groups), and micro (individual) levels, although recommended as future work within four other studies~\cite{scrievens18_signs_of_extremism_SIRA, araque20_radical_emotion_signals_similarity, hung16_fbi_radical, chandrasekharan17bag_of_com}. Moreover, we recommend expanding on DLAs as the target for future research based on their leading performance in RQ4. Neural language models offer a macro-level societal understanding due to their pre-trained corpus on academic sources, OpenWebText2 Reddit discussions, and Wikipedia~\cite{brown2020language}. Furthermore, transformer models beyond 764 million parameters are untested.

Bot, troll, meme, entity, dis/misinformation and satire detection remain underdeveloped–--which could lead to censorship or undermine democratic institutions. Five studies recommended multimedia detection as future work~\cite{chandrasekharan17bag_of_com, Preotiuc-Pietro_pol_ideology_twitter, abubakar19_lstm_gru_comparative_multiclass, benigni18communitymining_UNSUPERVISED, hall20_machines_unified_understanding_radicalizing}.

To protect user privacy from recreating user content from neural language models, we encourage \textit{privacy-by-design} software engineering through machine learning paradigms such as Differential Privacy (DP). DP-paradigm models and datasets reduce the potential for self-identification from trained models (i.e. data leakage, such as names or usernames in open-source datasets), as DP-paradigm models use pseudo-anonymised \textit{patterns} of groups and hate.

\subsection{Dataset Recommendations}
To investigate the roles of radicalisation, we recommend expanding on the dataset annotation approach by de Gibert et al.~\cite{de-gibert18-white-supremacy-dataset} by creating a \textit{conversation-level} dataset with public non-hateful replies to a post for context. Moreover, future benchmark datasets should consider pulling data across platforms to investigate macro-level radicalisation trends between platforms. We note that only two studies considered anti-Asian sentiment in COVID-related tweets, targeting a seldom explored topic and demographic~\cite{jia21_covid19_hs_ensemble, rustam21_covid_extremes} worthy of expansion given the ongoing COVID-19 pandemic.

Likewise, future datasets should consider the role of indigenous discussions and potential researcher biases given the Anglo-dominant field of ERH research. Given the rise of COVID extremism~\cite{covid_extremism_study}, far-right movements, and xenophobia in Oceania. Hence, we recommend geotargeted datasets to consider the differences for investigating ERH topics, which would demonstrate NZ's commitment to our Christchurch Call to Action Summit. Investigating unexplored and minority groups could also provide imperative insights for social scientists regarding the conversational dynamics, morphological mapping, and ideological isomorphism from radical minority groups towards the majority. Likewise, research on vulnerable communities (youth, gender and sexual minorities, religious, racial, and geographically distant peoples) would aid social media platforms in both identifying unique radicalising risks, as well as avenues for support and de-escalation. In the mental health end, we recommend building on Nouh et al.’s proposed approach of extracting textual, psychological and behavioural features~\cite{Nouh19_understanding_radical_mind}, both due to its performance, as well as its potential for analysing societal \textit{factors and ERH roots} such as correlations between mental health issues (isolation, depression etc.) and vulnerability to radicalisation towards violent extremism.

For any counter-extremism or de-radicalisation studies, we recommend work in ethical and legal guidelines to protect privacy, avoid backlash or inadvertent algorithmic amplification.

Investigating posts from periods of political, or social crisis (e.g., COVID health measures, post-terror attack discourse etc.) could also help identify cases of ERH on mainstream platforms before they are deplatformed/removed. Event-based datasets would provide unique sociological insights on the role of societal stress and emergencies on the human psyche and online group dynamics.

To reduce the cost, variability in inter-annotator agreement, and psychological impact of human annotation, we recommend unsupervised clustering-based research and propose using synthetic conversational agents to simulate extremist discourse. Simulating online radicalisation in a closed environment would present a safe, ethical, and non-invasive method to build benchmark datasets.

\bibliographystyle{ACM-Reference-Format}
\bibliography{Supplementary-Main}